\documentclass[twocolumn]{aastex631}  
\usepackage{graphicx}
\usepackage{txfonts}
\usepackage{xcolor}
   \newcommand{\rah}{$^{\mbox{\tiny h}}$}
   \newcommand{\ram}{$^{\mbox{\tiny m}}$}
   \newcommand{\ras}{$^{\mbox{\tiny s}}$}
   \newcommand{\decd}{$^{\circ}$}
   \newcommand{\decm}{$'$}
   \newcommand{\decs}{$''$}
   
   \newcommand{\beam}{$\theta_{\mbox{\tiny maj}}\times\theta_{\mbox{\tiny min}}$}
   
   \newcommand{\ujyperbeam}{$\mu$Jy\,beam$^{-1}$}

\shorttitle{A Multiwavelength Study of ELAN Environments (AMUSE$^2$)}
\shortauthors{Arrigoni Battaia et al.}

\begin{document} 

   \title{A Multiwavelength Study of ELAN Environments (AMUSE$^2$). Mass budget, satellites spin alignment and gas infall in a massive $z\sim3$ quasar host halo}

\correspondingauthor{Fabrizio Arrigoni Battaia}
\email{arrigoni@mpa-garching.mpg.de}

\author{Fabrizio Arrigoni Battaia}
\affiliation{Max-Planck-Institut fur Astrophysik, 
Karl-Schwarzschild-Str 1, 
D-85748 Garching bei M\"unchen, Germany}

\author{Chian-Chou Chen}
\affiliation{Academia Sinica Institute of Astronomy and Astrophysics, 
No.1, Section 4, Roosevelt Rd., Taipei 10617, Taiwan}

\author{Hau-Yu Baobab Liu}
\affiliation{Academia Sinica Institute of Astronomy and Astrophysics, 
No.1, Section 4, Roosevelt Rd., Taipei 10617, Taiwan}

\author{Carlos De Breuck}
\affiliation{European Southern Observatory, 
Karl-Schwarzschild-Stra{\ss}e 2, 
D-85748 Garching-bei-M\"unchen, Germany}

\author{Maud Galametz}
\affiliation{Astrophysics department, CEA/DRF/IRFU/DAp, 
Universit{\'e} Paris Saclay, UMR AI, 
91191 Gif-sur-Yvette, France}

\author{Michele Fumagalli}
\affiliation{Dipartimento di Fisica G. Occhialini, 
Universit\`a degli Studi di Milano Bicocca, 
Piazza della Scienza 3, 20126 Milano, Italy}

\author{Yujin Yang}
\affiliation{Korea Astronomy and Space Science Institute, 
776 Daedeokdae-ro, Yuseong-gu, 
Daejeon 34055, Republic of Korea}

\author{Anita Zanella}
\affiliation{INAF -- Osservatorio Astronomico di Padova, 
Vicolo dell'Osservatorio 5, I-35122, Padova, Italy}

\author{Allison Man}
\affiliation{Dunlap Institute for Astronomy and Astrophysics, University of Toronto, 
50 St George Street, Toronto ON, M5S 3H4, Canada}
\affiliation{Department of Physics \& Astronomy, University of British Columbia, 
6224 Agricultural Road, Vancouver, BC V6T 1Z1, Canada}

\author{Aura Obreja}
\affiliation{University Observatory Munich, 
Scheinerstra{\ss}e 1, D-81679 Munich, Germany}

\author{J. Xavier Prochaska}
\affiliation{Department of Astronomy \& Astrophysics, UCO/Lick Observatory, 
University of California, 1156 High Street, Santa Cruz, CA 95064, USA}
\affiliation{Kavli Institute for the Physics and Mathematics of the Universe (Kavli IPMU), 
5-1-5 Kashiwanoha, Kashiwa, 277-8583, Japan}

\author{Eduardo Ba\~nados}
\affiliation{Max-Planck Institute for Astronomy, 
K\"onigstuhl 17, D-69117 Heidelberg, German}

\author{Joseph F. Hennawi}
\affiliation{Department of Physics, Broida Hall, University of California, 
Santa Barbara Santa Barbara, CA 93106-9530, USA}
\affiliation{Leiden Observatory, Leiden University, 
PO Box 9513, NL-2300 RA Leiden, the Netherlands}

\author{Emanuele P. Farina} 
\affiliation{Max-Planck-Institut fur Astrophysik, 
Karl-Schwarzschild-Str 1, 
D-85748 Garching bei M\"unchen, Germany}
\affiliation{Gemini Observatory, NSF's NOIRLab, 670 N A'ohoku Place, Hilo, Hawai'i 96720, USA}

\author{Martin A. Zwaan}
\affiliation{European Southern Observatory, 
Karl-Schwarzschild-Stra{\ss}e 2, 
D-85748 Garching-bei-M\"unchen, Germany}

\author{Roberto Decarli}
\affiliation{INAF -- Osservatorio di Astrofisica e Scienza dello Spazio di Bologna, 
via Gobetti 93/3, I-40129, Bologna, Italy}

\author{Elisabeta Lusso}
\affiliation{Dipartimento di Fisica e Astronomia, Universit\`a di Firenze, 
via G. Sansone 1, 50019 Sesto Fiorentino, Firenze, Italy}
\affiliation{INAF -- Osservatorio Astrofisico di Arcetri, 
Largo Enrico Fermi 5, I-50125 Firenze, Italy}

\begin{abstract}
The systematic targeting of extended Ly$\alpha$ emission around high-redshift quasars resulted in the discovery of rare and bright Enormous Ly$\alpha$ Nebulae (ELANe) associated with multiple active galactic nuclei (AGN). We here initiate ``a multiwavelength study of ELAN environments'' (AMUSE$^2$) focusing on the ELAN around the $z\sim3$ quasar SDSS~J1040+1020, a.k.a. the Fabulous ELAN. We report on VLT/HAWK-I, APEX/LABOCA, JCMT/SCUBA-2, SMA/850$\mu$m, ALMA/CO(5-4) and 2mm observations and compare them to previously published VLT/MUSE data. The continuum and line detections enable a first estimate of the star-formation rates, dust, stellar and molecular gas masses in four objects associated with the ELAN (three AGNs and one Ly$\alpha$ emitter), confirming that the quasar host is the most star-forming (SFR~$\sim500$~M$_\odot$~yr$^{-1}$) and massive galaxy ($M_{\rm star}\sim10^{11}$~M$_{\odot}$) in the system, and thus can be assumed as central. All four embedded objects have similar molecular gas reservoirs ($M_{\rm H_2}\sim10^{10}$~M$_{\odot}$), resulting in short depletion time scales. This fact together with the estimated total dark-matter halo mass, $M_{\rm DM}=(0.8-2)\times10^{13}$~M$_{\odot}$, implies that this ELAN will evolve into a giant elliptical galaxy. Consistently, the constraint on the baryonic mass budget for the whole system indicates that the majority of baryons should reside in a massive warm-hot reservoir (up to $10^{12}$~M$_{\odot}$), needed to complete the baryons count. Additionally, we discuss signatures of gas infall on the compact objects as traced by Ly$\alpha$ radiative transfer effects and the evidence for the alignment between the satellites' spins and their directions to the central.
\end{abstract}


%

\section{Introduction}

The discovery of bright and extended Ly$\alpha$ nebulae at high redshift, detected either around  high-redshift radio galaxies (HzRGs; \citealt{MileyDeBreuck2008}) or as so-called Lyman-Alpha blobs (LABs; e.g., \citealt{matsuda04}), 
pinpoints the rarest overdensity peaks in the early Universe (e.g., \citealt{steidel00,Venemans2007,Yang2009,Yang2010,Badescu2017}). 
Indeed, HzRGs and LABs are extremely rare in the redshift range $2 < z < 5$, with number densities of a few times $10^{-8}$~Mpc$^{-3}$ (e.g., \citealt{Willott2001,Venemans2007}) and $\sim 10^{-6}-10^{-5}$~Mpc$^{-3}$ (e.g., \citealt{Yang2009}), respectively.
At these locations, in the so-called protoclusters, the formation and evolution of the progenitors of present-day ellipticals can take place thanks to violent bursts of star formation and mergers of coeval galaxies (e.g., \citealt{West1994,Kauffmann96}). 

Recently, systematic surveys of radio-quiet quasars uncovered an additional population of rare Ly$\alpha$ nebulae with observed surface brightness SB$_{\rm Ly\alpha}\gtrsim 10^{-17}$~erg~s$^{-1}$~cm$^{-2}$~arcsec$^{-2}$ on $\gtrsim 100$~kpc, maximum extents of $>250$~kpc, and total Ly$\alpha$ luminosities of $L_{\rm Ly\alpha}>10^{44}$~erg~s$^{-1}$ (\citealt{cantalupo14, hennawi+15, Cai2016, FAB2018, FAB2019}). These enormous Ly$\alpha$ nebulae (ELANe; \citealt{Cai2016}) are therefore outliers with respect to  
known nebulosities associated with radio-quiet objects. 
The current statistics show that only $4^{+3}_{-2}\%$ of relatively bright quasars ($M_{i} < -24$~AB mag) are associated with ELANe\footnote{At the moment of writing, $\approx 200$ quasars have been surveyed in the redshift range $2\lesssim z < 4$ down to similar depths able to detect ELANe, and with the specific aim of detecting extended Ly$\alpha$ emission (\citealt{cantalupo14,martin14a,hennawi+15,FAB2016,Borisova2016,FAB2019,Cai2019,Lusso2019,Husemann2018,OSullivan2020,Fossati2021}, see also discussion in \citealt{hennawi+15}).}.  
Converting the number density corresponding to the targeted quasars (e.g., \citealt{Shen2020}), this percentage translates to an ELAN number density of few times $10^{-6}$~Mpc$^{-3}$.

Interestingly, there are additional mounting lines of evidence suggesting that ELANe are located in overdense environments. Indeed, they are (i) all associated with multiple AGN, with up to four known quasars within the same structure (\citealt{hennawi+15}), (ii) frequently associated with exceptional overdensities of Ly$\alpha$ emitters on small (\citealt{FAB2018}) and on large scales (\citealt{hennawi+15,Cai2016}), and (iii) probably in fields characterized by high number counts of submillimeter sources (\citealt{FAB2018b}).
Despite these findings, a systematic study of the environment and nature of ELANe has not been conducted yet. For this reason, we initiated the project ``{\bf a} {\bf mu}ltiwavelength {\bf s}tudy of {\bf E}LAN {\bf e}nvironments'' (AMUSE$^2$) collecting datasets from the rest-frame ultraviolet out to the submillimeter regime with the specific aim of studying their astrophysics, while firmly locating these large-scale structures in the wide framework of galaxy formation and evolution.

In this paper of the series, we focus on the $z=3.164$ ELAN discovered with the Multi Unit Spectroscopic Explorer (MUSE; \citealt{Bacon2010}) around the bright quasar SDSS~J102009.99+104002.7 (hereafter QSO) by \citet{FAB2018}, {\it a.k.a.} the Fabulous ELAN. The same work reported additional four objects 
embedded in the ELAN: a faint companion quasar (QSO2), a faint obscured (type-II) AGN (AGN1) and two Ly$\alpha$ emitters (LAE1 and LAE2). The ELAN shows a coherent velocity shear of $\sim300$~km~s$^{-1}$ across its whole extent (
$\sim300$ projected kpc), which has been interpreted as the signature of inspiraling motions of accreting substructures within the bright quasar host halo (\citealt{FAB2018}). 

Here we report on our extensive campaign targeting this ELAN with VLT/HAWK-I, APEX/LABOCA, JCMT/SCUBA-2, SMA, and ALMA. Specifically, our observations target the $H$-band, 870~$\mu$m (single-dish), 850~$\mu$m (single-dish), 450~$\mu$m (single-dish), 850~$\mu$m (interferometer), 2~mm and the CO(5-4) rotational transition of the carbon monoxide, respectively. 

The paper is structured as follows. In section~\ref{sec:obs} we report on our observations and data reduction for each individual instrument/dataset. Section~\ref{sec:res} presents the observational results, quantifying the significance of the detections. The observational results allowed us to examine several aspects of the nature and astrophysics of this ELAN. 
In section~\ref{bar_budget}, we first estimate the star formation, dust, stellar and molecular gas masses,  
and infer the dark matter halo mass with two orthogonal methods.
In this way, we obtain a first-order mass budget of the whole system (section~\ref{sec:massBud}), which we use to forecast its evolution (section~\ref{sec:evo}). 
We discuss in section~\ref{sec:G_or} the evidence of alignment of the satellite spins with respect to their positional vector to the central quasar in the framework of the tidal torque theory. 
Section~\ref{sec:Mol_vs_Lya} then presents a comparison of the rotational transition CO(5-4) detected at the location of compact objects with the resonant Ly$\alpha$ line in their vicinity, discussing possible signatures of infall.  
Next, sections~\ref{sec:powering} and \ref{sec:extMol} briefly discuss the powering of the ELAN and the constraints on extended molecular gas, respectively. 
Finally, we summarize our findings in section~\ref{sec:summ}.

Throughout this paper, we adopt the cosmological parameters $H_0 = 70$~km~s$^{-1}$~Mpc$^{-1}$, $\Omega_{\rm M} =0.3$, and $\Omega_{\Lambda} =0.7$. In this cosmology, $1\arcsec$ corresponds to about 7.6 physical kpc at $z=3.164$ (redshift of the ELAN and the bright quasar from \citealt{FAB2018}). All distances reported in this work are proper.

\section{Observations and data reduction}
\label{sec:obs}

\subsection{APEX/LABOCA}

We used the Large APEX BOlometer CAmera (LABOCA; \citealt{Siringo2009}) on the APEX telescope to map a field of $\sim68$~arcmin$^{2}$ around the ELAN hosting QSO.
The 295 bolometers of LABOCA operate at an effective frequency of 345~GHz (or a wavelength of 870~$\mu$m), and the instrument is characterized by a main beam of 19$\arcsec$. 
The observations were conducted in service mode in October 2016 (ID: 098.A-0828(B); PI: F. Arrigoni Battaia) with zenith opacities between 0.2 and 0.4 at 870~$\mu$m. The field has been covered with a raster of spiral scanning mode, which optimizes the sampling of the field-of-view with the LABOCA instrument.
The total integration time on source resulted in 22 hours consisting of 176 scans of 7.5 minutes each.
The observations have been acquired with regular standard calibrations for pointing, focus and flux calibration (see e.g. \citealt{Siringo2009} for details).

The data reduction was performed with the Python-based BOlometer data Analysis Software package (\textsc{BoA}; \citealt{Schuller2012}) following the steps indicated in
\citet{Siringo2009} and \citet{Schuller2009}. Specifically, \textsc{BoA} processes LABOCA data including flux calibration, opacity correction, noise removal, and despiking of the timestreams.
We ran \textsc{BoA} using the default reduction script {\it reduce-map-weaksource.boa} which also filters out the low-frequency noise below 0.3~Hz.
The scans are then co-added after being variance-weighted. The final outputs are a beam-smoothed flux density and a noise maps.
The final map achieves a root-mean-square (rms) noise level of $2.6-3$~mJy~beam$^{-1}$ in its central part. We show the map for the full area covered in Appendix~\ref{app:LJmaps}.

\subsection{JCMT/SCUBA-2}

The SCUBA-2 observations for this ELAN field were conducted at JCMT during flexible observing in 
2018 February 12, and March 29 (program ID: M18AP054; PI: M. Fumagalli) under good weather conditions (band 1 and 2, $\tau_{225{\rm GHz}}\leq 0.07$).
The SCUBA-2 instrument observes simultaneously the same field at 850 and 450 $\mu$m, with an effective beam FWHM of $14.6\arcsec$ and $9.8\arcsec$, respectively (\citealt{Dempsey2013}).
The observations were performed with a Daisy pattern covering $\simeq13.7\arcmin$ in diameter, and were centered at the location of QSO (and thus the ELAN). 
To facilitate the scheduling we divided the observations in 5 scans/cycles of about 30 minutes, for a total of 2.5 hours.\\

For the data reduction we closely followed the procedures in \citet{TC2013a} and \citet{FAB2018b}. In brief, we reduced the data using the 
Dynamic Iterative Map Maker (DIMM) 
included in the Sub-Millimetre User Reduction Facility (SMURF) package from the STARLINK software (\citealt{Jenness2011,Chapin2013}). 
We adopted the standard configuration file dimmconfig\_blank\_field.lis for our science purposes. We thus reduced each scan and the MOSAIC\_JCMT\_IMAGES recipe in PICARD, 
the Pipeline for Combining and Analyzing 
Reduced Data (\citealt{Jenness2008}), was used to coadd the reduced scans into the final maps. 

We applied to these final maps a standard matched filter to increase the point source detectability, 
using the PICARD recipe SCUBA2\_MATCHED\_FILTER. We adopted the standard flux 
conversion factors (FCFs; 491 Jy pW$^{-1}$ for 450~$\mu$m and 537 Jy pW$^{-1}$ for 850~$\mu$m) 
with 10\% upward corrections for flux calibration. 
The relative calibration accuracy is shown to be stable and good to 10\% at 450~$\mu$m and 5\% at 850~$\mu$m (\citealt{Dempsey2013}). 

The final noise level at the location of the ELAN for our data is 1.01~mJy~beam$^{-1}$ and 10.97~mJy~beam$^{-1}$ at 850~$\mu$m and 450~$\mu$m, respectively.
Appendix~\ref{app:LJmaps} presents the SCUBA-2 maps for the whole field-of-view covered. In the reminder of this work we focus only on the ELAN location.

\subsection{SMA}
\label{sec:SMA}

We performed the SMA \citep{Ho2004} observations of this ELAN (Project code: 2017A-S015, PI: F. Arrigoni Battaia) on June 21 (UTC 3:00-8:30), June 27 (UTC 3:00-7:30), and July 10 (UTC 3:30-6:30) of 2017, in the compact array configuration.
However, for the observations on June 27, we only utilized the data taken after UTC 6:30 because we noticed a large antenna pointing error before then.
The atmospheric opacity at 225 GHz ($\tau_{\mbox{\tiny 225 GHz}}$) were  0.1-0.15, $\sim$0.1, and $\sim$0.05 during these three tracks of observations.

The observations were carried out in dual receiver mode by tuning the 345 GHz and 400 GHz receivers to the same observing frequencies.
These two receivers took left and right circular polarization, respectively, and covered the observing frequency 329-337 GHz in the lower sideband and 345-353 GHz in the upper sideband.
Correlations were performed by the SMA Wideband Astronomical ROACH2 Machine (SWARM) which sampled individual sidebands with 16384$\times$4 spectral channels.
The integration time was 30 seconds. 
Prior to data calibration, we binned every 16 spectral channels to reduce filesizes.
The observations on our target source cover the {\it uv} distance range of $\sim$8.5-88.5 $k\lambda$.

The target sources were observed in scans of 12 minute duration, which were bracketed by scans on the gain calibration quasar source 1058+015 with 3 minutes duration.
We observed Titan in the first two tracks, and observed Callisto in the last track for absolute flux calibrations.
We follow the standard data calibration strategy of SMA.
The application of Tsys information and the absolute flux, passband, and gain calibrations were carried out using the MIR IDL software package \citep{Qi2003}. 
The absolute flux scalings were derived by comparing the visibility amplitudes of the gain calibrators with those of the absolute flux calibrators (i.e., Titan and Callisto).
The derived and applied fluxes of 1058+015 were 2.5 Jy in the first two tracks, and 2.7 Jy in the last track.
We nominally quote the $\sim$15\% typical absolute flux calibration error of SMA.

After calibration, the zeroth-order fitting of continuum levels and the joint weighted imaging of all continuum data were performed using the Miriad software package \citep{Sault1995}. 
We performed zeroth-order multi-frequency imaging combining the upper- and lower-sideband data, to produce a sensitive continuum image at the central observing frequency (i.e., the local oscillator frequency).
Due to the different performance of the 345~GHz and 400~GHz receivers at the same observing frequency, it would be incorrect  
to treat half of the difference of the parallel hand correlations (i.e., $(LL-RR)/2$) as the thermal noise map.
Instead, we constructed the approximated noise map by first smoothing the upper-sideband image to the angular resolution of that of the lower-sideband image, and then take half of their difference.
Using natural weighting, we obtained  
a $\theta_{\mbox{\tiny maj}}$ $\times$ $\theta_{\mbox{\tiny min}}$ $=$ 2$\farcs$4 $\times$ 2$\farcs$0 (P.A.=67$^{\circ}$) synthesized beam, and a rms noise level of 1.4 mJy\,beam$^{-1}$.

\subsection{ALMA}
\label{sec:ALMA}

We performed four epochs of ALMA observations towards this ELAN (Project code: 2017.1.00560.S, PI: F. Arrigoni Battaia), on March 23, 24, 26, and 27 (UTC) of 2018 to constrain the CO(5-4) line emission ($\nu_{\rm rest}=576.267$~GHz) and its underlying $2$~mm continuum.
The pointing and phase referencing center is R.A. (J2000) = 10\rah20\ram09\ras.42, and Decl. (J2000) = 10\decd40\decm08\decs.71.
Combining all existing data yields an overall {\it uv} distance range covered of 12-740 meters.

The spectral setup of all our observations is identical.
There were two 2 GHz wide spectral windows (channel spacing 15.625 MHz) centered at the sky frequencies 149.514 GHz and 151.201 GHz, and two 1.875 GHz wide spectral windows (channel spacing 3.906 MHz) centered at the sky frequencies 137.784 GHz and 139.472 GHz. The latter two spectral windows with channel width of about 8.5~km~s$^{-1}$ are expected to encompass the CO(5-4) emission. 

For all four epochs of observations, the quasar J1058+0133 was chosen as the flux and passband calibrator.
We assume that J1058+0133 has a 3.09 Jy absolute flux and -0.46 spectral index at the reference frequency 144.493 GHz, which was based on interpolating the calibrator grid survey measurements taken in Band 3 ($\sim$91 and 103 GHz) on March 25, 2018, and in Band 7 ($\sim$343 GHz) on February 09, 2018.
Based on the results of the calibrator grid survey, we expect a nominal $\sim$10\% absolute flux error, and $\sim$0.1 in-band spectral index error as the grid survey measurements are sparsely sampled in time. 
We observed the quasar J1025+1253 approximately every 11 minutes for complex gain calibration.

We calibrated the data using the CASA software package \citep{McMullin2007} version 5.1.
The derived fluxes of the gain calibrator J1025+1253 were in the range $0.42 - 0.48$~Jy.
We fit the continuum baselines using the CASA task {\tt uvcontsub}. 
We jointly imaged all continuum data using the CASA task {\tt clean}, which produced the Stokes I image by averaging the parallel linear correlation data (i.e., I = $(XX+YY)/2$).
Our target sources are presumably weakly or not polarized.
Therefore, we regarded the $(XX-YY)/2$ image as an approximated thermal noise map.
The Briggs Robust = 2 weighted image achieved a synthesized beam of \beam = 0\farcs95$\times$0\farcs94 (P.A.=-5.3$^{\circ}$), and a rms noise of 4.7  \ujyperbeam.

For the spectral windows including the CO(5-4) emission, we generate the continuum from the channels not affected by line emission as identified from the  datacubes, and subtract it from the data. Continuum-subtracted datacubes were created with the CASA task {\tt tclean}, using Briggs cleaning with robustness parameter of 2 (corresponding to natural visibility weights). This approach maximises the signal-to-noise ratio and it is frequently used in observations of high-z quasars (e.g., \citealt{Decarli2018,Bischetti2021}).

\subsection{VLT/HAWK-I}
\label{sec:HAWKI}

We observed in $H$-band the ELAN around QSO with the wide-field near-infrared imager HAWK-I (\citealt{Casali2006}) on the Unit Telescope 4 (UT4, Yepun) of the Very Large Telescope (VLT) in service mode under the project 0102.C-0589(D) (PI: F.~Vogt). In this work we only focus on the $H$-band observations acquired with clear weather, i.e. 15, 23 February; 9, 22, 23 March 2019. 
HAWK-I has a field of view of $7\arcmin.5\times 7\arcmin.5$ covered by an array of $2\times2$ Hawaii-2RG detectors separated by $15\arcsec$ gaps. The observational strategy consisted in three fast 60s $H$-band exposures per observing block (OB) to which it is applied a dithering within a jitter box of 15$\arcsec$. The ELAN system was always acquired in the fourth quadrant, Q4, of the detector array. The total on-source time for our clear weather observations consists of 12 OBs, i.e. 36 minutes on source for the $H$-band. 

We reduced the data with the standard ESO pipeline version 2.4.3 for HAWK-I\footnote{\url{https://www.eso.org/sci/software/pipelines/hawki/hawki-pipe-recipes.html}}. In brief, the data are corrected for dark current and are flat-fielded. 
The sky subtraction is performed using the algorithm {\it pawsky$\_$mask}, which iteratively estimates the background by stacking with rejection the science frames and by constructing a mask for the objects in the data. The sky estimation ends once the number of masked pixels converges. The photometry of the images is calibrated with 2MASS stars in the field of view of our observations, achieving a $1\sigma$ AB magnitude limit of 26.0 mag in 1~arcsec$^2$ aperture. The intrinsic uncertainty on the photometric calibration is 0.1 mag. The astrometry is calibrated against the 2MASS catalogue (about 20 stars), with an average error in the coordinates fit of $\sim0.2\arcsec$. This astrometry calibration agrees well with the GAIA DR2 catalogue (\citealt{GAIA2018}). The seeing in the final combined image is of $0.5\arcsec$.


\section{Results}
\label{sec:res}

\begin{figure*}
\centering
\includegraphics[width=0.95\textwidth]{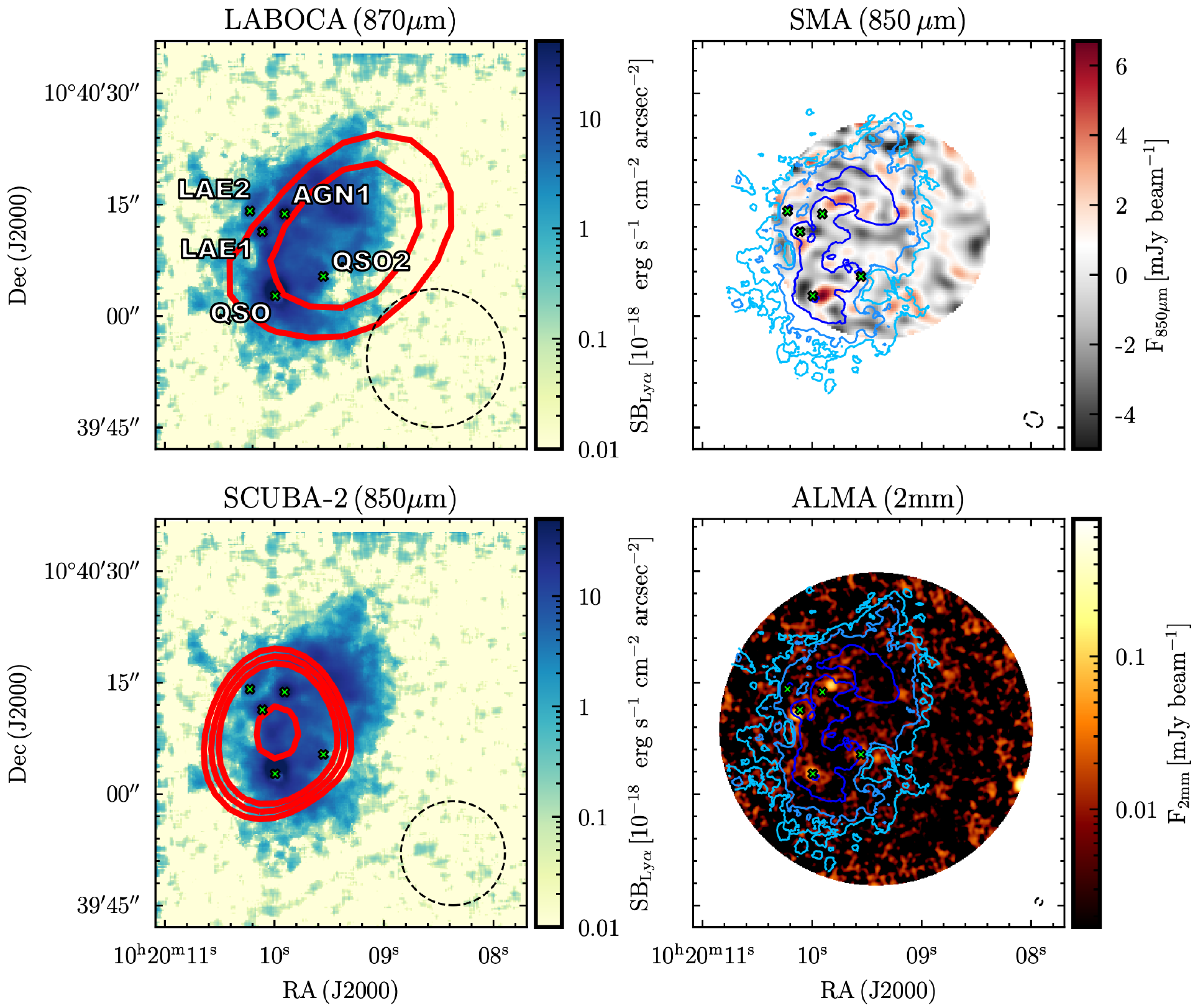}
\caption{\emph{Top left:} The detection obtained with LABOCA (red contours) is compared to the ELAN discovered in Ly$\alpha$ with MUSE (color map). The red contours indicate the isophotes at S/N=3, and 4. The dashed circle indicates the beam of LABOCA (FWHM=19$\arcsec$). \emph{Bottom left:} Same as the top left panel, but for the central detection for the SCUBA-2 instrument at 850~$\mu$m. The red contours indicate the isophotes at S/N=3, 4, 5, and 10. The dashed circle indicates the beam of SCUBA-2 at 850~$\mu$m (FWHM=14$\arcsec$). \emph{Top right:} Continuum map at $\sim850$~$\mu$m obtained with SMA (HPBW$\sim25\arcsec$). For comparison, the blue to turquoise contours indicate S/N = 2, 10, and 40 for the ELAN as detected with MUSE (\citealt{FAB2018}).  Five sources are detected in this map (see Section~\ref{sec:SMAcont} and Figure~\ref{SMA_cont_detections}). \emph{Bottom right:} Same as top right panel, but at $2$~mm using all the ALMA data (HPBW$=42.257\arcsec$). Eight sources are detected in this map (see Section~\ref{sec:ALMAcont} and Figure~\ref{ALMA_cont_detections}).} 
\label{continuum_detections}
\end{figure*}

\subsection{Single dish continuum detections}

The first data we acquired on this system, the 870~$\mu$m APEX/LABOCA data, revealed a 4.8$\sigma$ detection of ${12.5 \pm 2.6}$~mJy at the position of the ELAN (see top-left panel in Figure~\ref{continuum_detections}).
This surprisingly 
strong detection in an ELAN was then confirmed by the deeper 850~$\mu$m JCMT/SCUBA-2 observations, with a flux density of ${12.7\pm1.0}$~mJy (see bottom-left panel in Figure~\ref{continuum_detections}). We corrected this observed flux densities for flux boosting (see appendix \ref{app:LABOCADeboosting}), obtaining  f$_{\rm Deboosted} = 10.5\pm2.2$, and $11.7\pm0.9$~mJy, respectively for the LABOCA and SCUBA-2 detections (Table~\ref{tab:SingleDish}). 

Given the radio-quietness of all the sources embedded within the ELAN (\citealt{FAB2018}), the detected emission traces thermal dust emission from embedded starbusting galaxies. The strong detected fluxes would imply a star-formation rate of SFR$ = 1680\pm338$~M$_{\odot}$~yr$^{-1}$, following \citet{Cowie2017}.
These unexpectedly bright single-dish detections have been a fundamental stepping stone for the follow-up observations with interferometers.

The SCUBA-2 450~$\mu$m data are not deep enough to detect emission from the ELAN. We report the 450~$\mu$m upper limit at the location of the 850~$\mu$m detection in Table~\ref{tab:SingleDish}.

\begin{table}
\begin{center}
\caption{The continuum detections from LABOCA and SCUBA2.}
\scalebox{1}{
\footnotesize
\setlength\tabcolsep{4pt}
\begin{tabular}{lccc}
\hline
\hline
ID	    		&	f	&  SNR & f$_{\rm Deboosted}$\\
            		&  (mJy)        &      &    (mJy)           \\
\hline				       
LABOCA(870$\mu$m)	&   12.5    	& 4.8  & 	10.5$\pm2.2$	    \\
SCUBA2(850$\mu$m)	&   12.7	& 12.6 &        11.7$\pm0.9$         \\
SCUBA2(450$\mu$m)	&   $<33^{a}$	& -    &         -          \\ 	 	
\hline
\hline
\end{tabular}
}
\flushleft{\scriptsize $^a$ $3\sigma$ upper limit at the position of the SCUBA2 850~$\mu$m detection.\\ 
} 
\label{tab:SingleDish}
\end{center}
\end{table}

\subsection{SMA continuum at 850~$\mu$m}
\label{sec:SMAcont}

\begin{figure}
\centering
\includegraphics[width=1.0\columnwidth]{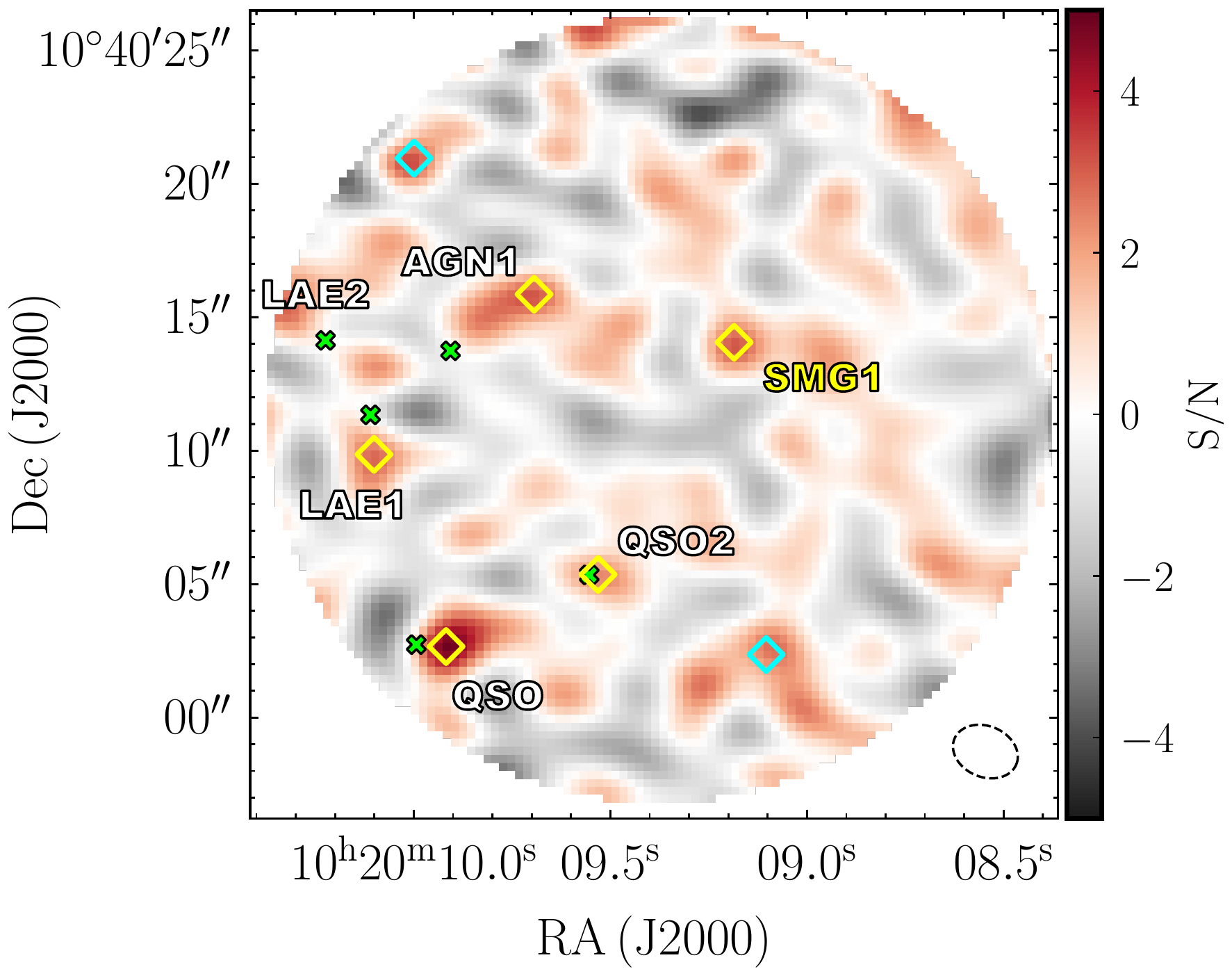}
\caption{S/N map at $\sim 850$~$\mu$m obtained with SMA. We show the location of (i) the detections down to S/N=2 (yellow diamonds), (ii) the spurious sources (cyan diamonds), and (iii) the location of known sources associated with the ELAN (green crosses; \citealt{FAB2018}). A newly discovered source is named SMG1. The beam of the observations is indicated in the bottom-right corner.} 
\label{SMA_cont_detections}
\end{figure}

We extracted continuum sources from the SMA continuum map (top right panel in Figure~\ref{continuum_detections}) using the same algorithm described in e.g., \citet{FAB2018b}, but working with the SMA beam of the current dataset.
Briefly, the algorithm iteratively searches for maxima in the S/N map (Figure~\ref{SMA_cont_detections}) while subtracting (at their locations) mock sources normalized to those peaks. The iterations are stopped once S/N$=2$ is reached. 
The S/N peaks found by the algorithm are included in a source candidate catalog.  In the current case, the algorithm found 7 sources. 
Subsequently, the same algorithm is applied to the negative dataset down to the same S/N threshold to estimate the number of spurious sources and clean the aforementioned catalog.
We find 
no spurious sources within a radius of $R=7\arcsec$ from the center of the map, 
one such source in the annuli within $7\arcsec<R<10.5\arcsec$, and four for $R>10.5\arcsec$.
Therefore, we consider reliable five of the seven sources detected in our map (yellow diamonds in Figure~\ref{SMA_cont_detections}). The two potentially spurious sources are indicated with cyan diamonds in Figure~\ref{SMA_cont_detections}, and are located respectively in  the $7\arcsec<R<10.5\arcsec$ and $R>10.5\arcsec$ regions. This analysis is confirmed by the absence of emission at these two locations in the ALMA continuum map (see section~\ref{sec:ALMAcont}), while all the other sources are very close to the positions of known sources associated with the ELAN or with ALMA detections (see section~\ref{sec:ALMAcont})\footnote{The alignment of each individual SMA S/N$>2$ source with the location of an ALMA detection strongly suggests that the reported sources are reliable. Within $R<12\arcsec$ from the phase center, where all our detections are located, we expect to have ~144 independent beam areas.
Following Gaussian noise, the chance to have a $>2\sigma$ positive noise peak is $\sim2.25$\%, so there could be $\sim 3$ noise peaks above $>2\sigma$. The probability of having a $>2\sigma$ noise peak at close location of an ALMA source is therefore $P_{\rm det} \sim 3/144=0.0225$. The probability of having 5 noise peaks (as the detected sources QSO, QSO2, LAE1, AGN1, SMG1) with SNR~$>2$ and at close location of ALMA detections is therefore very low. We estimated it to be $P_{\rm det}^5 \sim 5.8\times 10^{-9}$ (confirmed using Monte Carlo simulations).}. The detected sources are QSO, QSO2, AGN1, LAE1, and a newly discovered source SMG1.
The positions, S/N and fluxes for the five detections are listed in Table~\ref{tab:Interf}, together with the deboosted fluxes estimated as explained in Appendix~\ref{app:SMADeboosting}.
Summing up the deboosted fluxes of all detected sources, we find agreement within uncertainties with the detections in the single-dish datasets, $14.7\pm2.8$~mJy. Therefore, all the continuum emission detected by LABOCA and SCUBA-2 is ascribed to compact sources.

\begin{table*}
\begin{center}
\caption{The continuum measurements from ALMA, SMA, HAWK-I and MUSE.}
\scalebox{1}{
\footnotesize
\setlength\tabcolsep{4pt}
\begin{tabular}{l|ccccc|ccc|ccc}
\hline
\hline
ID	    &	f$_{\rm ALMA}$ 	&  SNR	& f$_{\rm ALMA}^{\rm Deboosted}$  &	 R.A.	   &	 Dec	&	 f$_{\rm SMA}$   &  SNR  & f$_{\rm SMA}^{\rm Deboosted}$&	    $H$          &  $i$  & $r$         \\
            &    (mJy)          &	& (mJy)           &    (J2000)     &	 (J2000)	&	      (mJy)	 &	 & (mJy)                  &	          (AB)   &  (AB) & (AB)   	\\
\hline
QSO	    &   0.24    	& 34.6  & 0.23$\pm$0.01 & 10:20:10.00  &    +10:40:02.7   &	    6.8$^{\rm b}$     &  4.9  & 5.7$\pm$1.8		  &    17.1$\pm$0.1	 &  17.98$\pm$0.01   &  17.67$\pm$0.01         \\
QSO2	    &	0.06 	 	& 8.9   & 0.06$\pm$0.01 & 10:20:09.56  &    +10:40:05.3   &	    3.3$^{\rm b}$     &  2.4  & 2.0$\pm$1.0		  &    23.5$\pm$0.1	 &  24.30$\pm$0.02   & 24.10$\pm$0.01	       \\
LAE1	    &   0.17    	& 23.9  & 0.17$\pm$0.01 & 10:20:10.15  &    +10:40:10.6   &	    3.6$^{\rm b}$     &  2.6  & 2.4$\pm$1.1		  &    24.6$\pm$0.6	 &  25.43$\pm$0.05   & 25.32$\pm$0.05	       \\
LAE2	    &	$<0.01^{\rm a}$ & -	& -		&	-      &	-	  &	    $<2.8^{\rm a}$    &  -    & -			  &    $>$26.8$^{\rm d}$ &  $>$28.6$^{\rm d}$	& $>$28.6$^{\rm d}$    \\
AGN1	    &	0.19    	& 18.2  & 0.13$\pm$0.01 & 10:20:09.83  &    +10:40:14.7   &	    4.0$^{\rm b}$     &  2.8  & 2.8$\pm$1.2		  &    24.7$\pm$0.7	 &  26.20$\pm$0.20   & 26.30$\pm$0.20	       \\
SMG1	    &	0.03 	 	& 4.4   & 0.02$\pm$0.01 & 10:20:09.18  &    +10:40:13.4   &	    3.1$^{\rm b}$     &  2.2  & 1.8$\pm$0.9		  &    24.6$\pm$0.3	 &  26.14$\pm$0.12   & 26.88$\pm$0.23	       \\
S1	    &	0.04 	 	& 6.6   & 0.04$\pm$0.01 & 10:20:09.41  &    +10:40:04.9   &	    $<2.8^{\rm c}$    &  -    & -			  &    24.8$\pm$0.3	 &  27.18$\pm$0.31   & $>$28.6$^{\rm d}$       \\
S2	    &	0.03 	 	& 4.1   & 0.03$\pm$0.01 & 10:20:10.16  &    +10:40:08.6   &	    $<2.8^{\rm c}$    &  -    & -			  &    23.2$\pm$0.1	 &  24.77$\pm$0.03   & 24.75$\pm$0.03	       \\
S3	    &	0.03 	 	& 4.1   & 0.03$\pm$0.01 & 10:20:08.78  &    +10:40:16.3   &	    $<2.8^{\rm c}$    &  -    & -			  &    $>$26.8$^{\rm d}$ &  $>$28.6$^{\rm d}$& $>$28.6$^{\rm d}$       \\
\hline
\hline
\end{tabular}
}
\flushleft{\scriptsize $^{\rm a}$ $2\sigma$ upper limit at the Ly$\alpha$ position from \citet{FAB2018}.\\ 
\scriptsize $^{\rm b}$ The coordinates of the SMA detections differ from the ALMA detections due to their lower SNR. We report them here for completeness for each source:  
[10:20:09.9, +10:40:03], [10:20:09.5, +10:40:05], [10:20:10.1, +10:40:10], [10:20:09.7, +10:40:16], [10:20:09.2, +10:40:14] for QSO, QSO2, LAE1, AGN1, and SMG1, respectively.\\
\scriptsize $^{\rm c}$ $2\sigma$ upper limit at the ALMA position.\\
\scriptsize $^{\rm d}$ $1\sigma$ limit at the ALMA position within an aperture of 2$\arcsec$ diameter.\\ 
} 
\label{tab:Interf}
\end{center}
\end{table*}

\subsection{ALMA continuum at 2~mm}
\label{sec:ALMAcont}

\begin{figure}
\centering 
\includegraphics[width=1.0\columnwidth]{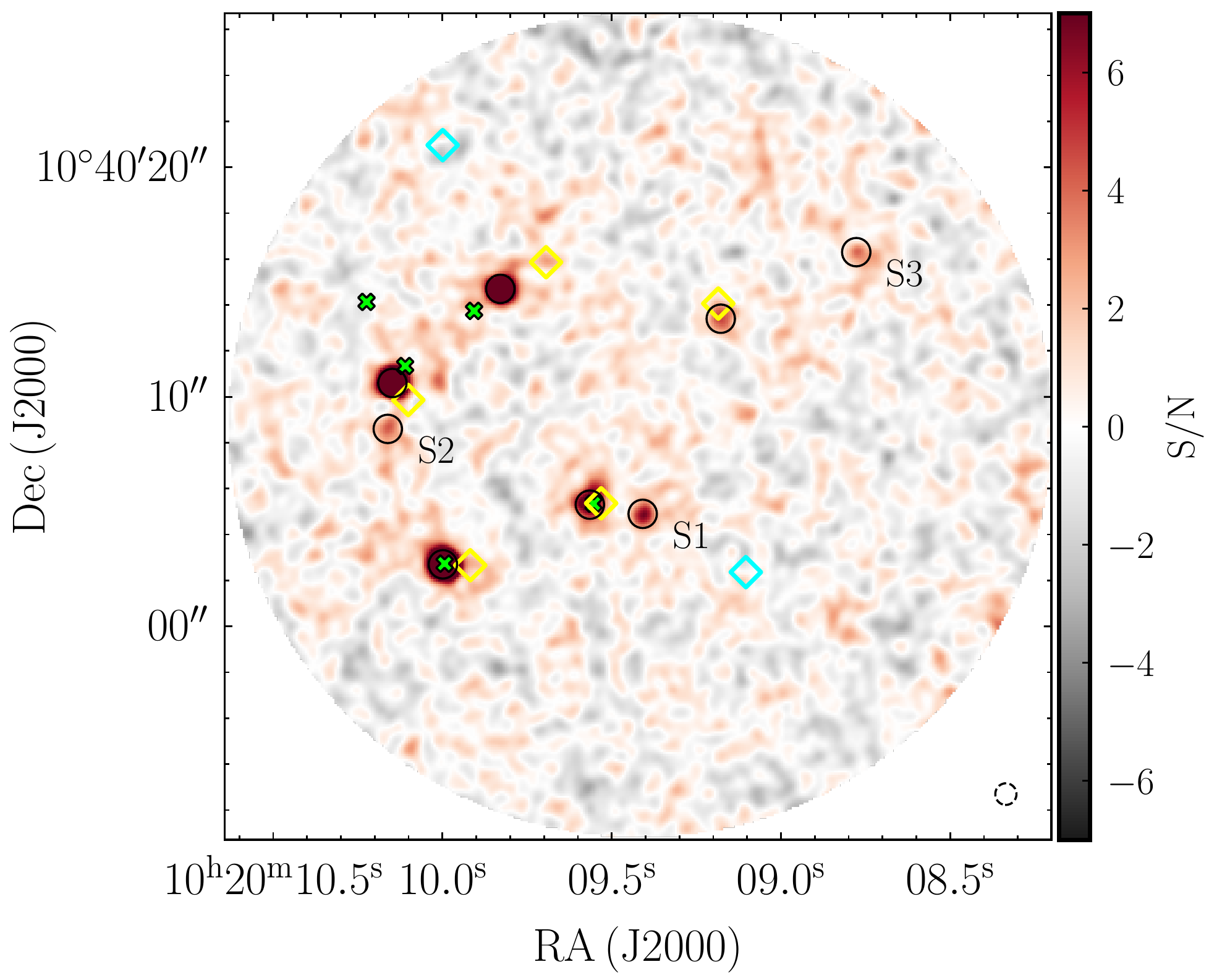}
\caption{S/N map at $\sim 2$~mm obtained with ALMA. We show the location of (i) the detections for S/N$>3.7$ (see section~\ref{sec:ALMAcont} for details), (ii) the SMA detections (yellow diamonds), (iii) the spurious SMA sources (cyan diamonds), and (iv) the location of known sources associated with the ELAN (green crosses; \citealt{FAB2018}). Three newly discovered sources are named S1, S2, and S3. The beam of the observations is indicated in the bottom-right corner.} 
\label{ALMA_cont_detections}
\end{figure}

We extracted sources from the ALMA continuum map at 2~mm (bottom right panel in Figure~\ref{continuum_detections}) following the same method as for the SMA data, but using the ALMA beam.
We considered as reliable only sources with S/N$>3.7$. Indeed, above this threshold we did not find any spurious source in the negative map.
Using this threshold, we found eight detections, shown as black circles in Figure~\ref{ALMA_cont_detections}:
(i) the known sources QSO, QSO2, LAE1, and AGN1, (ii) the source SMG1 discovered with SMA, and (iii) three additional sources which we dubbed S1, S2, and S3.
The coordinates, fluxes and S/N of all the sources, as well as their deboosted fluxes estimated as explained in Appendix~\ref{app:ALMADeboosting} are listed in Table~\ref{tab:Interf}.

As evident from Figure~\ref{ALMA_cont_detections}, some of the ALMA detections have shifts of $\sim$ few arcseconds with respect to the Ly$\alpha$ locations and to the SMA locations. While the latter is likely due to the low SNR of the SMA detections, we discuss in detail the shifts with respect to the Ly$\alpha$ locations in section~\ref{sec:Mol_vs_Lya}.

\subsection{HAWK-I and VLT/MUSE counterparts}
\label{sec:HAWKIresults}

\begin{figure}
\centering
\includegraphics[width=0.9\columnwidth]{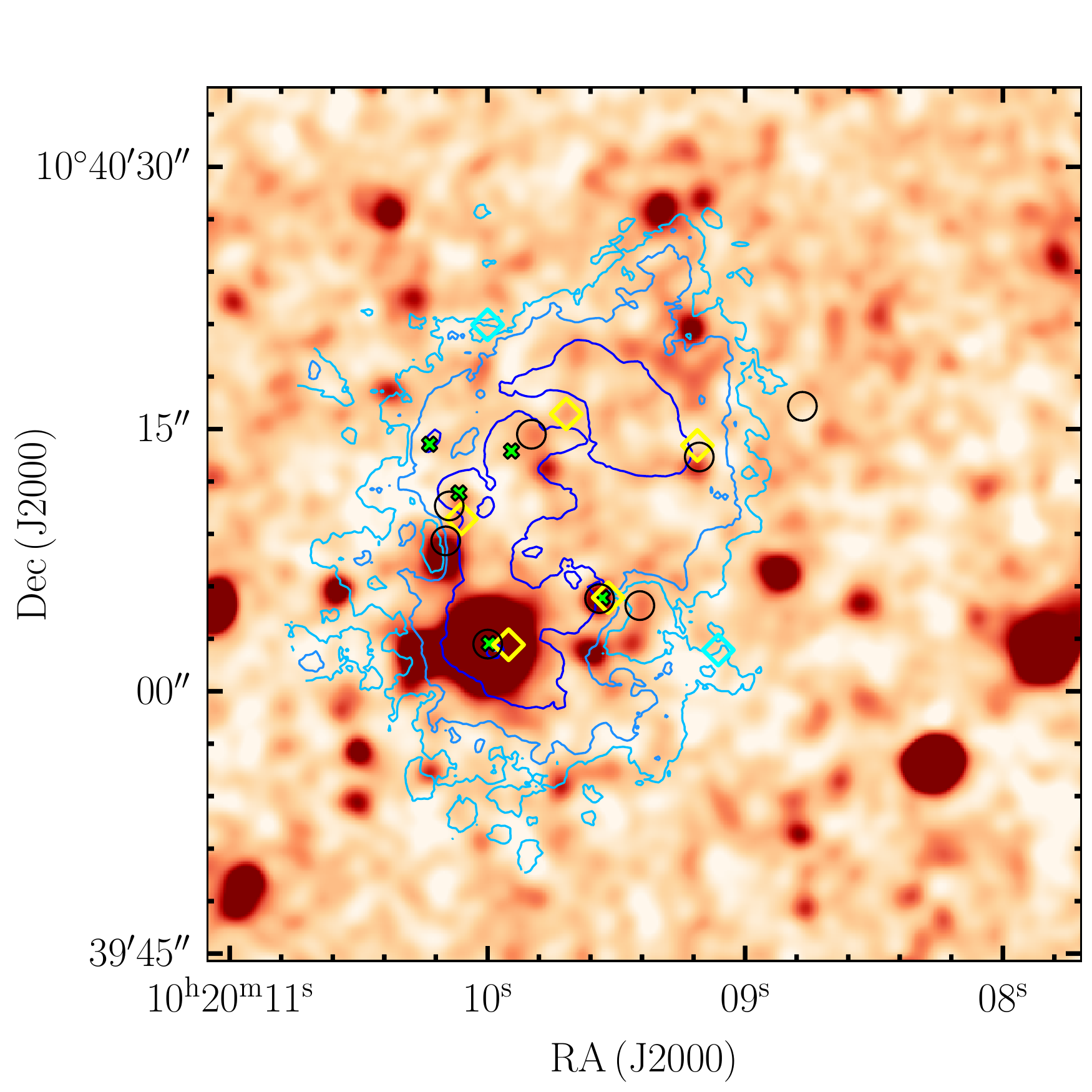}
\caption{$H$-band HAWK-I data within the field of view of Fig.~\ref{continuum_detections} smoothed with a 1$\arcsec$ Gaussian kernel. We highlight the location of (i) the ALMA detections (black circles), (ii) the SMA detections (yellow diamonds), (iii) the spurious SMA sources (cyan diamonds), and (iv) the location of known sources associated with the ELAN (green crosses; \citealt{FAB2018}). The ELAN isophotes are indicated as in Figure~\ref{continuum_detections}. 
} 
\label{HAWK-I_full_image}
\end{figure}

We inspected the $H$-band HAWK-I data presented in section~\ref{sec:HAWKI} at the location of the sources so far discussed.
As can be seen in Figure~\ref{HAWK-I_full_image}, we find clear detections for QSO, QSO2, S2, SMG1, and AGN1, while fainter emission at the location of LAE1 and S1.
LAE2 and S3 are undetected at the current depth. We extract magnitudes with apertures of 2$\arcsec$ diameter ($4\times$ the seeing) for all the sources except QSO.
Indeed, its flux is better captured by a 
3$\arcsec$ diameter aperture. 
We list the derived magnitudes in Table~\ref{tab:Interf}.

Further, we obtained the observed optical magnitude, $i$ and $r$, of all sources within the aforementioned apertures from the MUSE data presented in \citet{FAB2018}. These magnitudes are listed in Table~\ref{tab:Interf}.
We note that the MUSE $i$-band magnitude for AGN1 is different from the magnitude listed in \citet{FAB2018} because in that study the authors rely on the compact Ly$\alpha$ emission for determining the source position. However, there is an important shift between Ly$\alpha$ and the near and far-infrared continua detected in this work (Figure~\ref{HAWK-I_full_image}; see discussion in section~\ref{sec:Mol_vs_Lya}). 

\subsection{ALMA CO(5-4) line detections}
\label{sec:CO54results}

We rely on the publicly available code {\sc LineSeeker} (\citealt{GL2017}) to robustly identify sources with CO(5-4) line emission in the ALMA observations.  Indeed {\sc LineSeeker} has been 
developed to systematically search for line emissions in ALMA data and quantify their significance (e.g., \citealt{GL2019}). 
The code looks for signal in the ALMA datacubes on a channel by channel basis after convolving the data along the spectral axis with Gaussian kernels of different spectral widths. A list of line candidates is obtained by joining detected signal on different channels using the DBSCAN algorithm (\citealt{Ester1996}). {\sc LineSeeker} finally provides a S/N for each emission line candidate. This S/N is defined as the maximum value obtained from the different convolutions. Further,  
it runs the search algorithm also on the negative datacube and on simulated cubes to estimate the significance of the line candidates S/N, providing probabilities of the line being false.

\begin{figure}
\centering 
\includegraphics[width=1.0\columnwidth]{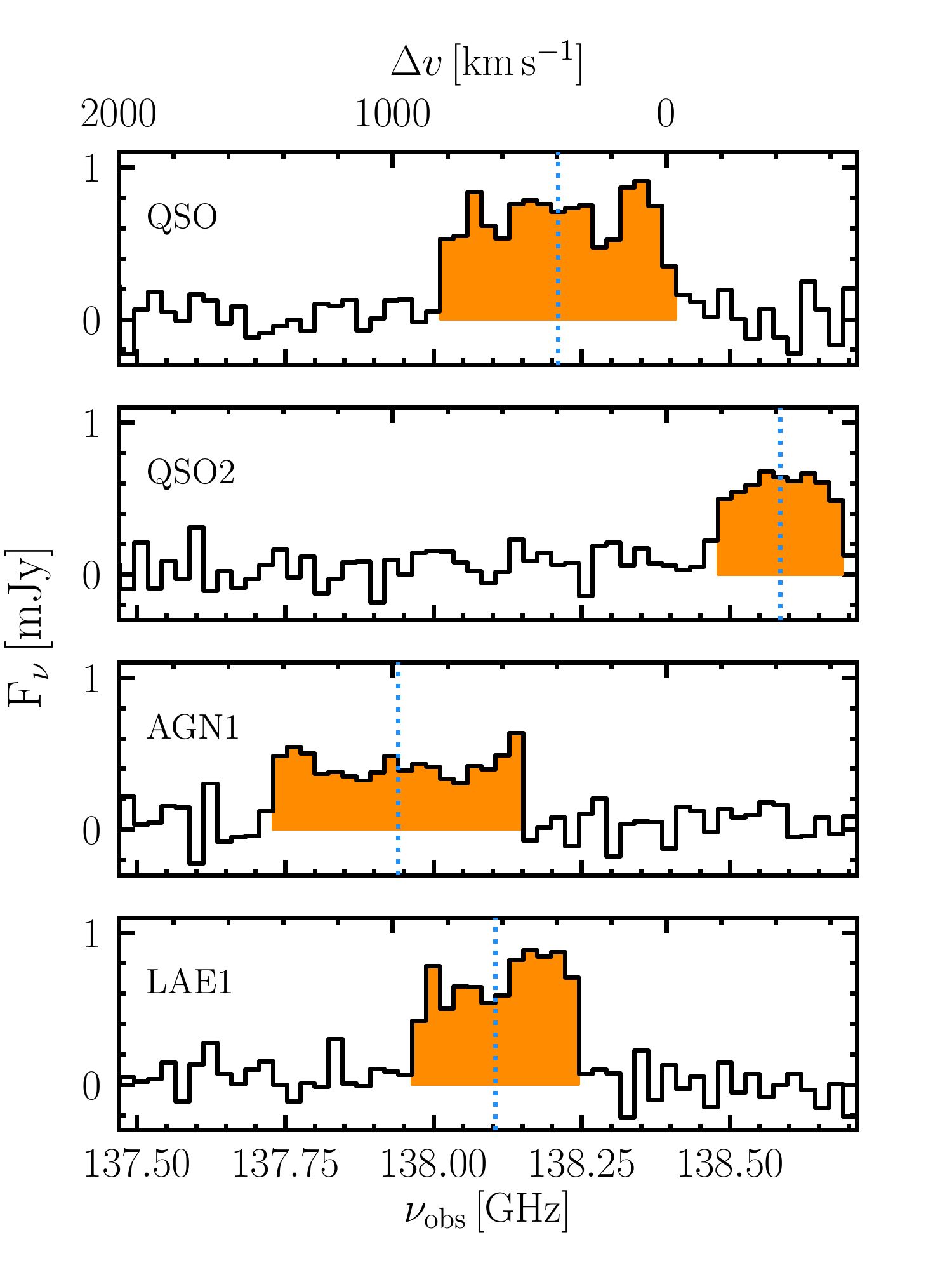}
\caption{The four ALMA CO(5-4) line detections. The spectra at the CO(5-4) location for QSO, QSO2, AGN1, and LAE1 are shown from top to bottom, where the channels were binned to a width of  23.4~MHz ($\sim 51$~km~s$^{-1}$). The rms noise for each spectra is listed in Table~\ref{tab:CO54}. The line emissions above $2\times {\rm rms}$ used to compute quantities in Table~\ref{tab:CO54} 
are highlighted in orange. The velocity shift of the top x-axis are computed with respect to the QSO 
systemic redshift ($z=3.164$) obtained from \ion{C}{4} (\citealt{FAB2018}). We consider the first moment of the CO(5-4) detections as new systemic redshifts.
In each panel, the vertical dotted blue line indicates such $z_{\rm CO(5-4)}$, which we use as reference redshift to derive the first moment maps (redshifts listed in Table~\ref{tab:CO54}).} 
\label{ALMA_CO54_spectra}
\end{figure}

Here we focus on 100$\%$ fidelity sources, i.e., sources whose probability of being false is 0 and whose S/N is larger than any detection in the negative data. In this way, we obtained four line detections at $>10\sigma$ from four sources detected also in the continuum, QSO, QSO2, AGN1, and LAE1\footnote{We looked for sources down to S/N=5 with {\sc LineSeeker}, but all additional detections are found at the edge of the primary beam with very low fidelity, and therefore they are not reliable.}. All the other known sources do not show evidence of CO(5-4) emission consistent with the system redshift. We then extracted the spectrum for each detected source using the minimum aperture that maximized the flux densities. 
An aperture $2\times$ the synthetized beam fulfilled this criterion. 
Figure~\ref{ALMA_CO54_spectra} shows the four spectra binned to channels of 23.4~MHz (or $\sim51$~km~s$^{-1}$), with the emission above $2\times$ the rms highlighted in each spectrum.
The spectra are shown indicating the velocity shift with respect to the QSO systemic redshift estimated from \ion{C}{4}
($z=3.164$; \citealt{FAB2018}) after correcting for the expected velocity shift between \ion{C}{4} and systemic (\citealt{Shen2016}). In the reminder of the paper we will consider the redshift of the CO emission as systemic redshift for each detected object. 

All line detected show velocity widths $> 200$~km~s$^{-1}$, when estimated using their second moment. Though their shape is relatively boxy and the widths of the highlighted velocity ranges in Figure~\ref{ALMA_CO54_spectra} are as high as $\sim1000$~km~s$^{-1}$ (see Table~\ref{tab:CO54}). The CO(5-4) line profile for QSO and AGN1 seems complex, with QSO presenting three tentative peaks, while AGN1 has a profile with higher flux densities at the edges of the line. The profile of AGN1 is suggestive of a molecular gas disk, similar to what is seen in other AGN host galaxies (e.g., in low-redshift radio galaxies; \citealt{Ocana2010}).  We will further discuss the Ly$\alpha$ and CO(5-4) velocity shifts and line shapes in Section~\ref{sec:Mol_vs_Lya}. 
Table~\ref{tab:CO54} lists the rms for these spectra and all the lines properties extracted using the highlighted velocities, i.e., redshift, line width, flux, line luminosities.   

\begin{table*}
\begin{center}
\caption{CO(5-4) detections from ALMA, and their derived galaxy properties.}
\scalebox{1}{
\footnotesize
\setlength\tabcolsep{4pt}
\begin{tabular}{l|cccc}
\hline
\hline
	    					            &	QSO 		          &  QSO2	           	   & AGN1  		   	   &	 LAE1	   	  	  \\
\hline\\[-0.2cm]
RMS noise per 23.4 MHz [${\rm \mu Jy}$]		            &	118	      	  	  & 	117	          	   &	133	           	   &	103	          	  \\
SNR$^{\rm a}$					            &	22.2			  &	12.4			   &	11.3			   &	18.0			  \\[0.1cm]
\hline\\[-0.2cm]
$z_{\rm CO(5-4)}^{\rm b}$	    		    	    &   $3.1695\pm0.0008$	  &	$3.1582\pm0.0006$          &	$3.1777\pm0.0009$	   &	$3.1727\pm0.0007$	  \\
CO(5-4) line width [km s$^{-1}$]$^{\rm c}$	            &	$418\pm101$               &	$211\pm103$		   &	$443\pm202$		   &	$246\pm96$		  \\
$I_{\rm CO(5-4)}$ [Jy km s$^{-1}$]	     		    &   $0.43\pm0.03$	          &	$0.22\pm0.02$		   &	$0.28\pm0.03$		   &	$0.34\pm0.02$		  \\
$L_{\rm CO(5-4)}$ [$10^7$~L$_{\odot}$]	    		    &	$4.5\pm0.3$		&     $2.3\pm0.2$	       &    $2.9\pm0.3$ 	     &    $3.6\pm0.2$		  \\
$L_{\rm CO(5-4)}^{\prime}$ [$10^9$~K km s$^{-1}$ pc$^2$]    &	$7.5\pm0.4$		&     $3.9\pm0.3$	       &    $4.8\pm0.5$ 	     &    $6.0\pm0.3$		  \\
$L_{\rm CO(1-0)}^{\prime}$ [$10^9$~K km s$^{-1}$ pc$^2$]$^{\rm d}$ &	$18.7\pm1.1$    &	$9.6\pm0.8$	           &	$12.1\pm1.3$             &	$14.9\pm0.8$            \\[0.1cm]
\hline\\[-0.2cm]
2mm major axis [$\arcsec $]$^{\rm e}$			    &	$1.07\pm0.04$	& $1.68\pm0.21$   & $1.12\pm0.07$    & $1.06\pm0.05$				 \\
2mm minor axis [$\arcsec $]$^{\rm e}$			    &   $0.98\pm0.03$   & $1.29\pm0.14$   & $1.04\pm0.06$    & $1.03\pm0.05$				 \\
2mm dec. major axis [$\arcsec$]$^{\rm f}$		            &	$0.52\pm0.09$ 	& $1.39\pm0.27$   & $0.64\pm0.16$    & $0.52\pm0.17$				 \\
2mm dec. minor axis [$\arcsec$]$^{\rm f}$			    &   $0.30\pm0.15$   & $0.89\pm0.22$   & $0.44\pm0.19$    & $0.43\pm0.14$				 \\
R$_{\rm 2mm}$ [{\rm kpc}]$^{\rm g}$			    &	$2.0\pm0.3$	& $5.3\pm1.0$     & $2.4\pm0.6$	     & $2.0\pm0.6$		   		 \\
CO(5-4) major axis [$\arcsec$]$^{\rm e}$			    &	$1.30\pm0.11$   & $1.67\pm0.15$   & $1.75\pm0.18$    & $1.39\pm0.11$	 			 \\
CO(5-4) minor axis [$\arcsec$]$^{\rm e}$			    &   $1.12\pm0.09$   & $1.36\pm0.11$   & $1.57\pm0.15$    & $1.29\pm0.10$				 \\
CO(5-4) dec. major axis [$\arcsec$]$^{\rm f}$ 	            &	$0.78\pm0.20$   & $1.33\pm0.20$   & $1.41\pm0.23$    & $0.95\pm0.19$				 \\
CO(5-4) dec. minor axis [$\arcsec$]$^{\rm f}$		    &   $0.49\pm0.27$   & $0.89\pm0.18$   & $1.21\pm0.22$    & $0.78\pm0.17$				 \\
R$_{\rm CO(5-4)}$ [{\rm kpc}]$^{\rm g}$			    &	$3.0\pm0.8$     & $5.1\pm0.8$     & $5.4\pm0.9$      & $3.6\pm0.7$		                 \\[0.1cm]
\hline\\[-0.2cm]
$L_{\rm bol}$ [erg~s$^{-1}$]$^{\rm h}$			    & 	$(4.6\pm0.2)\times10^{47}$ &  $(8.8\pm0.9)\times10^{45}$   &	$(1.0\pm0.5)\times10^{46}$ &	$(1.3\pm0.2)\times10^{45}$  \\
$L_{\rm IR}$ [$10^{12}$~L$_{\odot}$]$^{\rm i}$	            & 	$4.7\pm0.2$              &  $7.3\pm1.9$                &	$3.0\pm1.0$              &	$1.2\pm0.2$               \\
$L_{\rm IR}^{\rm total}$ [L$_{\odot}$]$^{\rm l}$	    & 	$(1.7\pm0.1)\times10^{14}$ &  $(7.8\pm2.3)\times10^{12}$   &	$(1.0\pm0.3)\times10^{13}$ &	$(1.2\pm0.2)\times10^{12}$  \\
SFR	[M$_{\odot}$~yr$^{-1}$]$^{\rm m}$	            &	$521\pm74$		   & $183\pm49$			   &	$104\pm21$		   &	$50\pm11$		    \\ 
SFR$_{\rm IR}$	[M$_{\odot}$~yr$^{-1}$]$^{\rm n}$	    &	$700\pm30$		   & $1087\pm283$		   &	$447\pm149$		   &	$179\pm30$		    \\
$M_{\rm dust}$ [M$_{\odot}$]$^{\rm h}$  		    &	 $(6.7\pm1.3)\times10^{8}$ &  $(1.5\pm0.4)\times10^{8}$  &	 $(5.7\pm1.3)\times10^{8}$  &	 $(1.6\pm0.4)\times10^{9}$  \\
$M_{\rm dust}$ [M$_{\odot}$]$^{\rm o}$                      &	 $(9\pm3)\times10^{8}$ &  $(3\pm2)\times10^{8}$	   &	 $(5\pm2)\times10^{8}$  &  $(4\pm2)\times10^{8}$   \\
\hline\\[-0.2cm]
$M_{\rm H_2,CO}$ [M$_{\odot}$]$^{\rm p}$	            &	$(1.5\pm0.1)\times10^{10}$&	$(7.7\pm0.7)\times10^9$    &	$(1.0\pm0.1)\times10^{10}$ &	$(1.2\pm0.1)\times10^{10}$\\
$r_{\rm H_2,dust}$ $^{\rm q}$	                            &	 $22\pm4$	          &	 $51\pm14$		   &		 $18\pm4$	           &	 $8\pm2$		  \\[0.1cm]
\hline
$M_{\rm star}$ [M$_{\odot}$]$^{\rm r}$			    &	$(1.2\pm0.8)\times10^{11}$ & 	$(1.4\pm0.4)\times10^{10}$     &   $(1.1\pm0.3)\times10^{10}$  &  $(4.0\pm0.5)\times10^{9}$   \\
$M_{\rm DM}$ [M$_{\odot}$]$^{\rm s}$			    &	$(6.9_{-5.6}^{+19.4})\times10^{12}$ & $(6.8\pm1.0)\times10^{11}$ & $(6.0_{-1.6}^{+0.7})\times10^{11}$ &  $(3.7_{-0.2}^{+0.3})\times10^{11}$ \\[0.1cm]
\hline
\hline
\end{tabular}
}
\flushleft{\scriptsize $^{\rm a}$ Integrated signal to noise ratio. \\ $^{\rm b}$ Redshift corresponding to the first moment of the line detected.\\ 
$^{\rm c}$ Line width computed from the second moment of each line detected following, e.g., equation 3 in \citet{Birkin2020}. 
Given the relatively boxy shape of the lines we report here for completeness the line width corresponding to the colored range in Figure~\ref{ALMA_CO54_spectra}: 
924, 513, 976, and 667~km~s$^{-1}$ for QSO, QSO2, AGN1, and LAE1, respectively.\\ 
$^{\rm d}$ CO(1-0) luminosities obtained assuming $I_{\rm CO(5-4)}$/$I_{\rm CO(1-0)}= 10$ (see Section~\ref{sec:molgasmass} for details).\\
$^{\rm e}$ Observed (i.e., beam-convolved) sizes of the 2mm continuum and CO(5-4) emitting region from 2D Gaussian fit of the ALMA maps (see Section~\ref{sec:CO54results} for details).\\
$^{\rm f}$ Beam-deconvolved sizes of the 2mm continuum and CO(5-4) emitting region from 2D Gaussian fit of the ALMA maps  (see Section~\ref{sec:CO54results} for details).\\
$^{\rm g}$ Effective radius of the 2mm continuum and CO(5-4) emitting region, defined as their major semiaxis (see Section~\ref{sec:CO54results} for details).\\
$^{\rm h}$ Obtained with the fit by {\sc CIGALE}.\\
$^{\rm i}$ Luminosity obtained by integrating only the dust emission of the SED due to star formation, in the rest-frame wavelength range 8-1000~$\mu$m.\\
$^{\rm l}$ Luminosity obtained by integrating the total SED in the rest-frame wavelength range 8-1000~$\mu$m.\\
$^{\rm m}$ SFR obtained with the fit by {\sc CIGALE}.\\
$^{\rm n}$ SFR obtained from the IR, assuming the relation (SFR$_{\rm IR}/[M_{\odot}~{\rm yr}^{-1}])=3.88\times10^{-44} (L_{\rm IR}/[{\rm erg~s^{-1}}])$ in \citet{Murphy2011}.\\ 
$^{\rm o}$ Dust mass obtained using a modified black body (see Section~\ref{sec:dustsfr} for details).\\
$^{\rm p}$ Molecular gas mass derived from the CO(5-4) line, assuming (i) a ratio $I_{\rm CO(5-4)}$/$I_{\rm CO(1-0)}= 10$ (or $r_{51}=0.4$), (ii) a luminosity-to-gas mass
conversion factor of $\alpha_{\rm CO} = 0.8$~M$_{\odot} ({\rm K\, km\, s^{-1}\, pc^2})^{-1}$. The reported error on these measurements only includes the error on $L_{\rm CO(5-4)}^{\prime}$. 
Large uncertainties are expected due to the unconstrained $r_51$ and the known uncertainties on $\alpha_{\rm CO}$ (see Section~\ref{sec:molgasmass} for details).\\
$^{\rm q}$ Molecular gas to dust mass ratio computed using the dust mass estimated by {\sc CIGALE}. 
The ratios obtained with the other dust mass estimates are consistent within 2$\sigma$. 
The errors on $r_{\rm H_2,dust}$ do not include the large uncertainties expected for $M_{\rm H_2,CO}$ (see Section~\ref{sec:molgasmass} for details).\\
$^{\rm r}$ Stellar mass estimated by {\sc CIGALE}. \\
$^{\rm s}$ Dark matter halo mass estimated assuming the stellar mass - halo mass relation in \citet{Moster2018}, and interpolating their models for the redshift of interest here.} 
\label{tab:CO54}
\end{center}
\end{table*}

\begin{figure*}
\centering
\includegraphics[width=1.0\textwidth]{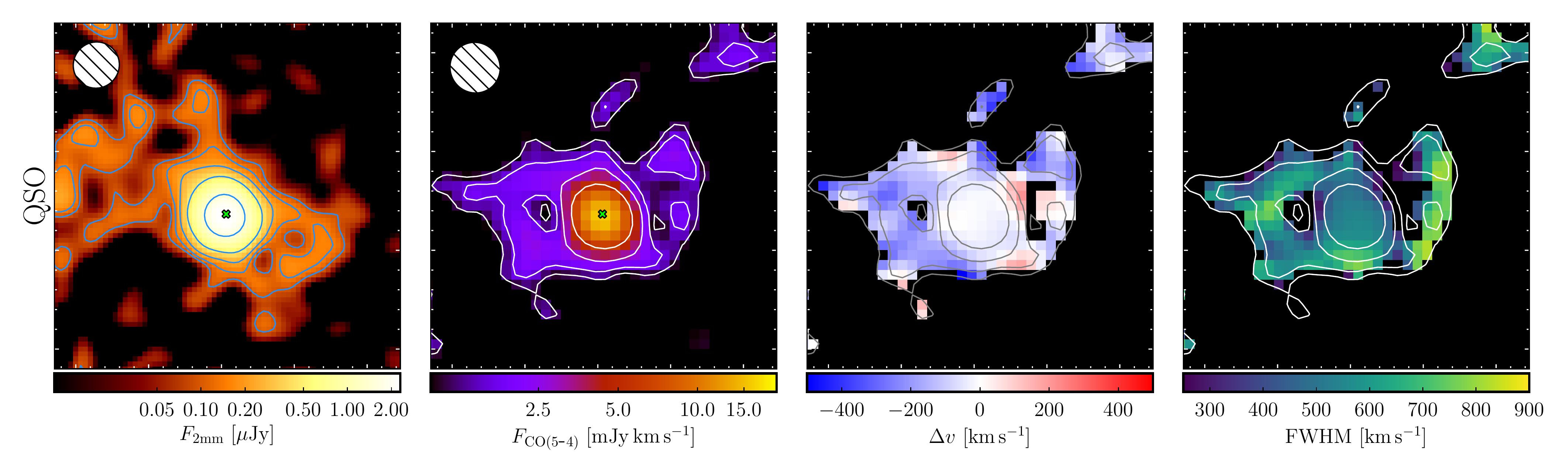}\\
\includegraphics[width=1.0\textwidth]{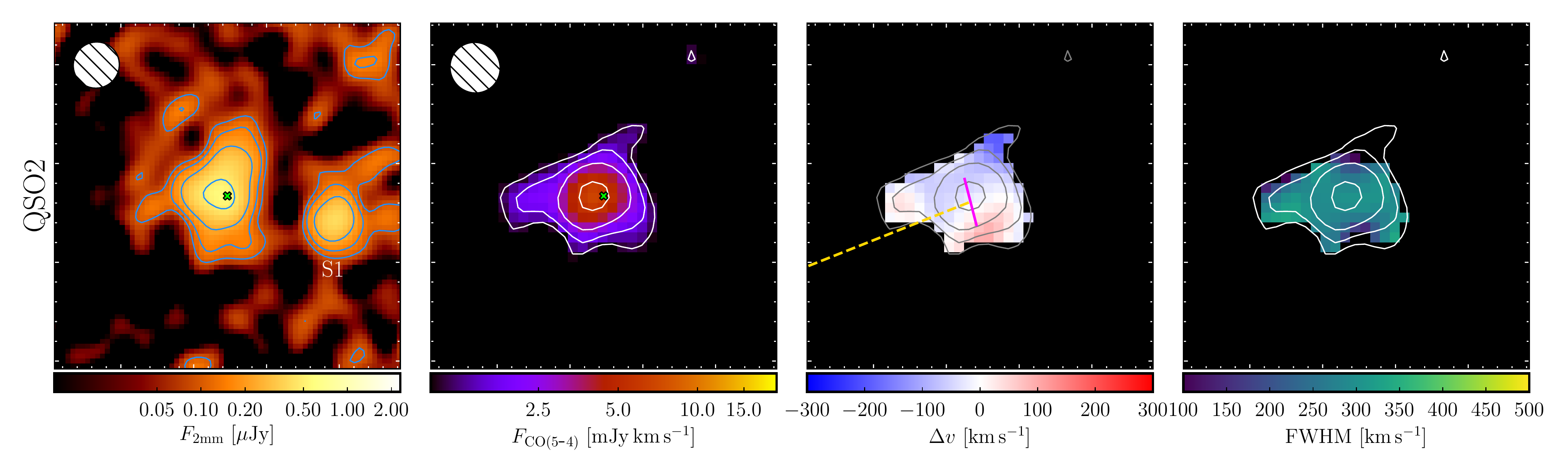}\\
\includegraphics[width=1.0\textwidth]{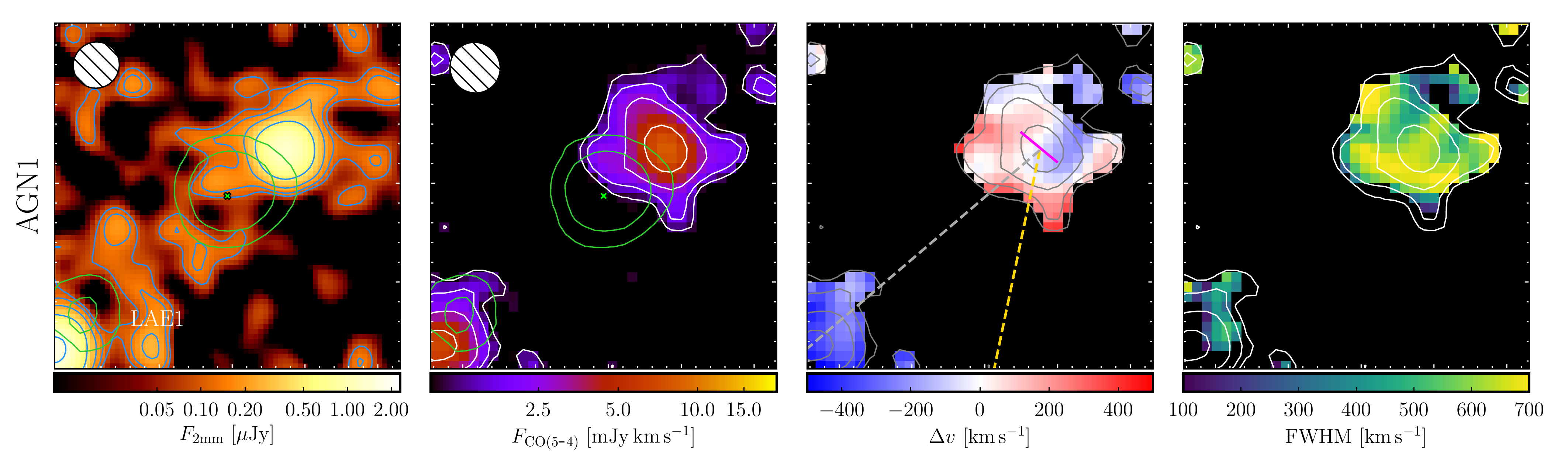}\\
\includegraphics[width=1.0\textwidth]{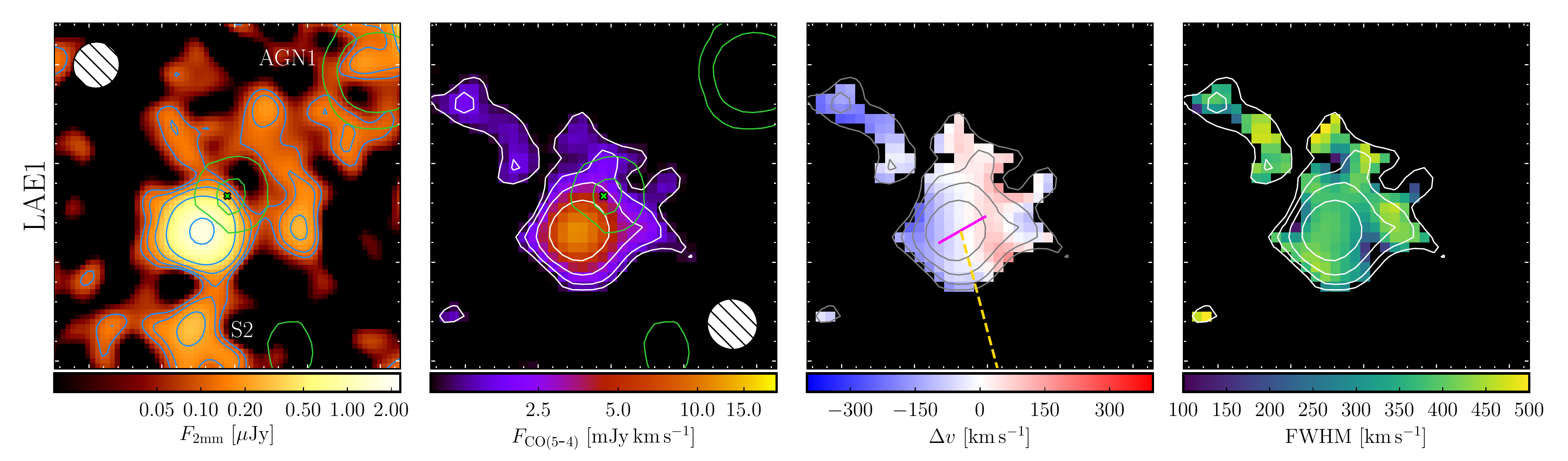}
\caption{ALMA continuum and CO(5-4) moment maps for each line detected source, QSO, QSO2, AGN1, and LAE1. Each panel is a cutout of $7\arcsec \times 7\arcsec$ (or about 53~kpc~$\times$~53~kpc). The contours indicate the 2, 3, 5, 10, and 30$\sigma$ isophotes, and the 2, 3, 5, and 10$\sigma$ isophotes, respectively for the continuum and the line moment zero maps. The contours on the first and second moments show the isophotes from the moment zero cutout. The green cross shows the Ly$\alpha$ location of each source as determined with the available MUSE data (\citealt{FAB2018}). In addition, for AGN1 and LAE1, which are affected by large offset between Ly$\alpha$ and ALMA observations, we indicate the 20 and 30$\sigma$ contours for the Ly$\alpha$ (green). The size of the synthesized beam is shown on the continuum and on the line moment zero maps. As can be seen from the variation of the velocity shift and FWHM, there are hints for marginally resolved kinematics. On the first moment maps of QSO2, AGN1, and LAE1, we highlight the line of nodes (magenta). 
For these sources we also indicate the direction to the QSO (yellow). For AGN1 we also show the direction to LAE1 (gray).} 
\label{cutouts_fig}\
\end{figure*}

Figure~\ref{cutouts_fig} shows cutouts of $7\arcsec \times 7\arcsec$ (or about 53~kpc~$\times$~53~kpc) of the moment zero, first\footnote{The first moment maps are computed with respect to each source CO(5-4) redshift, as listed in Table~\ref{tab:CO54} and shown in Figure~\ref{ALMA_CO54_spectra}.} and second moment maps, together with the continuum at each source position. As can be seen from this figure, the 2~mm continuum and the CO(5-4) emission are found at consistent sky locations within uncertainties. For this reason and to avoid confusion, in this work we only report the coordinates for the continuum (Table~\ref{tab:Interf})\footnote{Following Gaussian deconvolution theory, the position accuracy that can be achieved is ${\rm size}_{\rm beam} / {\rm SNR}$, where ${\rm size}_{\rm beam}$ is the beam size for our ALMA observations. Therefore, the faintest sources detected, those at SNR=4.1, have a position accuracy of 0.24\arcsec.}. 

The sizes of the 2mm continuum and CO(5-4) emitting regions are estimating by performing a 2D fit of the continuum and CO(5-4) moment zero maps. The fit is obtained using the task {\sc imfit} within CASA, selecting a rectangular region of $4\arcsec \times 4\arcsec$ around each source.
We fit a 2D Gaussian profile with the centroid, major and minor axis, position angle, and integrated flux as free parameters. 

All the observed sizes from the fits are in the range $1.1\times - 1.8\times$ of the synthesized beams, with all the observed CO(5-4) emitting regions on scales $\gtrsim 1.3$ the beam size\footnote{We also tested S\'ersic profile fits, finding that Gaussian profiles, i.e. S\'ersic profiles with $n\approx0.5$, are preferred at the current spatial resolution.}. Given the high S/N of the detections,
all the CO(5-4) emissions are therefore resolved (e.g., \citealt{Decarli2018}). 
The effective radius of each CO(5-4) emitting region, defined as the major semiaxis, is found to be $R_{\rm CO(5-4)}=3.0\pm0.8, 5.1\pm0.8, 5.4\pm0.9, 3.6\pm0.7$~kpc, respectively for QSO, QSO2, AGN1, and LAE1\footnote{The errors on the sizes take into account also correlated noise on beam scales following the formalism at \url{https://casa.nrao.edu/docs/taskref/imfit-task.html}.}.  
Hence QSO has likely the most compact host molecular reservoir down to the current depth of the observations.
QSO2 has also the continuum resolved on comparable sizes to its CO(5-4) emission, $R_{\rm 2mm}=5.3\pm1.0$~kpc. All these size measurements are reported in Table~\ref{tab:CO54}, and in section~\ref{sec:extMol} we discuss them in comparison with values from the current literature.

In addition, there are hints for resolved kinematics within each source. Indeed, 
there are symmetric blue and red shifts within the first moment maps of QSO2, AGN1, and LAE1 at the location of the highest S/N in the zero moment maps. Similar kinematic features have been reported in other high-$z$ quasars and have been interpreted as rotation (e.g., \citealt{Bischetti2021}). 
The line of nodes of these tentative rotation-like features was constrained by fitting a simple rotational curve to each object.
Specifically, we perform chi-square minimization to estimate the position angle of the major axis, defined as the angle taken in the anticlockwise direction between the north direction in the sky and the major axis of the galaxy. 
The rotational curve were assumed to follow the simplest function, the arctan (\citealt{Courteau1997}), which is flexible enough to reasonably describe $z \gtrsim 1$ rotating galaxies (e.g., \citealt{Miller2011,Swinbank2012}).  
We follow the procedures described in \citet{TC2017} to project the one-dimensional arctan function to two-dimensional, and run MCMC with the EMCEE Python package (\citealt{ForemanMackey2013}) to fit the velocity maps and obtain posterior probability distributions. Because the asymptotic velocity and inclination angle are essentially degenerate for our data quality, we treat these two as a single parameter in the fit\footnote{We stress that this fitting procedure is S/N weighted.}. As a result, 
the position angles of the major axis are $14^{+4}_{-5}$, $50^{+22}_{-22}$, and $120^{+9}_{-40}$ degrees for QSO2, AGN1, and LAE1, respectively.
We highlight the obtained line of nodes (magenta) in Figure~\ref{cutouts_fig} and list in Table~\ref{tab:Angles} the angles $\phi$ defining these directions in the reference frame East of North, together with their uncertainties. Table~\ref{tab:Angles} also lists the angles $\theta$ between the spin directions of QSO2, AGN1, and LAE1 with respect to the direction to the QSO on the projected plane. 
For each companion source we found that its line of nodes (spin) is almost perpendicular (parallel) to its direction to the QSO, though with large uncertainties.

\begin{table}
\begin{center}
\caption{\bf Angle measurements (in the reference frame East of North).} 
\scalebox{1}{
\footnotesize
\setlength\tabcolsep{4pt}
\begin{tabular}{lccc}
\hline
\hline
ID	&	$\phi^{\rm a}$	&  $\theta^{\rm b}$ 		&	$\phi^{\rm c}_{\rm off}$	\\
        &  (deg)        	&   (deg)   		&	(deg)			        \\
\hline\\[-2mm]				       
QSO2	&   14$^{+4}_{-5}$     & 8$^{+4}_{-5}$ 		&	14$\pm$39   			 \\[2mm] 
AGN1	&   50$^{+22}_{-22}$   & 28$^{+22}_{-22}$ 	&	50$\pm$76 		 \\[2mm]
LAE1	&   120$^{+9}_{-40}$   & 15$^{+9}_{-40}$ 	&	96$\pm$28 		 \\[2mm] 
\hline
\hline
\end{tabular}
}
\flushleft{{\scriptsize $^{\rm a}$ These major axis angles are obtained from the first moment maps as described in section~\ref{sec:CO54results}. \\ 
$^{\rm b}$ Angles $\theta$ between the spin vector of each object and the direction to QSO. \\  
$^{\rm c}$ These major axis angles are obtained from the offset positions in zero moment maps at negative and positive velocities, as described in section~\ref{sec:CO54results}.} }\\  
\label{tab:Angles}
\end{center}
\end{table}

\begin{figure}
\centering
\includegraphics[width=1.0\columnwidth]{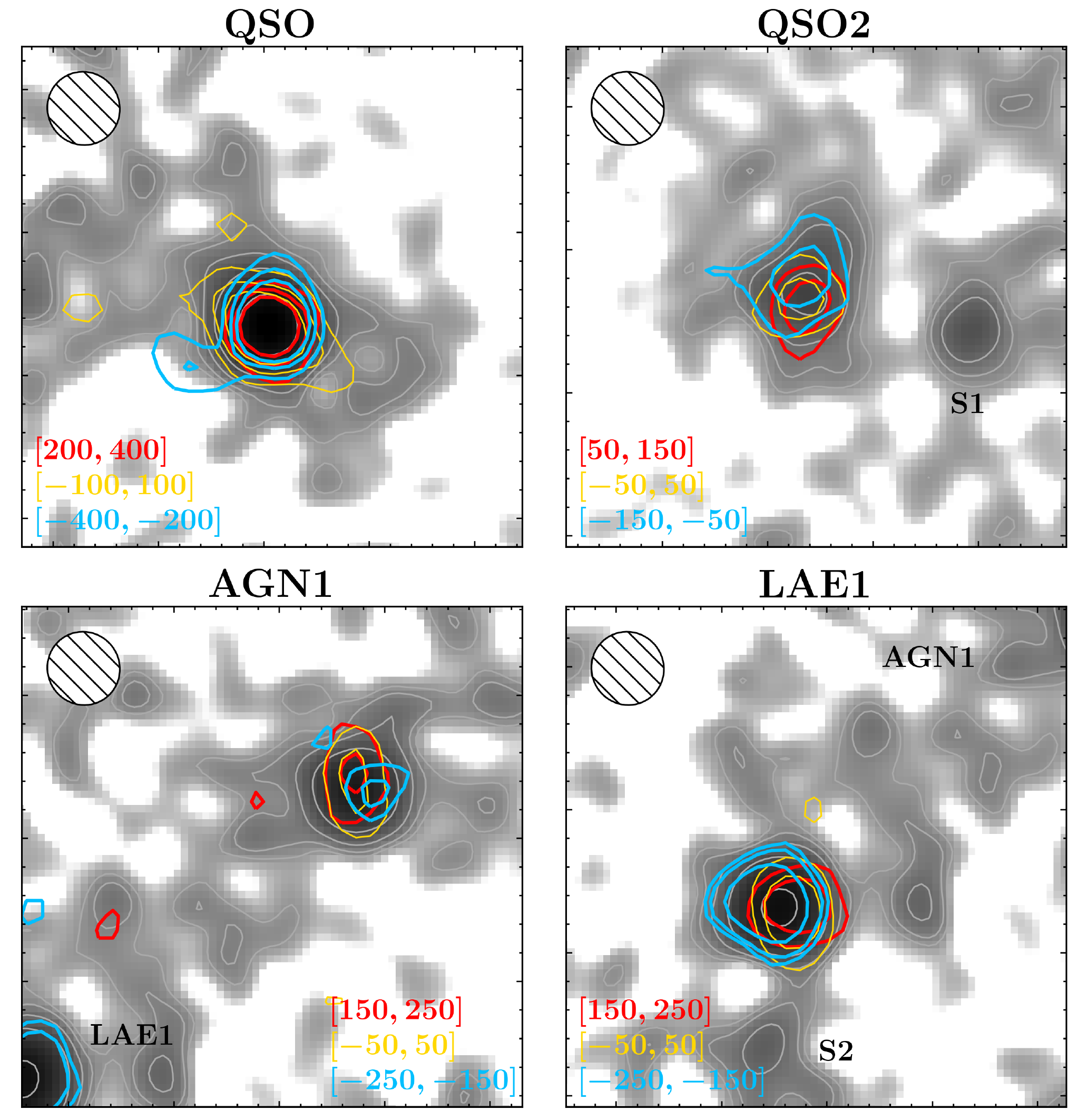}
\caption{CO(5-4) contours from zero moment maps within different velocity ranges (see colored legend on each panel) overlayed on the ALMA continuum maps for the sources detected in CO(5-4): QSO, QSO2, AGN1, and LAE1. Each panel is $7\arcsec \times 7\arcsec$ (or about 53~kpc~$\times$~53~kpc) as in Fig~\ref{cutouts_fig}. The contours are drawn at 3, 5, and 7$\sigma$, if possible.}
\label{vel_channel}\
\end{figure}

To further inspect these velocity shifts, we produced zero moment maps within several velocity ranges. Figure~\ref{vel_channel} shows an example of these maps, highlighting the location of the CO(5-4) emission at negative (blue contours), positive (red), and around zero (yellow) velocities with respect to the 2~mm continuum emission. This test further confirms our previous analysis. In particular, QSO does not show a spatial offset between the emission at positive and negative velocities. The three peaks in its integrated CO(5-4) spectrum seems therefore not associated with three distinct components at this spatial resolution. On the other hand, we find significant offsets between emission at negative and positive velocities of $0.4\pm0.2$, $0.4\pm0.2$, and $0.5\pm0.1$ arcsec for QSO2, AGN1, and LAE1, respectively. As a sanity check, the directions of these shifts are consistent with the major axis computed by fitting the moment maps. However, they have larger uncertainties as they are not based on a fit to the full data information. Specifically, the angles between the negative and positive peak are found to be $14\pm39$, $50\pm76$, and $96\pm28$ degrees for QSO2, AGN1, and LAE1, respectively. We also list these angles in Table~\ref{tab:Angles}. 

Finally, Figure~\ref{cutouts_fig} compares the location of the centroid of the Ly$\alpha$ emission of each source with the millimeter observations. While the QSO and QSO2 locations are consistent at different wavelengths, AGN1 and LAE1 show measurable shifts between the 2~mm continua (or CO(5-4) emission) and the Ly$\alpha$ emission. We estimate these to be 1.5$\arcsec$ (or $\sim11.4$~kpc) and 0.9$\arcsec$ (or $\sim6.8$~kpc), respectively for AGN1 and LAE1. These shifts are significant for both the ALMA and MUSE observations\footnote{The MUSE observations in \citet{FAB2018} have a seeing of $0.66\arcsec$.}. We stress that we have verified the astrometric calibration of the MUSE observations presented in \citet{FAB2018} against the two available sources (one is QSO) in the GAIA DR2 catalogue (\citealt{GAIA2018}) within the observations field-of-view. 
We found agreement between the astrometric calibration done using the SDSS DR12 catalogue (\citealt{sdssdr12}) and the few GAIA sources, confirming the precision of astrometry of about one pixel of the IFU data ($\sim0.2\arcsec$). We further note that the offsets of AGN1 and LAE1 are in opposite directions with respect to each other, which would require a rather weird distortion pattern throughout the data to cancel it out. Therefore, we are confident that the aforementioned shifts between millimeter observations and Ly$\alpha$ are real. 
We discuss these shifts in Section~\ref{sec:Mol_vs_Lya}.

\section{Estimated mass budget of the ELAN system}
\label{bar_budget}

In this section we attempt a first estimate of the mass budget of this ELAN system, specifically within the dark-matter (DM) halo expected to host the system in a $\Lambda$CDM universe. We start by estimating in first approximation the stellar masses, dust masses, star formation rates, and molecular gas masses for the sources with confirmed association with the ELAN, i.e., QSO, QSO2, AGN1 and LAE1 (sections~\ref{sec:dustsfr} and \ref{sec:molgasmass}). We then show that the derived masses for each source are consistent with the dynamical masses estimated from the CO(5-4) line emission under the assumption of a reasonable inclination angle (section~\ref{sec:DynMass}). In section~\ref{sec:dark_comp}, the estimated stellar masses are used to infer the DM halo mass of the system using the halo mass $M_{\rm DM}$ - stellar mass $M_{\rm star}$ relations in \citet{Moster2018}. The inferred DM halo mass is found in agreement with the estimate from an orthogonal method using the projected distances and redshift differences of the sources (\citealt{Tempel2014}). We then discuss in section~\ref{sec:bar_comp} the mass budget in the baryonic components, under several assumptions and also taking into account different baryon fractions. We conclude the section by forecasting the system evolution (section~\ref{sec:evo}).
In each section we discuss the limitations and assumptions of each method, and also indicate some of the needed datasets to refine our estimates.

\subsection{Dust and stellar masses, and star formation rates}
\label{sec:dustsfr}

\begin{figure*}
\centering
\includegraphics[width=1.0\textwidth]{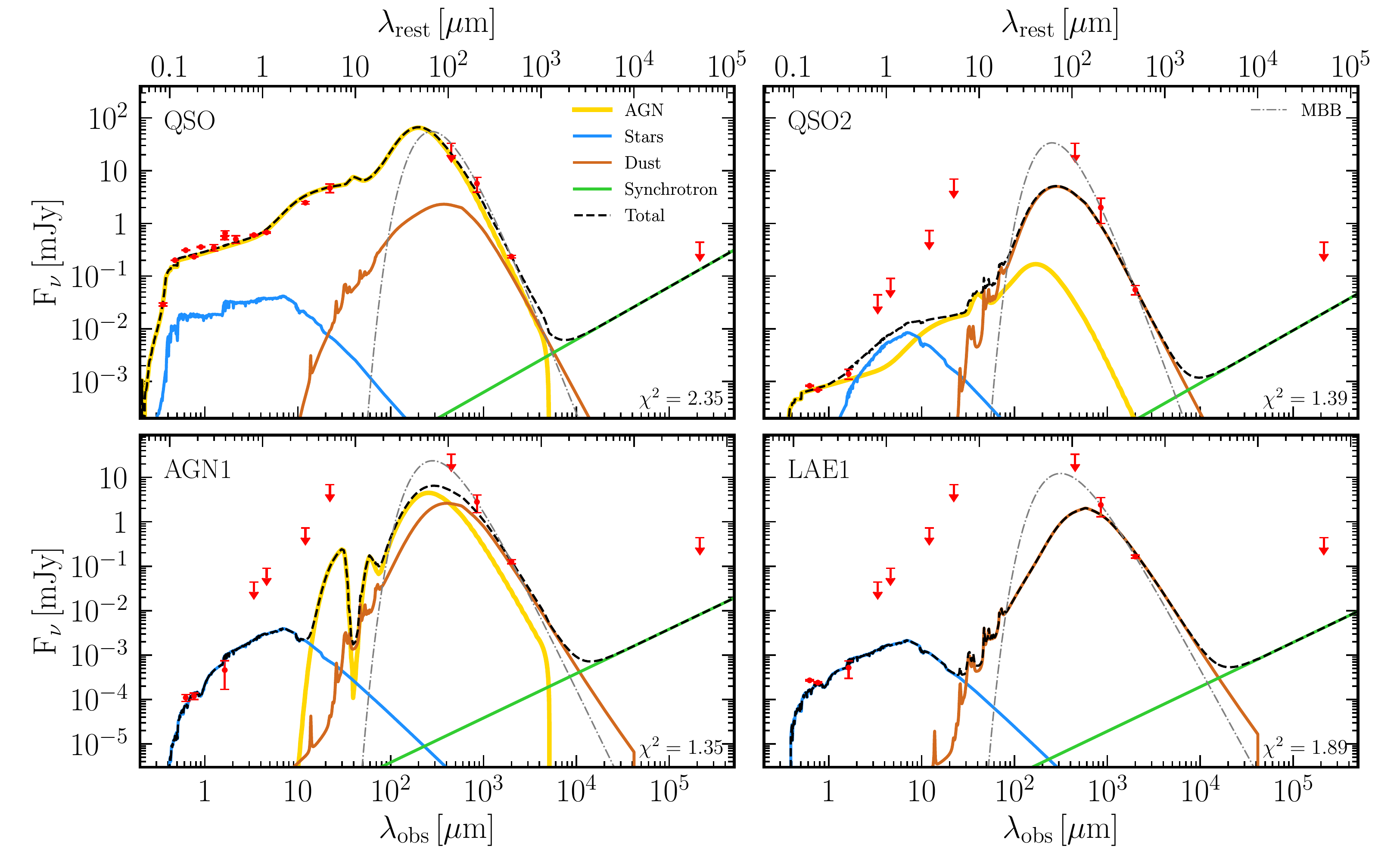}\\
\caption{Observed spectral energy distribution (SED) for each CO(5-4) detected source, QSO, QSO2, AGN1, LAE1. We fit the data-points (red)
with the SED fitting code {\sc CIGALE} (\citealt{Boquien2019}; see section~\ref{sec:dustsfr} for details). We show the best-fit model (black dashed line), and its individual components: stellar emission (blue), dust emission (brown), AGN emission (yellow), synchrotron radiation (green). We report the $\chi^2$ of the best-fit model in the bottom right corner. Table~\ref{tab:CO54} lists some of the properties obtained from these fits. For each source we also indicate a modified black-body with T=40~K for the FIR emission (dot-dashed gray line; see section~\ref{sec:dustsfr}).} 
\label{ALL_SEDs}\
\end{figure*}

The dust and stellar masses, as well as the star formation rates (SFRs) are estimated by fitting the spectral energy distribution (SED) of each source, as usually done in the literature (e.g., \citealt{CalistroRivera2016,Circosta2018}). The SEDs are built using the data described in the previous sections, together with the information at $3.4, 4.6, 12, 22$~$\mu$m from the AllWISE source Catalog\footnote{\url{https://wise2.ipac.caltech.edu/docs/release/allwise/}}, and at $1.4$~GHz from VLA FIRST (\citealt{Becker1994}). These additional data-points are listed in Appendix~\ref{app:dataSED} (Table~\ref{tab:app_phot}).

We rely on the SED fitting code {\sc CIGALE} (v2018.0, \citealt{Boquien2019}), which covers the full range of the current datasets, from rest-frame ultraviolet (UV) to radio emission. {\sc CIGALE} fits simultaneously all this wavelength range imposing energy balance between the UV and the infrared (IR) emission (reprocessed dust emission), while decomposing the SED into different physically motivated components. This energy balance is critical for getting meaningful stellar masses with few datapoints. For our specific case, we select (i) an AGN component (accretion disk plus hot dust emission; \citealt{Fritz2006}), (ii) dust emission from star-forming regions (\citealt{DraineLi2007,Draine2014}), (iii) radio synchrotron emission, and (iv) stellar emission from the host galaxy, which is modelled by an exponentially declining star formation history (SFH), the simple stellar population models of \citet{BruzualCharlot2003}, a Chabrier initial mass function (\citealt{Chabrier2003}), and a modified starburst attenuation law (based on \citealt{Calzetti2000} and \citealt{Leitherer2002}). Details on these specific models and a comparison with other models implemented in {\sc CIGALE} are discussed in \citet{Boquien2019}. The parameters available for each model using the code's notation and the ranges explored by our fit are:

\begin{itemize}
\item AGN emission: this model has seven parameters, five of which are left free to explore all the values allowed by {\sc CIGALE}. The remaining parameters are the AGN fraction ($fracAGN$; defined as the ratio of the AGN luminosity to the sum of the AGN and dust luminosities) and the angle between equatorial axis and line-of-sight ($psy$). We let $fracAGN$ vary between 0 and 1 in steps of 0.05. $psy$ is allowed to vary between 0.001 and 40.100 for type-2 AGN, and between 50.100 and 89.990 for type-1 AGN\footnote{In the case of LAE1, for which no AGN signature is present in the MUSE data (\citealt{FAB2018}), we neglect the AGN component during the fit.}.
\item dust emission: this model has four parameters (mass fraction of PAH, minimum radiation field, power-law slope, fraction of dust illuminated) which are left free to explore all the values allowed by {\sc CIGALE}.
\item synchrotron emission: this model has two parameters, the value of the FIR-to-radio coefficient (\citealt{Helou1985}) and the slope of the synchrotron power-law. Given the absence of tight constraints in the radio for any of the sources we fixed the slope to -1.0 (as observed in sources within other ELANe, e.g., \citealt{Decarli2021}) 
and the ratio to an arbitrary value satisfying the VLA FIRST upper limits. This portion of the SED has to be considered simply as illustrative.
\item stellar emission: the SFH is modelled with two parameters, age and e-folding time $\tau$. The age is allowed to vary between 0.1~Gyr and 2~Gyr (about the age of the universe at $z=3.164$) in step of 0.1 Gyr, while $\tau$ can vary between 0.1 and 10~Gyr in step of 0.1~Gyr. The attenuation model is set up so that the final $E(B-V)$ is between 0 and 3. All the other eight parameters are kept to the default values.
\end{itemize}

In addition, for high-redshift sources, {\sc CIGALE} applies a correction to rest-frame UV data for the attenuation from the foreground intergalactic medium following \citet{Meiksin2006}. 

In the continuum fit we do not include the nebular emission component, for which {\sc CIGALE} has built-in templates. Indeed, we find that the nebular $Ly\alpha$ line emission from some of these sources is displaced with respect to the continuum (e.g., section~\ref{sec:CO54results}). The best-fit models obtained following this procedure using the {\it pdf\_analysis} module in {\sc CIGALE} are shown in Figure~\ref{ALL_SEDs}, together with their $\chi^2$ values and the observed data-points. The likelihood-weighted output dust and stellar masses, and the SFRs together with their likelihood-weighted uncertainties are listed in Table~\ref{tab:CO54}. We stress that these uncertainties do not include systematic errors due to the models used, a priori assumptions on the nature of the sources, and the discrete coverage of the parameter space.

Specifically, we find dust masses in the range $M_{\rm dust}=1.5$-$16 \times 10^{8}$~M$_{\odot}$, with LAE1 being the most dust rich object in the system. As an additional check, we computed the dust masses 
using a modified black body model, assuming (i) a dust temperature $T_{\rm dust}=40$~K \footnote{This temperature is in the range of  $T_{\rm dust}$ for high-redshift quasars (e.g., \citealt{CarilliWalter2013}).}, (ii) a dust opacity at 850~$\mu$m of $\kappa_{\rm d}=0.43$~cm$^2$~g$^{-1}$ (\citealt{Li2001}), and (iii) a fixed dust emissivity power-law spectral index $\beta$ derived from the SMA and ALMA continuum.  
We find $\beta=2.4\pm0.1, 2.9\pm0.2,  2.3\pm0.2, 1.8\pm0.2,$ for QSO, QSO2, AGN1, and LAE1, respectively. The values for QSO, QSO2, and AGN1 are on the high side of 
the values usually found for high-redshift quasars (e.g., $\beta=1.95\pm0.3$, \citealt{PriddeyMcMahon2001}; $\beta=1.6\pm 0.1$, \citealt{Beelen2006}) or dusty star-forming galaxies (e.g., $\beta=2.0\pm0.2$ , \citealt{Magnelli2012}), and are consistent with the value of $\beta=2.5$ used to fit SEDs of HzRGs (\citealt{Falkendal2019}). 
The dust masses  
derived with this method are $M_{\rm dust} =  (9\pm3)\times 10^{8}$~M$_{\odot}$, $(3\pm2)\times 10^{8}$~M$_{\odot}$, $(5\pm2)\times 10^{8}$~M$_{\odot}$, and $(4\pm2)\times 10^{8}$~M$_{\odot}$, respectively for QSO, QSO2, AGN1, and LAE1\footnote{In this calculation we assumed dust to be optically thin in all four sources. Given the current source sizes estimated at $2$~mm, $\beta$ and the assumed dust opacity, this assumption is confirmed except for QSO, for which $\tau_{\rm dust} \geq 1$ at $\lambda < 145$~$\mu$m. Nevertheless, we do not have any information on the source sizes at $\lambda<2$~mm, and we decided to quote for QSO the $M_{\rm dust}$ value for the optically thin case for ease of comparison with the other sources. An optically thick calculation for QSO would give lower dust masses (e.g., \citealt{Spilker2016,Cortzen2020}) in better agreement with the {\sc CIGALE} fit.}. Hence, they agree with the {\sc CIGALE} output (LAE1 within 2$\sigma$). 
The obtained dust masses are (i) in the range usually derived in high-redshift quasars hosts from $z\sim2$ up to $z\sim7$ (e.g., \citealt{Weiss2003, Schumacher2012, Venemans2016}), (ii) within the typical range for high-redshift dusty, star-forming galaxies (e.g., \citealt{Casey2014, Dudzeviciute2020}), and (iii) similar to what is reported for HzRGs (few times $10^{8}$~M$_{\odot}$ assuming $T\sim50$~K; e.g., \citealt{Archibald2001}). 

The stellar masses are found to be in the range ${M_{\rm star}=4.0\times10^9}$~M$_{\odot}$ and $1.2 \times 10^{11}$~M$_{\odot}$, with QSO being hosted by the most massive galaxy in this system, as expected (\citealt{FAB2018}). The other associated galaxies are about 10$\times$ less massive (Table~\ref{tab:CO54}). Given that QSO greatly outshine its host galaxy the stellar mass derived with CIGALE has to be taken with caution. However, we show in section~\ref{sec:DynMass} that the dynamical mass derived from CO(5-4) is consistent with the presence of such a massive galaxy.
The obtained stellar masses agree well with literature values for quasar hosts (e.g., $\sim 10^{11}$~M$_{\odot}$, \citealt{Decarli2010}) and the more massive and IR detected LAEs (e.g., $10^9-10^{10.5}$~M$_{\odot}$, \citealt{Ono2010}) found at similar redshifts. 
The QSO host has a stellar mass consistent with the median stellar mass for the Spitzer $1<z<5.2$ HzRGs ($\sim10^{11}$~M$_{\odot}$; \citealt{DeBreuck2010}).

With these estimates, the dust-to-stellar mass ratios are 
$M_{\rm dust}/M_{\rm star} = 0.006\pm0.004, 0.010\pm0.004, 0.05\pm0.02, 0.4\pm0.1$, for QSO, QSO2, AGN1, and LAE1, respectively. These values are in agreement within uncertainties with observations of main-sequence and starburst galaxies at these redshifts (in the range $0.001-0.1$, e.g., \citealt{daCunha2015,Donevski2020}), except LAE1 which has a larger value. Given the similarity with the SED of AGN1, this tension could be solved by including an AGN component in the SED fit for LAE1. The current dataset does not show a clear evidence for  AGN activity in LAE1, but it could be obscured. X-rays observations are therefore needed to verify the nature of this source.

Further, {\sc CIGALE} determined the instantaneous SFR for each source, indicating ${\rm SFR}=521\pm74,183\pm49,104\pm21,50\pm11$~M$_{\odot}$~yr$^{-1}$ for QSO, QSO2, AGN1, LAE1, respectively. 
These SFRs are in line with values from the literature. In particular, the SFR in the QSO host agrees well with values usually found in $z\sim2-3$ quasars (e.g., \citealt{Harris2016}). We also computed the SFR from the total infrared (IR) emission using the rest-frame wavelength range $8-1000$~$\mu$m for the obtained SED. For this purpose, we used the relation (SFR$_{\rm IR}/[M_{\odot}~{\rm yr}^{-1}])=3.88\times10^{-44} (L_{\rm IR}/[{\rm erg~s^{-1}}])$ in \citet{Murphy2011}, usually employed for high-redshift quasars (e.g., \citealt{Venemans2017}). This relation has been computed using Starburst99 (\citealt{Leitherer1999}) with a fixed SFR and a Kroupa initial mass function (\citealt{Kroupa2001}), and assumes that the entire Balmer continuum is absorbed and re-radiated by dust in the optically thin regime. Applying this relation only to the IR luminosity $L_{\rm IR}$ computed excluding the AGN contribution (see Table~\ref{tab:CO54}),  
results in ${\rm SFR}_{\rm IR}=700\pm30,1087\pm283,447\pm149,179\pm30$~M$_{\odot}$~yr$^{-1}$ for QSO, QSO2, AGN1, LAE1, respectively. These values are higher than the instantaneous SFRs, reflecting the longer timescales probed by the IR tracers. 

\begin{figure}
\centering
\includegraphics[width=0.95\columnwidth]{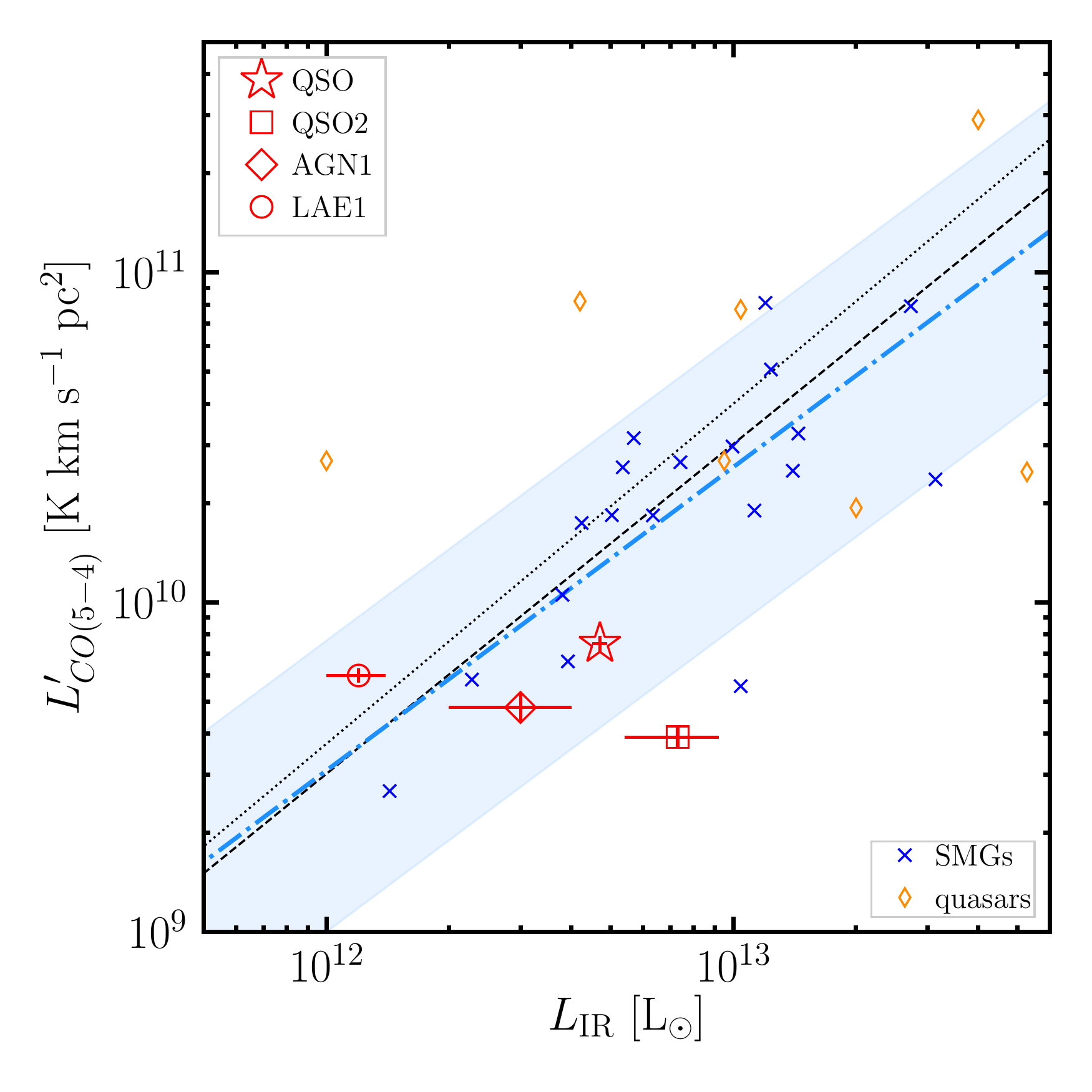}\\
\caption{$L^{\prime}_{\rm CO(5-4)}$ versus $L_{\rm IR}$ plot. We compare the location of QSO, QSO2, AGN1, and LAE1 (red) with SMGs (compilation in \citealt{CarilliWalter2013}) and high-redshift quasars ($z\sim2.5-6$; \citealt{Barvainis1997,Guilloteau1999,Weiss2003,Carilli2007,Wang2007,Wang2010,Schumacher2012}). We stress that $L_{\rm IR}$ for most of the high-redshift quasars is not corrected for a contribution from the AGN. We also indicate (i) the relation by \citet{Greve2014} obtained by fitting only low-redshift sources (no SMGs nor high-redshift quasars included; dotted line), (ii) the relation from \citet{Daddi2015} obtained by fitting from local spirals to SMGs (no high-redshift quasars considered; dashed line), and (iii) the relation from \citet{Valentino2020} (blue line with its intrinsic scatter), who updated the work by \citet{Daddi2015} using a larger galaxy sample (no high-redshift quasars considered).} 
\label{Lprime_CO}\
\end{figure}

Further, in Figure~\ref{Lprime_CO} we show the location of QSO, QSO2, AGN1, and LAE1 in the $L_{\rm IR}$ versus $L^{\prime}_{\rm CO(5-4)}$ plot in comparison to submillimeter galaxies (SMGs) and high-redshift quasars.  The CO(5-4) line is known to be linked to $L_{\rm IR}$ through a relation of the form log${L_{\rm IR}=\alpha {\rm log}L^\prime_{\rm CO(5-4)}+\beta}$ from low to high-redshift (e.g., \citealt{Greve2014, Daddi2015}, dotted and dashed lines). Our sources are consistent with the scatter of known high-redshift sources (e.g., \citealt{Valentino2020}, blue line with intrinsic scatter), ensuring that the AGN-corrected $L_{\rm IR}$ obtained from the SED fitting are reasonable. 

Summarizing, we find values for dust and stellar masses, and SFRs within the scatter of observations reported in the literature. Follow-up observations in the NIR and mm regimes are needed to lower the uncertainties on our estimates.

\subsection{Molecular gas masses derived from CO}
\label{sec:molgasmass}

It is common practice to obtain the molecular mass through the equation $M_{\rm gas}=\alpha L_{\rm CO(1-0)}^\prime$, with $\alpha$ being the CO luminosity-to-gas mass conversion factor and $L_{\rm CO(1-0)}^\prime$ the CO(1-0) luminosity in units of K~km~s$^{-1}$~pc$^2$ (e.g., \citealt{CarilliWalter2013,Aravena2019}). To use this equation, we assume $\alpha_{\rm CO}=0.8$~M$_{\odot}$~$({\rm K\, km\, s^{-1}\, pc^2})^{-1}$. This value has been derived for local ultra-luminous infrared galaxies (ULIRGs; e.g., \citealt{DownesSolomon1998}), and it is commonly used to calculate molecular gas masses in high-redshift quasars ($z \sim 2-7$; e.g., \citealt{Riechers2006,Coppin2008,CarilliWalter2013,Venemans2017}). As only 
the CO(5-4) line flux is available, we have to further assume a CO spectral line energy distribution (CO SLED) to derive how strong is the CO(1-0) transition. Current statistics show that the CO SLED of high-redshift quasars peaks at high-$J$ transition, CO(6-5) and CO(7-6), with a minimum flux ratio CO(5-4)/CO(1-0)$\sim10$ (\citealt{Weiss2007,CarilliWalter2013}). Therefore, 
we derive the corresponding $L_{\rm CO(1-0)}^\prime$ assuming this ratio.
The values obtained for the three AGN (QSO, QSO2, AGN1) are considered as possible upper limits (i.e. their CO SLED could be more excited; e.g., \citealt{Weiss2007a}), while for LAE1 
it is possibly a lower limit. For completeness we list the inferred CO(1-0) luminosities in Table~\ref{tab:CO54}. We note that the adopted CO(5-4)/CO(1-0) corresponds to a $r_{51}=L_{\rm CO(5-4)}^\prime/L_{\rm CO(1-0)}^\prime = 0.4$ (the range reported in \citet{CarilliWalter2013} is $r_{51}=0.69\pm0.3$). This value is also consistent within 2$\sigma$ with the values reported for the spectrum obtained for $z>2$ gravitationally lensed dusty star-forming galaxies by \citet{Spilker2014} ($r_{51}=0.67\pm0.20$), with the median value reported for 32 $z\sim1.2-4.1$ luminous submillimeter galaxies ($r_{\rm 51}=0.32\pm0.05$; \citealt{Bothwell2013}), the  average value for 8 $z>2$ star forming galaxies ($r_{51}=0.44\pm0.11$; \citealt{Boogaard2020}), and for the compilation of SMGs in \citet{Birkin2021} ($r_{51}=0.35\pm0.08$). 

Using the derived CO(1-0) luminosities and the assumed $\alpha_{\rm CO}$, 
the resulting gas masses $M_{\rm gas}$ are $(1.5\pm0.1)\times10^{10}$, $(7.7\pm0.7)\times10^9$, $(1.0\pm0.1)\times10^{10}$, and $(1.2\pm0.1)\times10^{10}$~M$_{\odot}$, respectively for QSO, QSO2, AGN1, and LAE1 (see also Table~\ref{tab:CO54})\footnote{We stress that the errors on the molecular gas masses here reported do not include systematics due to the uncertain $\alpha_{\rm CO}$, the CO excitation, and the possibility of having some CO-dark gas in these systems (e.g., \citealt{Balashev2017}). For example, if the conditions in the molecular gas are more similar to the Milky-Way, the molecular gas masses could be up to a factor of 5 larger, i.e. $\alpha_{\rm CO}\sim4$~M$_{\odot}$~$({\rm K\, km\, s^{-1}\, pc^2})^{-1}$ (e.g., \citealt{Bolatto2013}).}. These values, though uncertain, are within the ranges  
reported in the literature for high-redshift quasars
 and dusty star-forming galaxies ($M_{\rm gas}={\rm few} \times 10^{10}(\alpha_{\rm CO}/0.8)$~M$_{\odot}$; e.g., \citealt{CarilliWalter2013, Bothwell2013}), and are also at the low-end of masses reported for HzRGs ($10^{10}-10^{11}$~M$_{\odot}$; e.g., \citealt{MileyDeBreuck2008}). 

When comparing the obtained molecular gas masses to the dust masses derived in Section~\ref{sec:dustsfr}, we find molecular-to-dust mass ratios $r_{\rm H_2, dust}$ in the range 8 - 51 (see Table~\ref{tab:CO54}), which are very low in comparison to the usually assumed gas-to-dust mass ratios for local galaxies ($\sim100$; e.g., \citealt{Draine2007,Galametz2011})\footnote{Local galaxies, however, show molecular gas-to-dust ratios in a wide range dependent on metallicity, e.g.  $5\lesssim r_{\rm H_2, dust} \lesssim 4000$ with a median of 177 (\citealt{RemyRuyer2014}).} and high-redshift massive star-forming galaxies ($\sim100$; e.g., \citealt{Riechers2013}), even when correcting them for the fraction of gas in molecular form ($\sim 80$~\%; e.g., \citealt{Riechers2013}).  Interestingly, these values are more similar to what is seen in SMGs (e.g., $r_{\rm H_2, dust}=28^{+7}_{-6}$; \citealt{Kovacs2006}). Therefore, our measurement could be due to a real molecular gas deficiency or efficient dust absorption (i.e. larger dust opacities) in these sources, or could be related to the assumptions made to determine the molecular gas masses.  
Assuming a Milky-Way value of $\alpha_{\rm CO}\sim4$~M$_{\odot}$~$({\rm K\, km\, s^{-1}\, pc^2})^{-1}$, the molecular-to-dust mass ratios 
increase, but still two sources (AGN1 and LAE1) 
have values reflecting a depletion of molecular gas. Less excited CO ladders would then be needed to increase our molecular mass estimates and thus alleviate the tension for these sources. It is clear that our measurements need to be refined with follow-up observations of additional CO transitions (especially at lower $J$; ALMA, NOEMA, J-VLA) or other molecular gas tracers (e.g., [CI]), to at least remove  
the uncertainties on the CO excitation ladder.  
For completeness,  
the $r_{\rm H_2, dust}$ values are also listed in Table~\ref{tab:CO54}.

\subsection{Dynamical Masses from CO kinematics and sizes}
\label{sec:DynMass}

In this section we outline rough estimates of the dynamical masses for QSO, QSO2, AGN1, and LAE1, using the kinematics and sizes of the CO(5-4) emitting region. In turn, these dynamical masses can be compared to the galaxies mass budget derived in the previous sections. As  
rotation-like signatures are present in most of the CO(5-4) maps (Section~\ref{sec:CO54results}),
the bulk of the molecular gas is assumed to be in a disk with an inclination $i$. This approach is common practice in the study of unresolved line tracers of molecular gas associated with high-redshift quasars (e.g., \citealt{Decarli2018}). In this framework, the dynamical mass can be obtained as $M_{\rm dyn}=G^{-1}R_{\rm CO(5-4)}({\rm FWHM}/{\rm sin}(i))^2$ (\citealt{Willot2015}), where $G$ is the gravitational constant, $R_{\rm CO(5-4)}$ is the size of the CO(5-4) emitting region and FWHM is its line width\footnote{In this approximated calculation,  
the frequently used 0.75 factor to scale the line FWHM to the width of the line at 20\%
is omitted because the integrated CO(5-4) line shape is not a simple Gaussian.}.  
As the quality of our data does not allow an estimate of the inclination angle (Section~\ref{sec:CO54results}), we assume the mean inclination angle for randomly oriented disks $\langle {\rm sin}(i) \rangle =\pi/4$ (e.g., \citealt{Law2009}), obtaining $M_{\rm dyn}=(1.9\pm0.8)\times10^{11}, (8.5\pm6.7)\times10^{10}, (3.9\pm2.9)\times10^{11}, (8.2\pm5.2)\times10^{10}$~M$_{\odot}$ for QSO, QSO2, AGN1, and LAE1, respectively. These dynamical masses are  consistent within their large uncertainties to the sum of the galaxy different mass components (i.e., molecular, stellar, dust), also considering the contribution of a Navarro-Frenk-White (NFW; \citealt{nfw97}) dark-matter component within $R_{\rm CO(5-4)}$ assuming the concentration-halo mass relation in \citet{Dutton2014}.  

For QSO2, AGN1, and LAE1,  
the dynamical masses including some pressure support can be further assessed, using the observed asymptotic rotational velocities $v_{\rm rot}^{\rm obs}$ obtained from the 2D fit of their first moment maps described in section~\ref{sec:CO54results}, and the velocity dispersion $\sigma$ within their effective radii in their second moment maps. In this framework, $M_{\rm dyn}=2R_{\rm CO(5-4)}(v_{\rm rot}^{2}+\sigma^{2})/G$ (e.g., \citealt{Smit2018}), where $v_{\rm rot}=v_{\rm rot}^{\rm obs}/{\rm sin}(i)$, assuming again $\langle {\rm sin}(i) \rangle =\pi/4$. Using the computed values of $v_{\rm rot}^{\rm obs}=170^{+30}_{-60}, 190^{+60}_{-60}, 190^{+50}_{-40}$~km~s$^{-1}$ and $\sigma=119\pm89, 259\pm153, 156\pm138$~km~s$^{-1}$, we obtain $M_{\rm dyn}=1.4^{+0.6}_{-0.7}\times10^{11},3.1^{+2.1}_{-2.1}\times10^{11}, 1.4^{+0.8}_{-0.8}\times10^{11}$~M$_{\odot}$ for QSO2, AGN1 and LAE1, respectively. These masses are consistent with the previously determined values.

Notwithstanding the aforementioned fair agreement between the estimated dynamical masses and the mass budget for each galaxy, we note that the obtained dynamical masses are usually providing
larger mass estimates, 
especially for AGN1. This tentative mismatch can however be evidence of turbulence injection in the molecular reservoir due to different physical processes expected in galaxy evolution: infall of gas at velocities of hundreds of km~s$^{-1}$, stellar and AGN feedback. In other words, turbulence due to these processes could explain the large velocity dispersions seen in the four CO(5-4) detected objects (Figure~\ref{cutouts_fig}). Higher resolution observations, exploiting the ALMA longest baselines, are required to 
firmly assess the dynamical masses of these sources.

\subsection{The system mass budget}
\label{sec:massBud}

In this section we present an estimation of the mass budget of the whole system and compare that to the cosmic baryon fraction. To compute the dark matter and baryonic (stars, molecular gas, dust, atomic gas) components, we  
rely on the previously obtained values and on several additional assumptions which are needed to overcome the limitations of the current observations.

We assume that the sources detected by ALMA, i.e., QSO, QSO2, AGN1, LAE1, are the most massive objects in this system, 
and neglect the contributions from additional sources. It will be clear that additional satellite masses are well accommodated within the final error estimates of our discussion.

\subsubsection{The dark matter component}
\label{sec:dark_comp}

We used two methods to esimate the total dark matter (DM) halo mass for the system.
First, we interpolate the halo mass $M_{\rm DM}$ - stellar mass $M_{\rm star}$ relations in \citet{Moster2018} for the redshift of interest here, 
to obtain the expected $M_{\rm DM}$ for each of the sources.  
The resulting halo masses are in the range $3.7\times10^{11}$~M$_{\odot} \leq M_{\rm DM} \leq 6.9\times10^{12}$~M$_{\odot}$ (Table~\ref{tab:CO54}). To get a total halo mass, we then simply sum the obtained masses, finding $M_{\rm DM}^{\rm total} = (8.6^{+19.4}_{-5.6})\times10^{12}$~M$_{\odot}$, with 81~\% of the DM mass due to the QSO halo\footnote{The $M_{\rm DM}$-$M_{\rm star}$ relations in \citet{Moster2018} relate the stellar mass of the galaxy with its smooth dark-matter halo excluding subhalos. For the three companions of the quasar the $M_{\rm DM}$ estimates are likely upper limits since they might have already suffered some degree of tidal stripping.}. The large uncertainties here are  
due to the well-known challenges in assessing the stellar mass of the bright quasar hosts (e.g., \citealt{Targett2012}), and the larger scatter in halo mass for stellar masses close to the peak efficiency of star formation (\citealt{Moster2018}).

Secondly,  we estimate the dynamical mass of the system as done for low-redshift groups and clusters using the formalism of \citet{Tempel2014}. 
We apply this method to the studied ELAN using the ALMA data (systemic redshifts and positions) for each source.
This method assumes  
that (i) the system is already virialized\footnote{The system studied here might not be virialized. If this is the case, the mass computed assuming virialization is likely an overestimation of the true mass.}, (ii) dynamical symmetry, so that the true velocity dispersion $\sigma_v$ of the system is $\sqrt(3)\times$ the velocity dispersion along the line-of-sight, and (iii) a gravitational radius $R_{\rm g}$ obtained as in \citet{BinneyTremaine2008}, while assuming a DM density profile (here a NFW) and the observed spatial dispersion in the plane of the sky (equation 4 in \citealt{Tempel2014}).  
The total dynamical mass is then given by ${M_{\rm dyn}^{\rm tot}=2.325\times10^{12} (R_{\rm g}/{\rm Mpc})(\sigma_v/100\, {\rm km\, s^{-1}})^2}$~M$_{\odot}$. The observed projected distances and redshift differences result in $R_{\rm g} = 354\pm76$~kpc and $\sigma_v = 515\pm39$~km~s$^{-1}$, and therefore in $M_{\rm dyn}^{\rm tot}=(2.2\pm0.3)\times10^{13}$~M$_{\odot}$. If we then assume a maximum baryon fraction equal to the cosmic baryon fraction\footnote{In this work we assume a cosmic baryon fraction of 0.156 obtained as the ratio of the baryon density $\Omega_{\rm b}$ and matter density $\Omega_{\rm m}$ given by \citet{Planck2020}.}, the total DM mass for the system is $M_{\rm DM}^{\rm total}=(1.8\pm0.3)\times10^{13}$~M$_{\odot}$ \footnote{If we include in this calculation also LAE2, for which its position and redshift are known from \citealt{FAB2018}, the halo mass increases by 2\%.}. 

The two obtained values for the DM mass agree within uncertainties, with the dynamical mass on the high side of the first estimate possibly due to a lack of virialization in this system. Therefore, we can consider the stellar masses obtained in section~\ref{sec:dustsfr} to be reasonable. In the remainder of the analysis we will consider both estimates of DM masses, which overall suggest that this ELAN is sitting in a DM halo of $\sim10^{13}$~M$_{\odot}$.
Interestingly, this halo mass is on the high side of the halo mass measurements presented in the literature for quasars (usually between $10^{12}$ and $10^{13}$~M$_{\odot}$ at $z\sim3$; e.g., \citealt{Shen2007,KimCroft2008,Trainor2012,Eftekharzadeh2015}), possibly further confirming that ELANe are associated with the most massive and therefore overdense quasar systems (\citealt{hennawi+15,FAB2018}). In addition, this ELAN inhabits a DM halo as massive as those expected for HzRGs (e.g., \citealt{Stevens2003}), bright LABs (\citealt{Yang2010}), and SMGs (e.g., \citealt{Wilkinson2017,Lim2020}, but see \citealt{GarciaVergara2020}), revealing that it is among the most massive systems at its redshift.

\subsubsection{The baryonic components}
\label{sec:bar_comp}

For the baryon budget, we proceed by simply adding up the masses of each component for QSO, QSO2, AGN1, and LAE1 and propagating their errors, finding ${M_{\rm star}^{\rm total}=(1.5\pm0.8)\times10^{11}}$~M$_{\odot}$, ${M_{\rm dust}^{\rm total}=(3.0\pm0.4)\times10^9}$~M$_{\odot}$, $M_{\rm H_2}^{\rm total}={(4.5_{-0.2}^{+17.9})\times10^{10}}$~M$_{\odot}$, for the total stellar, dust, and molecular masses.
In the error budget for the molecular mass we include the large uncertainty (a factor of 5) on $\alpha_{\rm CO}$. This large uncertainty should also include the possibility of molecular gas extending on scales larger than the body of galaxies as seen for example in HzRGs environments (e.g., \citealt{Emonts2016}; see Section~\ref{sec:extMol} for discussion). From the mass budget we then miss the atomic gas components at different temperatures, i.e., cold ($\sim 100$~K), cool ($\sim 10^{4}$~K), and warm-hot ($>10^5$~K). 

We can predict the amount of cold atomic gas by assuming that the interstellar-medium molecular gas fraction is $\sim80$~\% at high-redshift (e.g., \citealt{Riechers2013}), and in turn that the cold atomic gas fraction is therefore $f_{\rm cold}\sim20$~\%.  This is also consistent with current estimates for such massive halos at $z\sim3$ from semi-empirical models of galaxy evolution (e.g., \citealt{Popping2015}). We therefore include in the budget a total cold atomic mass of $M_{\rm HI}^{\rm cold}=(0.9_{-0.1}^{+3.6})\times10^{10}$~M$_{\odot}$. This prediction could be tested by targeting the [CII] emission at 158~$\mu$m with e.g., ALMA (e.g., \citealt{Fujimoto2020}).

To derive a total cool gas mass, we can instead rely on the large-scale Ly$\alpha$ emission detected with VLT/MUSE in \citet{FAB2018}. 
Given that the projected maximum distance of the Ly$\alpha$ emission is comparable with 
the obtained virial radii, 
we assume that all the Ly$\alpha$ nebula sits within the halo. This is also in agreement with the discussion in \citet{FAB2018}, who argued that the Ly$\alpha$ emission traces the motions of substructures accreting within the bright quasar massive halo. We then assume that the visible Ly$\alpha$ emitting gas is the densest cool gas in the halo, and thus the one contributing to most of its mass. The cool gas mass can then be estimated as $M_{\rm cool}^{\rm total} = A\, m_{\rm p}/X\, f_C\, N_{\rm H}$ (\citealt{qpq4}), where $A$ is the projected area on the sky covered by the ELAN in cm$^2$ ($609.36$~arcsec$^2$ within the $2\sigma$ isophote in \citealt{FAB2018}), $m_{\rm p}$ is the proton mass, $X=0.76$ is the hydrogen mass fraction (e.g., \citealt{Pagel1997}), $f_{\rm C}$ is the cool gas covering factor within the ELAN, and $N_{\rm H}$ is the total hydrogen column density of the emitting gas. We assume (i) $f_{\rm C}=1$ as it has been shown that the observed morphology of extended Ly$\alpha$ nebulae can be reproduced if $f_{\rm C} \gtrsim 0.5$ (\citealt{fab+15a}), and (ii) a constant ${\rm log}(N_H/{\rm cm^{-2}}) = 20.5 \pm 1.0$, which is the median value found by \citet{QPQ8} for optically thick absorbers in $z\sim2$ quasar halos (see their Figure 15). For this latter value we assume an error which encompasses the large uncertainties in some of those authors data-points. Inserting these values in the aforementioned relation gives  
$M_{\rm cool}^{\rm total} = 1.2_{-1.0}^{+10.5}\times10^{11} $~M$_{\odot}$. We stress that this calculation neglects additional cool gas within the halo not emitting Ly$\alpha$ above the sensitivity of the observations in \citet{FAB2018}. 
However, the current area of the nebula covers 42\% or 25\% of the projected halo for the Moster et al. or the Tempel et al. calculation, respectively. Even if we assume the full halo to be covered by Ly$\alpha$ emission the total mass would increase by a factor of 2.4 or 4, respectively, thus falling within the errors of the previous measurement.
Interestingly, the obtained $M_{\rm cool}^{\rm total}$ agrees well with cool gas masses reported for $z\sim2-3$ quasars halos ($M_{\rm cool}^{\rm total}>10^{10}$~M$_{\odot}$; e.g., \citealt{QPQ5,QPQ8}), and reported for other ELANe ($1.0\times10^{11}$~M$_{\odot} < M_{\rm cool}^{\rm total} < 6.5\times10^{11}$~M$_{\odot}$; \citealt{hennawi+15}).

\begin{figure}
\centering
\includegraphics[width=1.0\columnwidth]{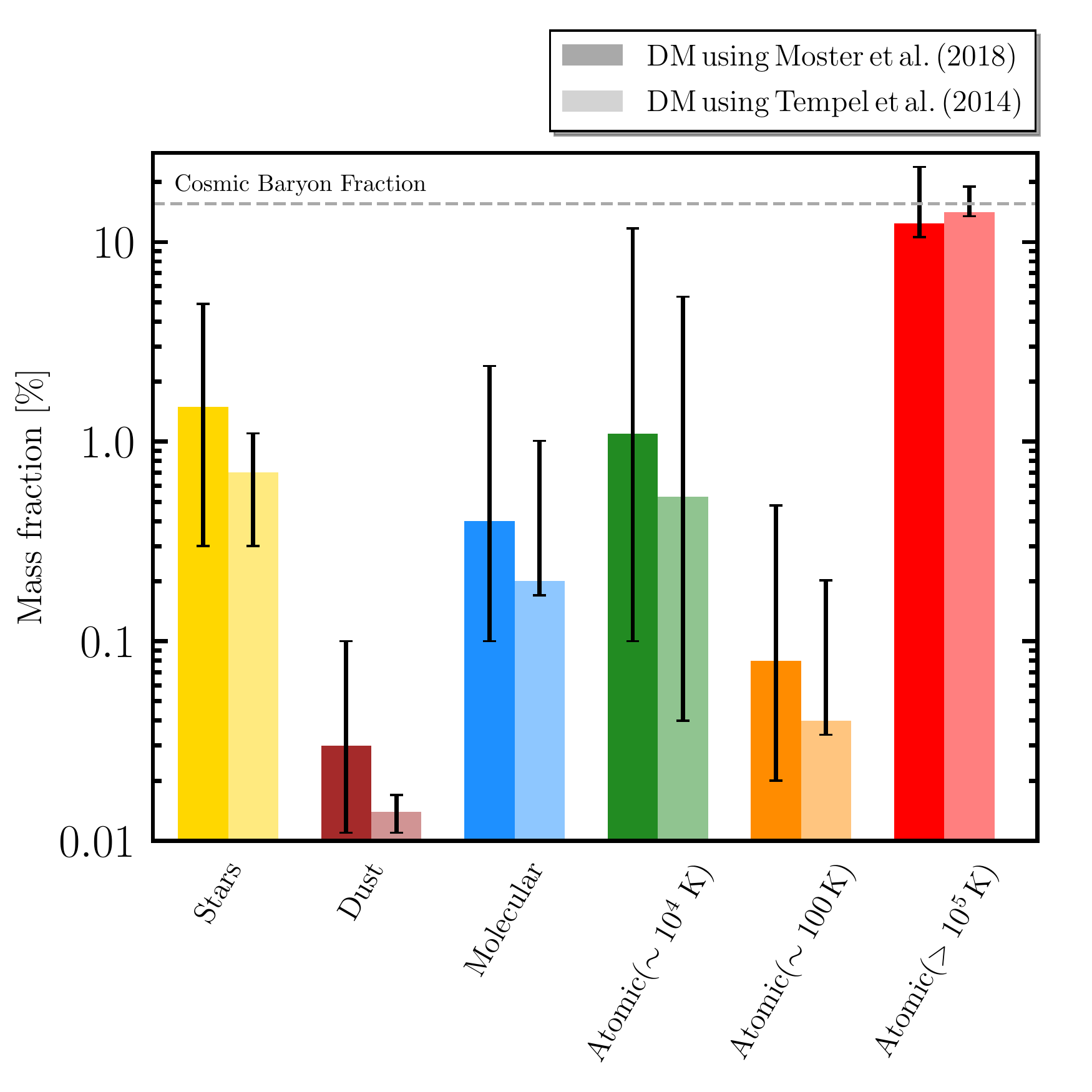}\\
\caption{The baryonic mass budget for the whole system within its virial radius, divided in its stellar, dust, molecular, atomic (cold, cool, and warm-hot) components. 
Each component has been calculated assuming a halo baryon fraction equal to the cosmic value of 15.6~\% (\citealt{Planck2020}). The fractions are computed for both $M_{\rm DM}$ values obtained in this work: using the relations in \citet{Moster2018} and following the formalism in \citet{Tempel2014}. See Section~\ref{sec:massBud} for details.} 
\label{MassBud}
\end{figure}

Figure~\ref{MassBud} summarizes the discussed baryonic components as fractions of the total mass of the system, which has been derived by assuming a halo baryon fraction equal to the cosmic value (15.6~\%).
It is clear that the stars, dust, molecular, cold and cool atomic components add up to a small fraction of the cosmic value, with 21~\% or 10~\% of the baryons in these constituents depending on the DM mass considered, Moster et al. or Tempel et al., respectively.  These values, though uncertain\footnote{If all the halo is filled with Ly$\alpha$ emitting gas at low surface brightness as explained previously, these fractions would go up to 32\% and 20\% assuming the same constant $N_{\rm H}$, respectively.}, are lower than the estimated value reported for $z\sim2$ quasars ($56\%$; \citealt{QPQ8}).
We can easily explain these differences with the larger halo masses derived in this work with respect to the assumed halo mass in \citet{QPQ8} ($M_{\rm DM}=10^{12.5}$~M$_\odot$). In other words, we find similar baryonic masses but in a halo which is $2.7\times$ or $5.9\times$ more massive. If all the quasars in \citet{QPQ8} inhabit DM halos as massive as the one of this ELAN, they would have a similar baryon budget.

As expected from galaxy formation theories (e.g., \citealt{DekelBirnboim2006}), our analysis suggests that the rest of the baryonic mass is in a warm/hot phase which permeates the halo of this massive system. Assuming a halo baryon fraction equal to the cosmic value, this warm/hot phase would represent a reservoir as massive as $M_{\rm warm-hot}^{\rm total} = (1.3_{-0.2}^{+1.2}) \times 10^{12} $~M$_\odot$ or $M_{\rm warm-hot}^{\rm total} = (3.1_{-0.1}^{+1.1})\times 10^{12} $~M$_\odot$, for Moster et al. and Tempel et al. DM calculations, respectively (Figure~\ref{MassBud}). The warm-hot phase together with the cool phase would then represent 87~\% or  94~\% of the baryon fraction. 

We can further gain some intuition on the halo gas physical properties by assuming the cool and warm-hot phases to coexist in pressure equilibrium. This assumption is likely not valid in turbulent massive halos (e.g., \citealt{Nelson2020}), but it is useful as first order approximation. We can therefore derive the physical properties of the two phases, namely temperature ($T_{\rm cool}$, $T_{\rm warm-hot}$), volume density ($n_{\rm H}^{\rm cool}$, $n_{\rm H}^{\rm warm-hot}$) and volume filling factors ($f_{\rm V}^{\rm cool}$, $f_{\rm V}^{\rm warm-hot}$).  
To do so, the following three relations have to be considered simultaneously: (i) the Ly$\alpha$ surface brightness in an optically thin scenario ${\rm SB_{\rm Ly\alpha}}\propto \alpha_{\rm A}(T_{\rm cool}) N_{\rm H}^{\rm cool} n_{\rm H}^{\rm cool} f_{\rm C}$ (\citealt{qpq4}), where $\alpha_{\rm A}(T_{\rm cool})$ is the temperature-dependent coefficient for case A recombinations (e.g., \citealt{HuiGnedin1997}); (ii) the pressure balance $n_{\rm H}^{\rm cool}T_{\rm cool}=n_{\rm H}^{\rm warm-hot}T_{\rm warm-hot}$; (iii) the mass ratio of the two phases ${M_{\rm warm-hot}/M_{\rm cool} = (n_{\rm H}^{\rm cool} / n_{\rm H}^{\rm warm-hot}) (f_{\rm V}^{\rm warm-hot} / f_{\rm V}^{\rm cool})}$. Using the observed average ${\rm SB}_{\rm Ly\alpha} = 6.08\times10^{-18}$~erg~s$^{-1}$~cm$^{-2}$~arcsec$^{-2}$, $T_{\rm warm-hot}=T_{\rm virial}=GMm_{\rm p}/(3k_{\rm B}R_{\rm vir})$ (e.g., \citealt{WhiteRees1978}), and the mass ratio obtained by assuming a halo baryon fraction equal to the cosmic value, we find $T_{\rm cool} = 10^{4.4}\ (10^{4.7})$~K, $n_{\rm H}^{\rm cool}= 3.1\ (3.1)$~cm$^{-3}$, $n_{\rm H}^{\rm warm-hot} = 10^{-2.3}\ (10^{-2})$~cm$^{-3}$, $f_{\rm V}^{\rm cool} = 2.6\times10^{-4}\ (1.2\times10^{-4})$, where we quote in brackets the value for the DM calculation following the formalism in \citet{Tempel2014}. 
A scale length for the structures in the cool gas responsible for the Ly$\alpha$ emission can then be computed as $l_{\rm cool} = N_{\rm H}^{\rm cool}/n_{\rm H}^{\rm cool} = 56\ (33)$~pc. This simple calculation agrees with previously reported properties for cool dense gas in ELANe (\citealt{cantalupo14,hennawi+15,fab+15b}; see also discussion in \citealt{Pezzulli2019}). 

Several recent works have studied the survival of cool clouds against hydrodynamic instabilities while moving throughout the hot halo with velocities of the order of few hundreds km~s$^{-1}$ (e.g., \citealt{Gronke2018,Kanjilal2020}). Specifically, it has been shown that if the cool dense gas falls out of pressure balance (e.g., due to radiative processes), it could shatter in smaller structures (\citealt{McCourt2018}) and entrain in the warm-hot medium without being destroyed by Kelvin-Helmholtz instabilites if the original cloud sizes are larger than a critical scale (equation 2 in \citealt{Gronke2018}). The gas properties we found (scale length, temperature, density), if translated to cloud properties, fulfil this survival criterion, and given the density contrast, might therefore lead to those clouds whose fragments are able to coagulate into larger cloudlets and therefore survive within the harsh environment of a quasar hot halo (\citealt{Gronke2020}). Therefore, these small-scale processes could be the reason why there is enough dense gas resulting in the bright ELAN emission seen
(see also \citealt{Mandelker2020}). 

The aforementioned pressure balance calculation assumes that all Ly$\alpha$ emission is due to recombination, but as we will show in section~\ref{sec:Mol_vs_Lya} there is evidence for resonant scattering at least on scales of $\sim 15$~kpc (or 2 arcsec) around compact sources. For this reason, we re-compute the aforementioned quantities by assuming that only a fraction of the observed ${\rm SB}_{\rm Ly\alpha}$ is due to recombination. As expected, lowering the ${\rm SB}_{\rm Ly\alpha}$ signal due to recombination decreases $n_{\rm H}^{\rm cool}$ (and therefore increases $f_{\rm V}^{\rm cool}$), while increasing $T_{\rm cool}$. For example, assuming a recombination fraction of 50~\%, we find $T_{\rm cool} = 10^{4.6}\ (10^{4.9})$~K, $n_{\rm H}^{\rm cool}= 1.9\ (1.9)$~cm$^{-3}$, $f_{\rm V}^{\rm cool} = 4.1\times10^{-4}\ (1.9\times10^{-4})$, $l_{\rm cool} = 89\ (53)$~pc, where 
the values in brackets correspond to the DM calculation following the formalism in \citet{Tempel2014}. The warm-hot densities are not affected because they are linked to the two phases mass ratio. Also in this scenario the aforementioned cloud survival scenario holds.  

We further conduct this calculation assuming the possibility that the halo baryon content is only a small fraction of the cosmic value. Indeed, current cosmological hydrodynamical simulations of structure formation implement strong feedback (supernova and AGN) recipes to match the observed properties of galaxies (e.g., \citealt{Schaye2015, Springel2018}). These feedbacks are able to eject large fractions of baryons from the halos of interest here, making the halo baryon fraction only $\sim1/3$ of the cosmic value (e.g., \citealt{Dave2009, Wright2020}).  In this framework, the hot reservoir within the virial radius will decrease, resulting in $M_{\rm hot}/M_{\rm cool}= 2.9$ or $7.6$ for the Moster et al. and Tempel et al. calculations, respectively.
In other words, the hot phase is 74~\% or 88~\% of the halo baryons, which would be in agreement with recent results from cosmological simulations ($\sim80\%$; e.g., \citealt{Gabor2015,Correa2018}). Assuming 50~\% Ly$\alpha$ emission from recombination, we then find $T_{\rm cool} = 10^{4.2}\ (10^{4.5})$~K, $n_{\rm H}^{\rm cool}= 0.7\ (1.2)$~cm$^{-3}$, $n_{\rm H}^{\rm warm-hot} = 10^{-2.9}\ (10^{-2.6})$~cm$^{-3}$, $f_{\rm V}^{\rm cool} = 6.5\times10^{-4}\ (2.9\times10^{-4})$, $l_{\rm cool} = 141\ (83)$~pc, where 
again the values in brackets correspond to 
the DM calculation following the formalism in Tempel et al..
Also in this scenario the aforementioned cloud survival scenario holds. Direct observations of the warm-hot phase are therefore crucial for constraining the warm-hot fraction, and ultimately galaxy formation models.

\subsection{The system evolution}
\label{sec:evo}

Here we briefly discuss the fate of this ELAN system in light of the ensemble of our findings presented in previous sections.
We first focus on the halo mass. Following the expected evolution of DM halos in cosmological simulations (e.g., \citealt{vandenBosch2014}), the obtained $M_{\rm DM}$ values at $z\sim3$ would then result in halo masses $M_{\rm DM}> 
10^{14}$~M$_{\odot}$ at $z=0$. This result is a first evidence that this ELAN could be considered the nursery of a local elliptical galaxy. Using then the molecular depletion time scale $t_{\rm depl} = M_{\rm H2}/{\rm SFR}$ (e.g., \citealt{Tacconi2020} 
and references therein), we can assess how long the current star formation can be sustained in the ELAN system without any additional fuel from gas recycling or infall. Assuming the SFRs obtained with {\sc CIGALE} for the sources within the ELAN, we find ${t_{\rm depl}^{\rm QSO} = 29-144}$~Myr, ${t_{\rm depl}^{\rm QSO2} = 42-210}$~Myr, ${t_{\rm depl}^{\rm AGN1} = 96 - 480}$~Myr, ${t_{\rm depl}^{\rm LAE1} = 240 - 1200}$~Myr, where these conservative ranges take into account the uncertainty on 
$\alpha_{\rm CO}$ (i.e. a factor of 5). Without the help of recycling and infall, these objects are thus not able to sustain their current SFR for long periods, with the longest t$_{\rm depl}$ allowing to possibly reach $z\sim2$ if the merger between QSO and LAE1 does not happen by then. 

To fuel the system for longer periods, a net mass infall is therefore required. To compute a rough estimate for the mass inflow rate, we assume gas infall velocities constant with radius and given by the first order approximation ${v_{\rm in}=0.9 v_{\rm vir}}$ (\citealt{Goerdt2015}), and use the Ly$\alpha$ emission as a mass tracer. Assuming the total cool gas mass ${M_{\rm cool}^{\rm total}=1.2\times10^{11}}$~M$_{\odot}$ (section~\ref{sec:bar_comp}) and that all the mass will end up onto the central object, we then find a mass inflow rate of ${\dot{M}_{\rm in}^{\rm cool}(z=3.164)=320}$~M$_{\odot}$~yr$^{-1}$ for both DM halos obtained in section~\ref{sec:dark_comp}. This fresh fuel for star formation is able to delay the depletion of the central object by a factor of $2.6$ at fixed SFR, but is not able to 
keep up with the star formation rate of the central object. As the gas accretion rate is expected to decrease at lower redshifts (e.g., \citealt{McBride2009}) a corresponding decrease in the SFR is expected with a certain delay, with the system activity almost shut down by $z\sim1$. 

We can indeed compute a $z=0$ stellar mass by assuming (i) the gas accretion rate as a function of $z$ in \citet{McBride2009} (i.e., ${\dot{M}_{\rm in}\propto(1+z)^{2.5}}$ for $z\geq1$ and ${\dot{M}_{\rm in}\propto(1+z)^{1.5}}$ for $z<1$) normalized to ${\dot{M}_{\rm cool}(z=3.164)}$ \footnote{We stress that we are not differentiating between ``cold'' and ``hot'' mode accretion, which should be respectively dominant at high and low redshift for this system (e.g., \citealt{DekelBirnboim2006}). The accretion rate here should include both modes as it is a scaled down version of the DM accretion rate.}, and (ii) that all cold and cool material, and satellites now in the QSO halo will end up in the central object by then. We find that the halo accretion down to $z=0$ contributes $7.2\times10^{11}$~M$_{\odot}$, while the latter $2\times10^{11}$~M$_{\odot}$, if outflows are not effective in removing mass from such a massive galaxy/halo (i.e., the wind material rains back onto the galaxy; e.g., \citealt{OppenheimerDave2008}). This is certainly true at low redshift, while at high-redshift winds could push material out of the halo virial radius. Nonetheless, this material may have fall back into the central or accreted satellites by $z=0$. We further assume that the accreted mass is translated to stellar mass with a star formation efficiency per free-fall time that scales as (1+z) (\citealt{Scoville2017})\footnote{For this rough calculation we use an average free-fall time of 10~Myr for molecular clouds (\citealt{Chevance2020}).}. By $z=0$ the stellar mass due to large-scale accretion is $1.5\times10^{11}$~M$_{\odot}$.
Taking into account all these assumptions and adding together (i) the stellar mass due to large-scale accretion, (ii) the mass from the satellites, and (iii) the $z=3.164$ stellar mass of QSO host, we obtain a $z=0$ stellar mass ${M_{\rm star}(z=0) = 4.7\times10^{11}}$~M$_{\odot}$, 94\% of which has been built before $z\sim1$.
The obtained $M_{\rm star}(z=0)$ is  
similar to the masses of local  
giant elliptical galaxies, e.g., NGC~4365 and NGC~5044 ($M_{\rm star}^{\rm NGC~4365}=(4.3\pm0.7)\times10^{11}$~M$_{\odot}$ and $M_{\rm star}^{\rm NGC~5044}=(5.7\pm0.7)\times10^{11}$~M$_{\odot}$, e.g.,\citealt{Spavone2017}). 
Computing the expected halo mass using the relations in \citet{Moster2018}, we find $M_{\rm DM}=10^{14.7}$~M$_{\odot}$, which could represent a local galaxy cluster. This calculation is in agreement with the evolutionary tracks in \citet{Chiang2013} (see e.g., their Figure~2).

This rough calculation once again points to the fact that this ELAN should be part of a large-scale proto-cluster, as found for other ELANe (e.g., \citealt{hennawi+15, Cai2016}, see also discussion in \citealt{FAB2018}). Wide field coverage is therefore needed to confirm this hypothesis and further pin down the evolution of this system.

\begin{figure}
\centering
\includegraphics[width=1.0\columnwidth]{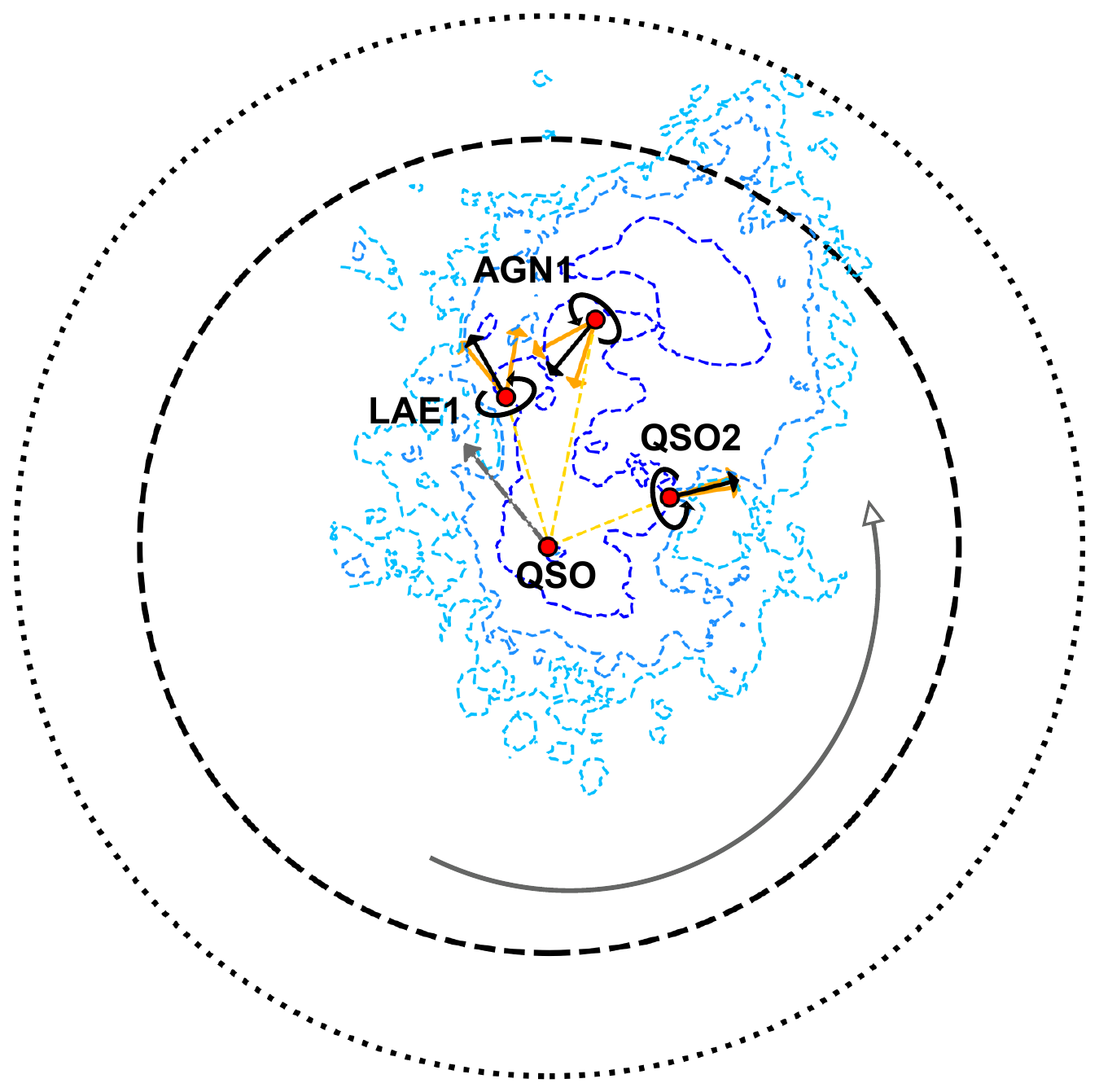}\\
\caption{Representation of the spin vectors for the QSO DM halo (gray) and the inspiraling satellites (black with uncertainty range in orange) obtained from the first moment maps of the Ly$\alpha$ and CO(5-4) emissions, respectively. The observed satellite spin vectors are almost aligned with the projected direction to QSO (yellow dashed lines). The inferred spin vector of the DM halo is directed towards the reader with an angle $\eta \approx 40^{\circ}$ (see section~\ref{sec:G_or} for details). For the sake of clarity we indicate with circular arrows indicate the rotation direction for each object. The blue dashed contours show the Ly$\alpha$ emission isophotes as in e.g., Fig~\ref{continuum_detections}. The dashed and dotted circle indicates $R_{\rm vir}=163$~kpc or $212$~kpc for the halo mass computed using Moster et al. relations, or Tempel et al. formalism, respectively.} 
\label{AngMom_sketch}
\end{figure}

\section{Satellites' spins alignment}
\label{sec:G_or}

Here we discuss, in the framework of the tidal torque theory, the evidence for the alignment between the satellites' spins and their position vectors to QSO, as reported in section~\ref{sec:CO54results}.
The tidal torque theory (\citealt{Hoyle1951, Peebles1969, Doroshkevich1970, White1984}) is at the basis of the current understanding of galaxies spin acquisition (i.e. angular momentum). In this framework, a net angular momentum is generated in collapsing protogalaxies by tidal torques due to neighbouring perturbations, resulting in a correlation between the galaxy spin direction and the principal axes of the local tidal tensor. Correlations between galaxies spin and large-scale structures (knots, filaments, sheets, voids) are therefore expected in the absence of strong non linear processes.  In particular, DM only simulations have shown that halo spins tend to be perpendicular to the closest large-scale filament if their mass is above a critical mass, while low-mass halos are preferentially aligned with the closest filament (e.g., \citealt{AragonCalvo2007,Codis2012,Veena2020}). This picture also holds for cosmological hydrodynamical simulations that include baryon physics, and feedback processes (e.g., \citealt{Dubois2014,Wang2018}), usually showing that galaxies whose stellar mass is $M_{\rm star}\gtrsim10^{10.5}$~M$_{\odot}$ have their spin perpendicular to the closest filament, while lower mass galaxies show a parallel spin (e.g., \citealt{Codis2018,Soussana2020,Kraljic2020}). A consequence of these alignments is that outer satellites (i.e., at $R \gtrsim 0.5R_{\rm vir}$) should show a spin preferentially aligned with their position vector relative to the central object (\citealt{Welker2018}). In other words, outer low-mass infalling galaxies should have their angular momenta still aligned with the filament they are coming from and parallel to the direction to the central object.

The ELAN here studied, with bright satellites and Ly$\alpha$ emission that reaches the edge of the virial halo, is a perfect laboratory to investigate these theoretical expectations. We first focus on the extended Ly$\alpha$ emission which has been shown to trace the DM halo motions, though with a lag (\citealt{FAB2018}). If we then interpret the Ly$\alpha$ emission as a tracer of halo rotation and that gas and DM spins are almost aligned\footnote{It has been shown that the gas and DM angular momenta could be misaligned, with a median misalignment angle of 30$^{\circ}$ (e.g, \citealt{vandenBosch2002}).}, the halo spin is likely North-West directed and pointing towards the observer, with an inclination angle $\eta \approx 46^{\circ}$ or $33^{\circ}$ that makes the observed gas circular velocities ($\sim 380$~km~s$^{-1}$) smaller than the velocities expected for the obtained massive DM halo assuming a NFW profile ($V_{\rm c}^{\rm max} = 490$~km~s$^{-1}$ or $V_{\rm c}^{\rm max} = 635$~km~s$^{-1}$).  
The fact that we do not detect rotation in the host galaxy of QSO could further suggest that the galaxy spin and inner halo spin are slightly different than the whole halo spin (e.g., \citealt{Bullock2001}) with the host galaxy spin almost aligned with our line of sight. We note however that strong AGN feedback processes on ~10~kpc scales could hinder a weak rotation signal.
Secondly, we focus on the satellites (QSO2, AGN1, LAE1) spin alignment with respect to the direction to the central galaxy (QSO). As shown in section~\ref{sec:CO54results}, the three satellites (which all have $M_{\star}<10^{10.5}$~M$_{\odot}$) have their major axis defined by the CO(5-4) line of nodes consistent with being perpendicular to the quasar direction. Their spins are therefore almost aligned with the projected position vector to QSO (Table~\ref{tab:Angles})\footnote{The probability that the current configuration happens by chance is very low. Indeed, if we give an equal probability for any angle within 180 degrees, the angles spanned by our estimates will have a probability of 9/180=0.05, 49/180=0.27, 44/180=0.24 for QSO2, LAE1, and AGN1, respectively (1sigma case). The global probability of this alignment happening by chance is therefore 0.3\% (1sigma), 2.6\% (2sigma), 8.7\% (3sigma).}. AGN1 is the source with the largest misalignment, possibly due to tidal torques exerted by LAE1, which sits in projected close proximity. AGN1 spin vector is indeed in between the directions to LAE1 and QSO (see cutout in Figure~\ref{cutouts_fig}). As a last remark, we also notice that AGN1's spin is anti-aligned with the spins of both LAE1 and QSO2 which are sitting at closer projected distances to QSO.

Overall, all these findings, summarized in Figure~\ref{AngMom_sketch},  
are a tentalizing evidence of the theoretical expectations for the angular momentum alignment. 
The infalling satellites are inspiraling within QSO DM halo with their spins still almost aligned to the large-scale filaments they are coming from, and likely perpendicular to the spin of the QSO DM halo. Follow up observations at a higher spatial resolution (to better resolve the host galaxies and their kinematics; e.g., HST, JWST, and ALMA observations), and at a deeper sensitivity (to detect additional satellites; e.g., MUSE and ALMA observations) are needed to confirm this framework.

\section{Ly${\rm \alpha}$ emission versus CO(5-4): signatures of gas infall}
\label{sec:Mol_vs_Lya}

In this section we compare the CO(5-4) emission to the Ly$\alpha$ emission in the vicinity of QSO, QSO2, AGN1, and LAE1 to ascertain whether strong radiative transfer effects and/or illumination effects are in place in the vicinity of these sources, ultimately unveiling gas motions for the $T\sim10^4~K$ gas.
Indeed, while CO(5-4) is a rotational transition which traces well the gas kinematics, the Ly$\alpha$ photons are known to undergo a walk in frequency and space due to resonant scattering in most astrophysical environments (e.g., \citealt{Neufeld_1990,Laursen2009,Dijkstra2017}). Only knowing the systemic redshift of a source allows one to understand the Ly$\alpha$ line shape in the light of emitting gas' kinematics (e.g., \citealt{Yang2011,Yang2014,Yang2014b,Ao2020}). We first discuss the spatial location of the CO(5-4) and Ly$\alpha$ emission, 
and then compare the line emission shapes. 

\begin{figure}
\centering
\includegraphics[width=1.0\columnwidth]{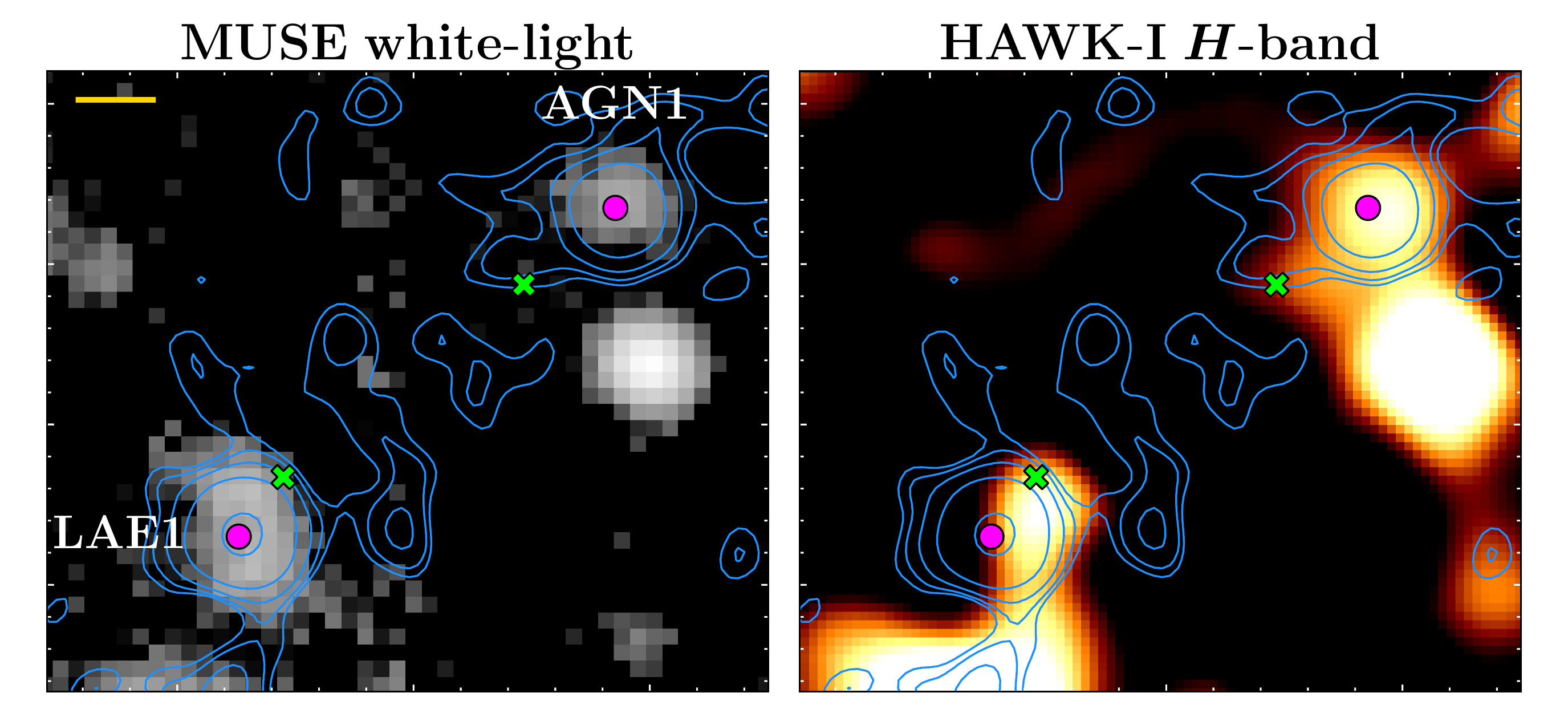}
\caption{Zoom-in comparison of LAE1 and AGN1 positions in different observations. \emph{Left:} $9\arcsec \times 7\arcsec$ (or $\sim$68~kpc~$\times$~53~kpc) portion of the MUSE white-light image with highlighted the position of the compact Ly$\alpha$ emission (green crosses) and the position of the CO(5-4) emission (magenta circles) and the ALMA continuum (blue contours; see also Figure~\ref{cutouts_fig}). The yellow scale in the top-left corner indicates $1\arcsec$. \emph{Right:} same as the left panel, but for the HAWK-I $H$-band data smoothed with a $1\arcsec$ Gaussian kernel.} 
\label{AGN1-LAE1_HAWKI_MUSEwhite}
\end{figure}

As already mentioned in Section~\ref{sec:CO54results}, while  
for the QSO and QSO2 the location of the 2~mm continuum, CO(5-4) and the Ly$\alpha$ emission coincide, 
the AGN1 and LAE1 show shifts between the millimeter and the Ly$\alpha$ emission (see Figure~\ref{cutouts_fig}). Specifically, the centroids of each emission are at distances of 1.5$\arcsec$ (or $\sim11.4$~kpc) and 0.9$\arcsec$ (or $\sim6.8$~kpc), respectively for AGN1 and LAE1. For AGN1, a similar shift is found when comparing the location of the Ly$\alpha$ emission and the $H$-band emission from HAWK-I. While for LAE1 the $H$-band emission is in between the Ly$\alpha$ and millimeter emission. We zoom on these shifts in Figure~\ref{AGN1-LAE1_HAWKI_MUSEwhite}. As reported in sections~\ref{sec:HAWKI} and \ref{sec:CO54results}, our astrometry in the MUSE and HAWK-I data has been checked against GAIA and it is therefore assumed to be correct within small uncertainties. Also, the SMA map shows shifts on the location of its detections especially with respect to MUSE, but we do not consider their positions as well-defined given the low significance of the SMA detections. 

The observed shifts between Ly$\alpha$ emission and the infrared continuum could be due to a combination of  
different effects: (i) presence of a path of least resistance for the Ly$\alpha$ and/or UV photons in these directions (e.g., small scale winds) or dust obscuration (e.g., \citealt{Hodge2015}), (ii) presence of gas displaced from the host galaxy of LAE1 and AGN1 due to e.g., gas infall, ram pressure or tides, (iii) interaction between two galaxies. A way to disentangle these scenarios is to resolve these systems with high resolution imaging in the UV and NIR wavelength ranges (e.g., HST, JWST) to get their morphologies and inclinations. Another possibility is to compare their CO(5-4) and Ly$\alpha$ line shapes, ultimately constraining the kinematics of the Ly$\alpha$ emitting gas.

\begin{figure*}
\centering
\includegraphics[width=1.0\textwidth]{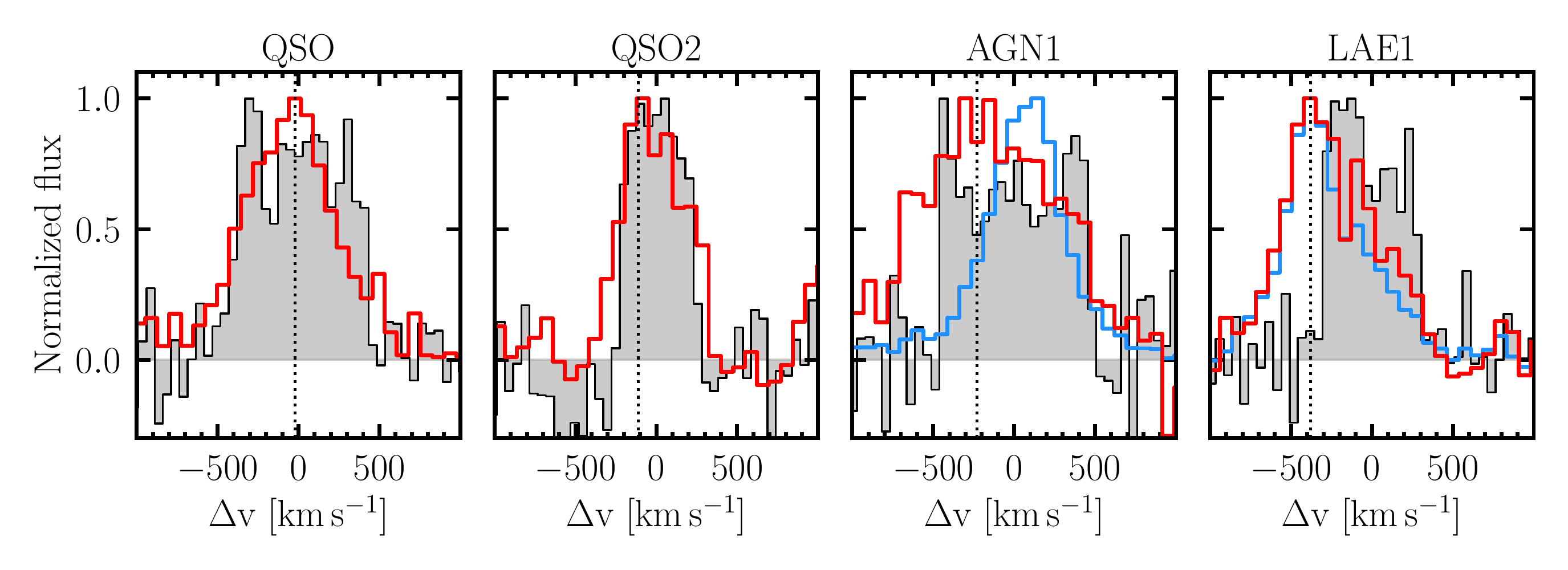}
\caption{Comparison of the line shapes for CO(5-4) (black-gray) and Ly$\alpha$ emission (red) at the location of QSO, QSO2, AGN1, and LAE1. Zero velocity is the redshift from the CO(5-4) line emission. For AGN1 and LAE1 we also show the Ly$\alpha$ spectrum (blue) at the compact peak in Ly$\alpha$ emission in their vicinity (see section~\ref{sec:Mol_vs_Lya}; green crosses in Figures~\ref{cutouts_fig} and \ref{AGN1-LAE1_HAWKI_MUSEwhite}). The vertical dotted lines indicate the Ly$\alpha$ velocity shift computed as the first moment of its flux distribution. The ALMA spectra are binned as in Figure~\ref{ALMA_CO54_spectra}, while the MUSE data are at the instrument resolution.} 
\label{Lya_vs_CO54}
\end{figure*}

Figure~\ref{Lya_vs_CO54} shows the comparison of the normalized CO(5-4) line profile (black-gray) together with the normalized Ly$\alpha$ line profile found in the same aperture after continuum subtraction (red) for QSO\footnote{The central $1\arcsec \times 1\arcsec$ portion of the aperture for QSO has been masked as it is used for the normalization during its PSF subtraction (\citealt{FAB2018}).}, QSO2, AGN1 and LAE1. 
In this figure we use $z_{\rm CO(5-4)}$ as reference velocity (Table~\ref{tab:CO54}). The two line profiles clearly differ. 
The most striking features  
are (i) the Ly$\alpha$ peak is always blueshifted with respect to the reference velocity of the CO(5-4) line emission, making all Ly$\alpha$ profiles blue-skewed  
in contrast to most of current observations of high-redshift star-forming galaxies that show a redshifted Ly$\alpha$ with respect to systemic (e.g., \citealt{Verhamme2018}), (ii) the Ly$\alpha$ blueshift is larger for less massive objects or for smaller SFRs, $\Delta v_{\rm Ly\alpha \ peak} = -20.0\pm10, -112\pm7, -250\pm12, -380\pm10$~km~s$^{-1}$, respectively for QSO, QSO2, AGN1, and LAE1, and (iii) an overall similar width for the two emission lines. 

All the effects visible in Figure~\ref{Lya_vs_CO54}
could be easily explained by the radiative transfer of Ly$\alpha$ emission. Indeed, blue-skewed Ly$\alpha$ profiles are expected for infalling gas because photons redward of the line center, that would otherwise escape the medium, are seen in resonance in the reference frame of the infalling atoms, while blue photons easily escape (e.g., \citealt{Zheng2002, Dijkstra2006a, Verhamme2006, Laursen2009}). Specifically, models of Ly$\alpha$ emission from collapsing shells showed that the peak of the resultant Ly$\alpha$ profile depends on the velocity of infall, with the peak progressively displaced towards negative velocities for increasing infall velocities (see e.g., Figure~7 in \citealt{Verhamme2006}). However, for high enough infall velocities, the peak position is expected to move back closer to line center (i.e., systemic redshift). We argue that the spectra around LAE1, AGN1, QSO2, QSO show this trend, with LAE1 and QSO having respectively the smallest and the largest gas infall velocities in this sample. This can be verified by computing, in first order approximation, the expected infall velocities for these objects. We assume that the gas inflow velocities are constant within the halo, $\sim0.9 v_{\rm vir}$ (e.g., \citealt{Goerdt2015}), till very close to galaxies (e.g., the aperture for our spectra), with gas then accreting in free-fall onto the galaxies. We also assume as galaxy sizes the observed effective  radii of the CO(5-4) emitting regions (Table~\ref{tab:CO54}). 
Integrating the infall velocity formula (e.g., \citealt{Goerdt2015}) from 14.4~kpc down to the galaxy sizes, we find 
$v_{\rm inflow}= 276, 301, 310, 812$~km~s$^{-1}$, for LAE1, AGN1, QSO2, and QSO, respectively. Considering also differences in opacities, a factor of $\sim 3$ larger infall velocity for QSO with respect to LAE1 could explain why the Ly$\alpha$ peak near QSO bounces back towards line center because of radiative transfer effects. This would be also in agreement with the expectation of higher accretion rates onto the galaxies with larger SFRs. If the SFR is in steady state with the accretion rate, the inflowing mass at the radius $R$ would be roughly $M_{\rm in}={\rm SFR}/(v_{\rm inflow}/R)$, resulting in about few times $10^{9}$~M$_{\odot}$ for each object. We note that such large inflow velocities for cool gas (up to $2\times v_{\rm vir}$) are seen in cosmological simulations of quasar host galaxies in the inner portions of the halo where baryons dominate (\citealt{Costa2015}).  

Instead, the fact that the Ly$\alpha$ and CO(5-4) emissions have similar line widths 
could be ascribed to resonant scattering effects of Ly$\alpha$ in the presence of dust. Indeed, it has been shown that dust preferentially absorbs Ly$\alpha$ photons in the wings of the line profile because these photons are produced in the dustier regions within galaxies (e.g., \citealt{Laursen2009b}). Following these predictions, Ly$\alpha$ line profiles in dusty environments should have narrower profiles than in dust-free objects. 

Lastly, we checked the Ly$\alpha$ profile also at the location of the bright compact Ly$\alpha$ emission displaced from
LAE1 and AGN1, and thanks to which \citet{FAB2018} discovered these sources. We show also these spectra (blue) in
Figure~\ref{Lya_vs_CO54}. We have two different configurations. LAE1 presents the same Ly$\alpha$ profile at both
locations, i.e. at LAE1 and at $\sim 6.8$~kpc. This implies that we are probing the same infalling material through
different sightlines (e.g., \citealt{Ao2020}). Conversely, AGN1 shows a completely different Ly$\alpha$ profile at $\sim
11.4$~kpc, with its shape slightly red-skewed. This configuration, together with the galaxy kinematics traced by CO(5-4)
(section~\ref{sec:G_or}) and the fact that \ion{He}{2} and \ion{C}{4} emission have been detected at this displaced location (\citealt{FAB2018}) favors a scenario in which we are seeing outflows or material displaced from AGN1 and ionized by its radiation along its minor axis,  
and inflows along its major axis.

Summarizing, we find signatures of gas infall as traced by Ly$\alpha$ resonant scattering in all of the four sources for which we have a CO(5-4) detection. Detailed resonant scattering calculations are needed to confirm this scenario and test it against contaminations from (i) the large-scale Ly$\alpha$ emission and (ii) the interplay between inflow and outflow signatures within the same observational aperture.

\section{Powering of the ELAN}
\label{sec:powering}

Ly$\alpha$ nebulae around quasars are usually explained by invoking photoionization from the embedded AGN (e.g., \citealt{heckman91a, Weidinger05}). However, there could be additional powering sources, like star formation in the quasar host galaxy and satellites (e.g., \citealt{Ao2015}), and/or resonant scattering to large scales of the Ly$\alpha$ photons generated within galaxies/AGN (e.g., \citealt{Hayes2011,Geach2016,Kim2020}), and/or Ly$\alpha$ cooling radiation powered by gravitational collapse (e.g., \citealt{Haiman2000,Dijkstra2006}), and/or fast winds (e.g., \citealt{Taniguchi2000}). In this section we briefly discuss these processes in the framework of the ELAN studied here. The interplay between these mechanisms could be very different in specific systems especially due to geometry (e.g., misalignment between surrounding gas distribution and quasar ionizing cones or quasar obscuration) and the phase in which the system is seen (e.g., ongoing strong wind/outflow or presence of active companions).

Regarding this ELAN, it has been shown that QSO is able to power the entire Ly$\alpha$ emission (${L_{\rm Ly\alpha}^{\rm total}=3.2\times10^{44}}$~erg~s$^{-1}$). Specifically, if the whole nebula is within the halo virial radius, the photoionization scenario implies the emitting gas is optically thin to ionizing photons, requiring very high, interstellar-medium densities 
to explain the absence of \ion{He}{2}$\lambda1640$ down to current observational limits (\citealt{FAB2018}). Cosmological simulations are starting to approach these densities showing ubiquitous cool dense gas throughout the halo of simulated galaxies (e.g., \citealt{Hummels2019}). Nonetheless, the  
high densities predicted in a photoionization scenario could be lower if (i) a fraction of the Ly$\alpha$ is due to scattering of photons produced by compact sources, and (ii) star-formation and/or collisional excitation is powering a fraction of the Ly$\alpha$ emission.

We compute a rough estimate for the fraction of Ly$\alpha$ emission powered by star formation by assuming the SFR obtained with CIGALE. We first converted the SFRs to Ly$\alpha$ luminosities assuming (i) case-B recombination, $L_{\rm Ly\alpha}=8.7L_{\rm H\alpha}$ (e.g., \citealt{Osterbrock2006}), and (ii) ${\rm SFR}=7.9\times10^{-42}L_{\rm H\alpha}$~M$_{\odot}$~yr$^{-1}$ (e.g., \citealt{kennicutt98}), finding
$L_{\rm Ly\alpha}^{\rm QSO}=5.7\times10^{44}$~erg~s$^{-1}$, $L_{\rm Ly\alpha}^{\rm QSO2}=2.0\times10^{44}$~erg~s$^{-1}$, $L_{\rm Ly\alpha}^{\rm AGN1}=1.1\times10^{44}$~erg~s$^{-1}$, $L_{\rm Ly\alpha}^{\rm LAE1}=5.5\times10^{43}$~erg~s$^{-1}$. The sum of these values clearly exceed the total observed Ly$\alpha$ emission for the ELAN. However, this calculation assumes that all ionizing photons impinge on the gas. This is likely not the case as it has been shown that the average escape fraction at similar redshifts is $\sim5\%$ (e.g., \citealt{Matthee2017})\footnote{We stress that the escape fraction of ionizing photons is a rather debated measurement. The estimate in \citet{Matthee2017} is consistent with escape fractions as high as $10\%$ for $z\sim3-4$.}. Assuming this escape fraction, the total SFR of the four objects could therefore power only $\sim14\%$ of the observed Ly$\alpha$ emission. At a fixed cool gas mass, this contribution would lower the predicted $n_{\rm H}$ from recombination by a similar amount.

On top of this, a comparable fraction of Ly$\alpha$ photons produced in the body of galaxies due to SFR and/or AGN could escape and thus scatter in the surrounding gas distribution and eventually escape towards the observer (e.g., \citealt{Duval2014}).
Because of the absence of compact Ly$\alpha$ emission at the location of the ALMA continuum for LAE1 and AGN1, we think that the escape of Ly$\alpha$ photons and/or UV photons from these objects is highly directional. Indeed we find that  
the displaced compact Ly$\alpha$ emission in close vicinity to the location of LAE1 and AGN1 (see section~\ref{sec:Mol_vs_Lya}; \citealt{FAB2018}) corresponds to $8.7\%$ and $7.7\%$ of their Ly$\alpha$ emission expected from SFR, respectively. 
These values might represent an upper limit on the fraction of Ly$\alpha$ photons scattered outside each galaxy. 
Therefore, this calculation  
suggests that all the compact sources within the ELAN may contribute up to at least $\sim30\%$ considering both photoionization from SFR and resonant scattering.

We can further compute a conservative estimate for the Ly$\alpha$ photons available for scattering and produced 
by the QSO. 
To this aim, 
the Ly$\alpha$ line is convolved with the line shape of the observed ELAN, and integrated to obtained  
$(L_{\rm Ly\alpha}^{\rm QSO})_{\rm obs}=6.4\times10^{44}$~erg~s$^{-1}$, which is $\sim2\times$ the total ELAN luminosity. As also QSO photons outside of this range can in principle interact with the ELAN gas after some scattering, the available QSO Ly$\alpha$ photons abundantly pass the total Ly$\alpha$ luminosity of the ELAN. Nevertheless, because of the physics inherent to the propagation of resonant scattering photons (e.g., \citealt{Dijkstra2017}) and because of the large distances within the halo, scattered QSO photons likely dominate the nebula powering preferentially in the inner halo (see discussion in \citealt{FAB2018} and references therein). 

In addition, a fraction of the Ly$\alpha$ emission in this massive system could be due to collisional excitation (e.g., \citealt{Furlanetto05}). This contribution is notoriously difficult to predict as it strongly depends on temperature (exponential dependence) and gas density squared (\citealt{Osterbrock1989}). Analytical and numerical studies considering this powering mechanism did reproduce the observed Ly$\alpha$ luminosities of high-redshift nebulae, underlying the possible importance of this mechanism (\citealt{Dijkstra2009_hd,Rosdahl12}).  
For example, cosmological simulations of structure formation for massive halos showed that the Ly$\alpha$ emission from the gas with $n_{\rm H}\gtrsim0.3$~cm$^{-3}$ (the cool halo gas for those simulations) is dominated by collisional excitation and accounts for 40\% of the total luminosity (\citealt{Rosdahl12}). These simulations did not include AGN photoionization and therefore their results need to be taken with caution when compared to the system studied here.

Lastly, given the active nature (both AGN and SFR) of the sources within the ELAN, fast winds or even outflows are expected to be present at some times during its evolution. Fast shocks generated in this scenario would produce a strong ultraviolet radiation field (e.g., \citealt{Allen2008}), which can contribute to the powering of the ELAN.
\citet{FAB2018} discussed how the $300$~kpc velocity shear and small Ly$\alpha$ velocity dispersion across this ELAN run counter to the presence of an halo-scale wind. However, small scales ($\sim 10$~kpc) winds in proximity of the compact sources within the ELAN could not be excluded. Our observations seem to confirm this picture, showing possible evidence for winds around at least AGN1 (section~\ref{sec:Mol_vs_Lya}). A certain portion of the ELAN could be therefore powered by such small-scales winds.

Overall, a complex interplay between AGN and SFR ionization, Ly$\alpha$ resonant scattering, fast winds, and collisional excitation is needed to fully comprehend the powering of ELANe. Their large and often asymmetric extents are likely due to the presence of active companions which help in illuminating the surrounding gas distribution. Similar conclusions have been reported when studying the widespread Ly$\alpha$ emission in a $z\sim3$ protocluster field (\citealt{Umehata2019}).

\section{Note on molecular gas extending outside the body of galaxies}
\label{sec:extMol}

Widespread large reservoirs ($\sim10^{11}$~M$_{\odot}$) of cold molecular gas have been found across $\sim 40 -70$~kpc in two $z\sim2$ HzRGs fields. In one case the reservoir was found in proximity of the HzRG itself (\citealt{Emonts2016,Emonts2018}), while in the second case around an H$\alpha$ emitter (\citealt{Dannerbauer2017}). Because of the similarities between the ELAN studied here and the halo expected to host an HzRG, and the high densities 
usually invoked to explain the Ly$\alpha$ emission, it is possible that ELANe are multiphasic, at least on tens of kpc close to embedded sources (see also \citealt{VidalGarcia2021}).
Interestingly, an extended molecular gas reservoir as massive as those found in HzRGs would increase the molecular gas budget for the targeted ELAN to values similar to the stars budget (see Fig.~\ref{MassBud}). We note that this occurrence is within our current conservative error estimate for the molecular phase.

The CO(5-4) ALMA observations presented in section~\ref{sec:CO54results} provide sizes for the CO(5-4) emitting region for QSO, QSO2, AGN1 and LAE1 in the range $R_{\rm CO(5-4)}\sim 3-5$~kpc. These values are in overall agreement with current molecular sizes reported for different high-redshift galaxy populations (e.g., SMGs, \citealt{Ivison2011,Spilker2015,TC2017,CalistroRivera2018}; quasars, \citealt{Decarli2018,Stacey2020}; HzRG disks, \citealt{Man2019}; star-forming galaxies, \citealt{Kaasinen2020}), but possibly on the high-side, especially for QSO2 and AGN1. However, there are no statistical observations of CO(5-4) sizes in high-redshift objects, nor many CO size estimates at $z\sim3$, hampering any firm conclusion. 

Nevertheless, the ALMA observations did not unveil any large-scale, widespread molecular reservoir in the system studied here. 
Assuming that the excitation in a hypothetical extended molecular reservoir is similar to within the compact sources ($r_{51}=0.4$, $\alpha_{\rm CO}=0.8$~M$_{\odot}$~(K~km~s$^{-1}$~pc$^{-2}$)$^{-1}$), and considering a line width of $\sim 300$~km~s$^{-1}$, 3$\times$ the noise rms ($\approx167$~$\mu$Jy) corresponds to a molecular gass limit $3.4 \times 10^8$~M$_\odot$~beam$^{-1}$, which implies a beam-averaged surface molecular gas mass $\Sigma_{\rm H2} < 19$~M$_{\odot}$~pc$^2$. This value is consistent with similar non detections in other two ELANe (\citealt{Decarli2021}), excluding molecular gas surface densities similar to starbursting environment in the whole extent of the ELAN. 
We further checked our data by applying a taper, up to considering only the range of uv-distances sensitive to scales $\sim12-30$ arcsec. These tapered data have a noise ${\rm rms}\sim594$~$\mu$Jy~beam$^{-1}$ in $\sim 50$~km~s$^{-1}$. Again, we did not find any detection (corresponding to a beam-averaged surface molecular gas mass $3\sigma$ limit of $\Sigma_{\rm H2} < 69$~M$_{\odot}$~pc$^2$), confirming that CO(5-4) traces excited molecular gas within the body of galaxies. Deeper observations of ELAN systems targeting low-J CO transitions or additional molecular gas tracers (e.g., [CI] and [CII] emission) 
are therefore needed to unveil the extended molecular gas reservoir, if any exists.

\section{Summary}
\label{sec:summ}

We initiated the project ``a multiwavelength study of ELAN environments'' (AMUSE$^2$)
to investigate several aspects of the astrophysics of high-redshift massive systems associated with quasars, which ELANe seem to trace.
In this  
paper, we report on VLT/HAWK-I, APEX/LABOCA, JCMT/SCUBA-2, SMA/850$\mu$m, ALMA/CO(5-4) and 2mm observations targeting the ELAN around the $z\sim3$ quasar SDSSJ~1040+1020 (\emph{a.k.a} QSO) discovered by \citet{FAB2018}. This ELAN was known to host three AGN and two LAEs (LAE1 and LAE2). Specifically, VLT/MUSE unveiled that the bright central quasar QSO has a companion quasar QSO2 and a companion type-II AGN, AGN1.

The single dish observations resulted in a surprisingly strong detection at 850~$\mu$m ($\sim 11$~mJy) at the ELAN location. However, the interferometric observations confirmed that this emission accounts for multiple sources associated with the ELAN. Our multiwavelength observations added several continuum data-points to the four brightest sources SED (QSO, QSO2, AGN1, LAE1), and unveiled their relatively boxy CO(5-4) emission with integrated flux in the range $0.22 \lesssim I_{\rm CO(5-4)} \lesssim 0.43$~Jy~km~s$^{-1}$. This emission is spatially resolved and shows evidence of kinematics reminiscent of rotation-like patterns in three sources, QSO2, AGN1, and LAE1. Further, the Ly$\alpha$ emission in the vicinity of AGN1 and LAE1 is found to peak at a displaced position with respect to the continuua and the CO(5-4) emission, by $\sim11.4$~kpc and $\sim6.8$~kpc, respectively. A comparison of the Ly$\alpha$ emission extracted from the same aperture of the CO(5-4) emission revealed blue-skewed Ly$\alpha$ spectra for all four sources, and comparable line widths.

We use this dataset to attempt a first calculation of the total mass of the system and forecast its evolution.
First, stellar and dust masses, and star formation rates are obtained through SED fitting for the sources with confirmed association, i.e., QSO, QSO2, AGN1, and LAE1. While their molecular gas masses are obtained from the CO(5-4) detections. The estimated stellar, dust, and molecular gas masses are consistent with the dynamical masses obtained from CO(5-4) under the assumption of a reasonable inclination angle. Further, two orthogonal methods are used to infer the total DM halo mass hosting this ELAN system: the halo mass - stellar mass relation, and the use of radial velocity dispersion and group extent. Both methods give a consistent answer, this ELAN 
likely inhabits a massive DM halo of M$_{\rm DM}=(0.8-2)\times10^{13}$~M$_{\odot}$. 
Following this methodology, our main findings are as follows.

   \begin{enumerate}
      \item The total current SFR in the system is at least SFR~$\sim860$~M$_{\odot}$~yr$^{-1}$, with the QSO host being the most star-forming galaxy SFR$\sim500$~M$_{\odot}$~yr$^{-1}$ (section~\ref{sec:dustsfr}).
      \item The molecular gas mass estimated through the CO(5-4) emission for QSO, QSO2, AGN1 and LAE1 is $M_{\rm H_2}\sim 10^{10}$~M$_{\odot}$ for each system.  The dust content in these galaxies is found to be in the range
      $M_{\rm dust}=(0.3 - 4)\times10^9$~M$_{\odot}$ (Table~\ref{tab:CO54}; section~\ref{bar_budget}).
      \item The total baryonic mass budget for the whole system considering the stellar, dust, molecular, cool gas ($T\sim10^4$~K), and cold gas ($T\sim100$~K) masses, sums up to $10-21$\% of the cosmic baryon fraction. Therefore a hot reservoir as massive as $M_{\rm warm-hot} \sim 10^{12}$~M$_{\odot}$ is needed to complete the cosmic baryon budget for such a halo. Assuming baryon fractions seen in current cosmological simulations (i.e., $\sim1/3$ of the cosmic value), the hot phase would instead represent $74-88$\% of the baryons in this system (section~\ref{sec:massBud}).
      \item The fate of the system is predicted by estimating the molecular depletion time scale for each object and considering its halo mass and expected mass accretion rates. The targeted ELAN is likely the progenitor of an elliptical galaxy as massive as giant local ellipticals (e.g., NGC~4365, NGC~5044), and with its DM halo expected to achieve by $z=0$ a DM mass as high as $10^{14.7}$~M$_{\odot}$ (section~\ref{sec:evo}).
   \end{enumerate}

The observations further probe several aspects of the interplay between the galaxies embedded in the ELAN and the large-scale gas distribution. The main points we discussed are as follows.   
   
   \begin{enumerate}
      \item The first-moment maps of the CO(5-4) emission show rotation signatures in QSO2, AGN1, and LAE1. Their projected angular momentum vectors, though uncertain, are found to be almost parallel to the projected position vector to the central QSO. This finding hints to a scenario in which the infalling QSO satellites have their spins still almost aligned to the large-scale filaments they come from. Further, the spin of the QSO DM halo, inferred by assuming that the Ly$\alpha$ signal traces halo motions with a certain lag (\citealt{FAB2018}), is roughly perpendicular to those of the satellites.  
      These results would be in line with the theoretical expectations from the tidal torque theory (section~\ref{sec:G_or}).

      \item Tentalizing signatures of gas infall onto QSO, QSO2, AGN1, LAE1 are evident when comparing the Ly$\alpha$ emission shape with respect to the redshift obtained from the CO(5-4) emission. Indeed, 
      the observed line shapes could be explained by Ly$\alpha$ radiative transfer effects in infalling gas in dusty environments. Further, the velocity shifts of the Ly$\alpha$ peaks decrease with increasing stellar mass or SFR, with QSO (LAE1) having the smallest (largest) negative shift in the sample of four sources. This effect could be due to higher infall velocities onto more massive systems, with the infall velocities onto the QSO host galaxy being the largest. This picture agrees with the SFRs in these systems, i.e. highest (lowest) SFR in QSO (LAE1) (section~\ref{sec:Mol_vs_Lya}).

      \item Additional likely signatures of Ly$\alpha$ resonant scattering are the large displacements ($\sim10$~kpc) of the peak emission around LAE1 and AGN1 with respect to their ALMA and HAWK-I detections. Resonant scattering of Ly$\alpha$ photons seems therefore in place around each source on scales as large as 10-15~kpc (section~\ref{sec:Mol_vs_Lya}).
      \item SFR and Ly$\alpha$ resonant scattering from all the compact sources within the ELAN may contribute up to at least $\sim30$\% of the total luminosity of the ELAN, with the fraction of Ly$\alpha$ scattered photons from the quasar being a large incognita. Future radiative transfer calculations with high-resolution cosmological simulations of similar massive systems may shed light on the powering of ELANe (section~\ref{sec:powering}).
      \item No large-scale molecular reservoir is found as traced by CO(5-4) down to $\Sigma_{H2}<19$~M$_{\odot}$~pc$^2$, confirming that high $J$-transitions trace highly excited CO gas on kiloparsecs scales (section~\ref{sec:extMol}). Observations at lower $J$ transitions or additional tracers are needed to unveil extended molecular reservoirs in ELANe (if any).
  \end{enumerate}

Overall these observations confirm the richness of information encoded in ELAN systems. These rare objects can be used as laboratories to study several open questions regarding the high redshift universe, from the angular momentum accretion onto galaxies, to gas infall from large-scales to quasar scales, to the cool gas phase and its coexistence with a warm-hot phase. Current and future top-notch facilities (e.g., BlueMUSE, \citealt{Richard2019}; JWST, \citealt{Gardner2006}) will allow us to address in increasing details the astrophysics of these massive systems, and ultimately pin down the physics regulating their baryons flow and galaxy evolution.

\begin{acknowledgements}
{\small We thank Ian Smail and Francesco Valentino for providing comments on an early version of this work.
CCC acknowledges support from the Ministry of Science and Technology of Taiwan (MoST 109-2112-M-001-016-MY3). H.B.L. is supported by the Ministry of Science and Technology (MoST) of Taiwan (Grant Nos. 108-2112-M-001-002-MY3).
Y.Y. was supported by Basic Science Research Program through the National Research Foundation of Korea (NRF) funded by the Ministry of Science, ICT \& Future Planning (NRF-2019R1A2C4069803).
A.M. is supported by a Dunlap Fellowship at the Dunlap Institute for Astronomy \& Astrophysics, funded through an endowment established by the David Dunlap 
family and the University of Toronto. The University of Toronto operates on the traditional land of the Huron-Wendat, the Seneca, and most recently, 
the Mississaugas of the Credit River. A.M. is grateful to have the opportunity to work on this land.
AO is funded by the Deutsche Forschungsgemeinschaft (DFG, German  Research Foundation) –- 443044596.
This project has received funding from the European Research Council
(ERC) under the European Union Horizon 2020 research and innovation
programme (MagneticYSOs project, grant agreement No 679937).
This project has received funding from the European Research Council (ERC) under the European Union's Horizon 2020 research and innovation programme (grant agreement No 757535). This work has been supported by Fondazione Cariplo, grant No 2018-2329.
Based on observations collected at the European Organisation for Astronomical Research in the Southern Hemisphere under ESO programmes 094.A-0585(A), 096.A- 0937(A), 098.A-0828(B), and 0102.C-0589(D).
This paper makes use of the following ALMA data: ADS/JAO.ALMA\#2017.1.00560.S. ALMA is a partnership of ESO (representing its member states), NSF (USA) and NINS (Japan), together with NRC (Canada), MOST and ASIAA (Taiwan), and KASI (Republic of Korea), in cooperation with the Republic of Chile. The Joint ALMA Observatory is operated by ESO, AUI/NRAO and NAOJ.
The Submillimeter Array is a joint project between the Smithsonian Astrophysical Observatory and the Academia Sinica Institute of Astronomy and Astrophysics and is funded by the Smithsonian Institution and the Academia Sinica.
The James Clerk Maxwell Telescope is operated by the East Asian Observatory on behalf of The National Astronomical Observatory of Japan; Academia Sinica Institute of Astronomy and Astrophysics; the Korea Astronomy and Space Science Institute; Center for Astronomical Mega-Science (as well as the National Key R\&D Program of China with No. 2017YFA0402700). Additional funding support is provided by the Science and Technology Facilities Council of the United Kingdom and participating universities and organizations in the United Kingdom and Canada. Additional funds for the construction of SCUBA-2 were provided by the Canada Foundation for Innovation. 

The authors wish to recognize and acknowledge the very significant cultural role and reverence that the summit of Mauna Kea has always had within the indigenous Hawaiian community. We are most fortunate to have the opportunity to conduct observations from this mountain.}
\end{acknowledgements}

\facilities{ALMA, APEX(LABOCA), JCMT(SCUBA-2), SMA, VLT(MUSE, HAWK-I)}

\software{astropy \citep{Astropy2013,Astropy2018}}

\bibliographystyle{AASJournal} 
\bibliography{allrefs}

\begin{appendix} 

\section{The APEX/LABOCA and JCMT/SCUBA-2 maps}
\label{app:LJmaps}

\begin{figure}
\centering
\includegraphics[width=0.5\columnwidth]{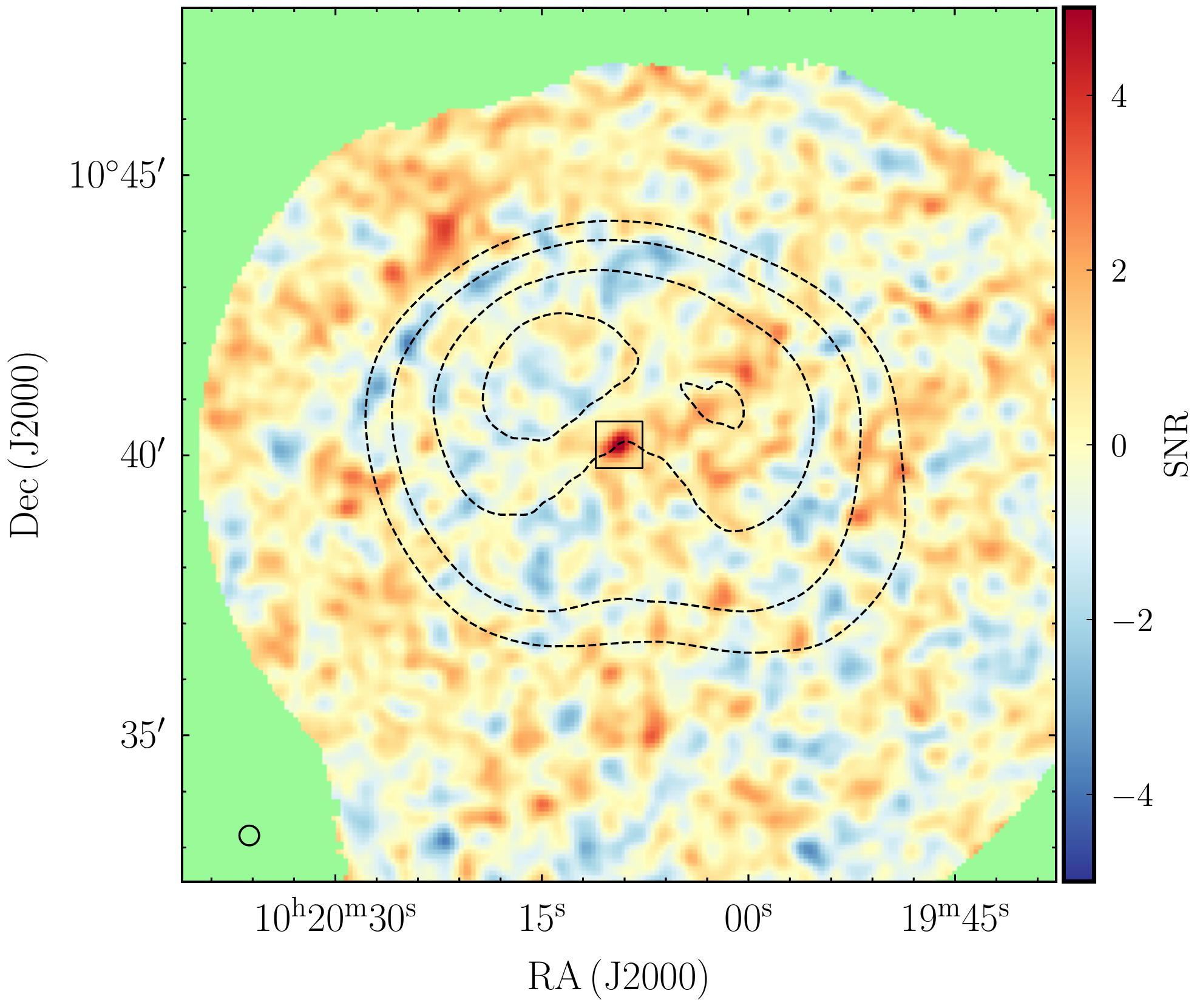}
\caption{870~$\mu$m LABOCA S/N map of the field around the ELAN. Dashed contours indicate the noise at 4.0, 3.5, 3.0, and 2.7 mJy~beam$^{-1}$, from the outer to the inner portion of the map. The black box shows the location of the ELAN and has the size of the cutout shown in Figure~\ref{continuum_detections}. The region within a noise of 4 mJy~beam$^{-1}$ represents a field of 68~arcmin$^2$ (or $\sim 14$~Mpc$^2$). In the green region there are no data. The main beam of LABOCA is shown in the bottom left corner.} 
\label{LABOCA_detection}
\end{figure}

\begin{figure}
\centering
\includegraphics[width=0.5\columnwidth]{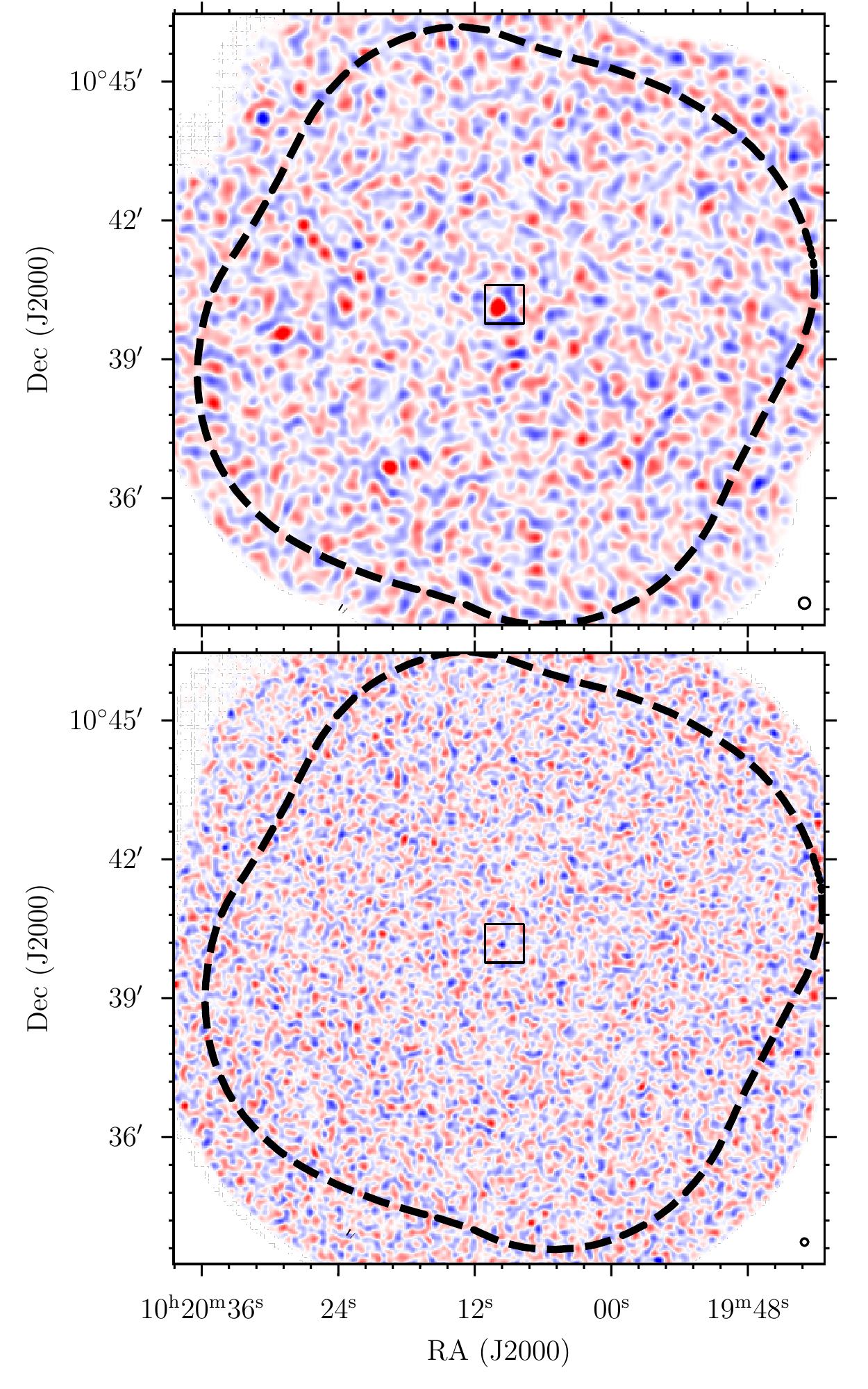}
\caption{SCUBA-2 S/N maps at $850$~$\mu$m (\emph{top panel}) and $450$~$\mu$m (\emph{bottom panel}) for the field around the ELAN. The black square indicates the field of view of Figure~\ref{continuum_detections}.
The maps are shown with a linear scale from -5 (blue) to 5 (red). For both fields, we indicate the noise contour (black dashed) for 3$\times$ the central noise and the effective beam of the observations in the bottom right corner.} 
\label{JCMT_detection}
\end{figure}

In this appendix we show, for completeness, the APEX/LABOCA and JCMT/SCUBA-2 maps for the full area covered by the observations (Figures~\ref{LABOCA_detection},\ref{JCMT_detection}). In this work we focus only on the ELAN location (the black box in each map) and we defer to an other paper of this series (Nowotka et al. 2021) the characterization of the ELAN large-scale environment (e.g., \citealt{FAB2018b}). 

\section{Flux deboosting for JCMT/SCUBA-2 and APEX/LABOCA observations}
\label{app:LABOCADeboosting}

\begin{figure}
\centering
\includegraphics[width=0.5\columnwidth]{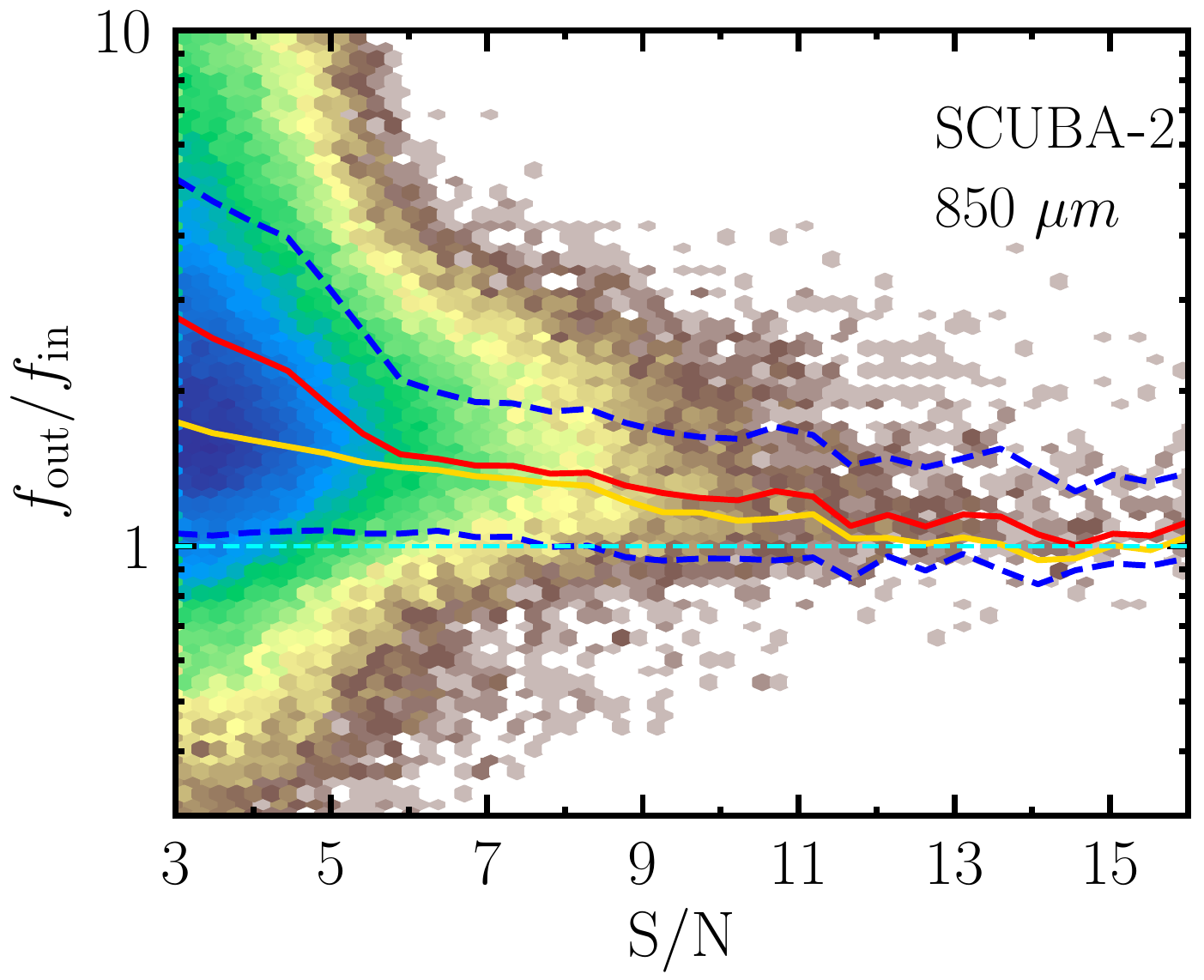}
\caption{The ratio between the output and the input flux densities of mock sources as a function of the input S/N for the SCUBA-2 850~$\mu$m observations.
The data-points obtained from 1500 realizations are shown in 2D hexagonal bins (high density of points in blue; low density in brown).  
We show the mean (red) and the median (yellow) values of the flux ratio. The blue dashed curves enclose the $1\sigma$ range relative to the mean curve. To help guiding the eye, the cyan line indicates the ratio $f_{\rm out}/f_{\rm in}=1$. In this work we correct the flux densitiy of the detected source at the ELAN position using the mean curve (see Table~\ref{tab:SingleDish}).} 
\label{SCUBA2_fluxBoosting}
\end{figure}

\begin{figure}
\centering
\includegraphics[width=0.5\columnwidth]{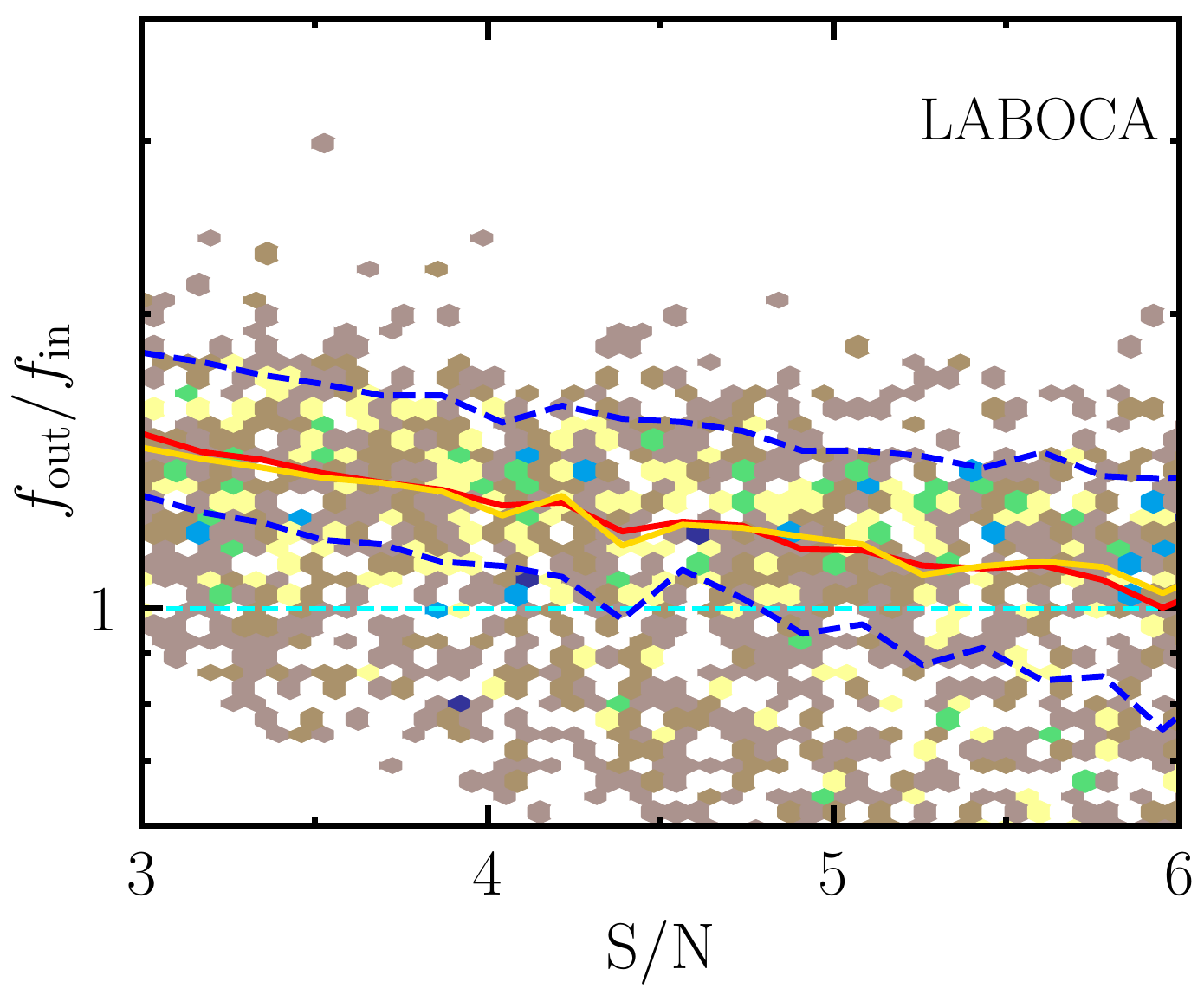}
\caption{The ratio between the output and the input flux densities of mock sources as a function of the input S/N for the LABOCA observations. 
The data-points obtained from 1500 realizations are shown in 2D hexagonal bins (high density of points in blue; low density in brown).
We show the mean (red) and the median (yellow) values of the flux ratio. The blue dashed curves enclose the $1\sigma$ range relative to the mean curve. To help guiding the eye, the cyan line indicates the ratio $f_{\rm out}/f_{\rm in}=1$. In this work we correct the flux densities of the detected source at the ELAN position using the mean curve (see Table~\ref{tab:SingleDish}).} 
\label{LABOCA_fluxBoosting}
\end{figure}

The flux densities of detections at low S/N in sub-millimeter observations are usually boosted due to the presence of noise fluctuations (e.g., \citealt{Eales2000, Coppin2006, Simpson2015}).
We quantify the flux boosting affecting the SCUBA-2 and LABOCA data by comparing the fluxes of injected mock sources with their recovered fluxes. Specifically, we proceed as follows. 
First, we created jackknife maps by inverting half of the scans during the coadding, keeping all processing steps as for the normal data reduction. Being thus free of any astronomical signal, these jackknife maps serve as noise maps.
For each instrument, we then created 1500 mock maps by injecting sources in the respective jackknife map, assuming a broken power-law for the counts with parameter values as in \citet{FAB2018b}. We then extracted the mock sources with a similar algorithm as in Section~\ref{sec:SMAcont}, but using the psf of the 850$\mu$m/SCUBA-2 (e.g., \citealt{TC2013b}) and LABOCA (e.g., \citealt{Weiss2009}) instruments. Figure~\ref{SCUBA2_fluxBoosting}, and \ref{LABOCA_fluxBoosting} show the results for SCUBA-2 and LABOCA, respectively. We note that we have a much larger number of detected sources in the SCUBA-2 850~$\mu$m maps because of their better sensitivity.

At the S/N of the detected emission at the ELAN position we found an average flux boosting of 1.09 and 1.19, respectively for SCUBA-2 and LABOCA. These values are in agreement within uncertainties with previous estimates of flux boosting for these instruments, 1.05 (\citealt{TC2013a}) and 1.13 (\citealt{Weiss2009}), respectively for SCUBA-2 at 850~$\mu$m and LABOCA. We use the found average values to correct for this effect (Table~\ref{tab:SingleDish}).

\section{Flux deboosting for SMA continuum sources}
\label{app:SMADeboosting}

\begin{figure}
\centering
\includegraphics[width=0.5\columnwidth]{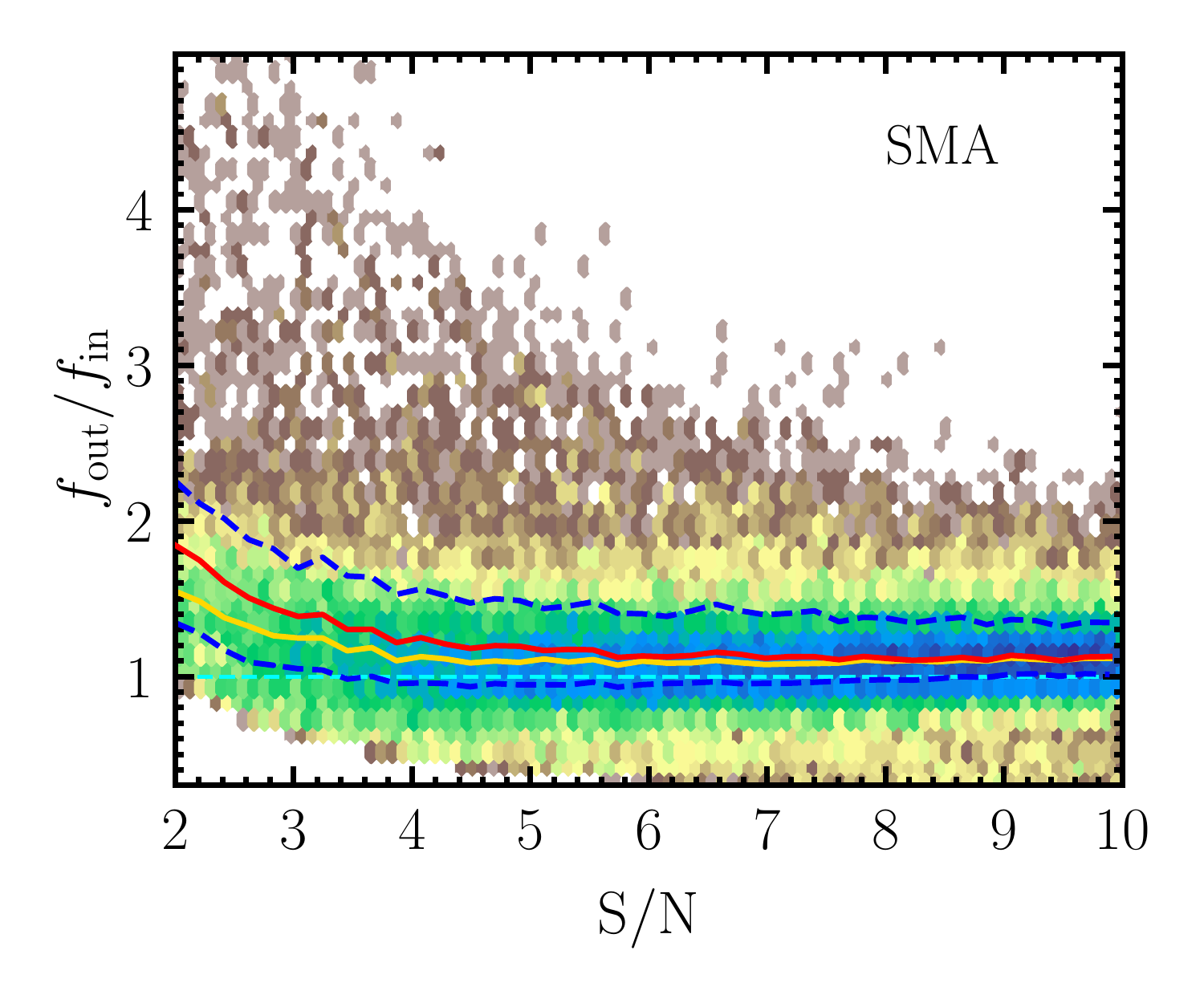}
\caption{The ratio between the output and the input flux densities of mock sources as a function of the input S/N for the SMA observations. 
The data-points obtained from 5000 realizations are shown in 2D hexagonal bins (high density of points in blue; low density in brown).
We show the mean (red) and the median (yellow) values of the flux ratio. The blue dashed curves enclose the $1\sigma$ range relative to the mean curve. To help guiding the eye, the cyan line indicates the ratio $f_{\rm out}/f_{\rm in}=1$. In this work we correct the flux densities of the detected sources using the mean curve (see Table~\ref{tab:Interf}).} 
\label{SMA_fluxBoosting}
\end{figure}

To quantify the level of flux boosting in the SMA data we proceeded as follows.
First, we constructed a Jackknife map, i.e., a noise map, as described in Section~\ref{sec:SMA}.
We then created 5000 SMA observations by injecting in each map 10 mock point sources with uniformly distributed random fluxes between 1 and 10 times the noise level.
The sources are introduced at random locations within an area equal to the primary beam of the SMA observations.
We then extracted sources from these 5000 realizations by using the same algorithm used for the detection of sources in Section~\ref{sec:SMAcont}. 
A source is considered to be recovered if it is detected with S/N$\geq 2$ and within one beam width from the position of injection.
The recovered and input flux densities are then compared to constrain the flux boosting at different input S/N. Figure~\ref{SMA_fluxBoosting} shows the results of this comparison.
Sources with input S/N$=2$ (5) are boosted on average (red curve) by 84\% (20\%), while the flux boosting is only about 10\% for sources at S/N$>7$.
We corrected the flux densities in our SMA catalogue based on the average curve shown in Figure~\ref{SMA_fluxBoosting}.

\section{Flux deboosting for ALMA continuum sources}
\label{app:ALMADeboosting}

\begin{figure}
\centering
\includegraphics[width=0.5\columnwidth]{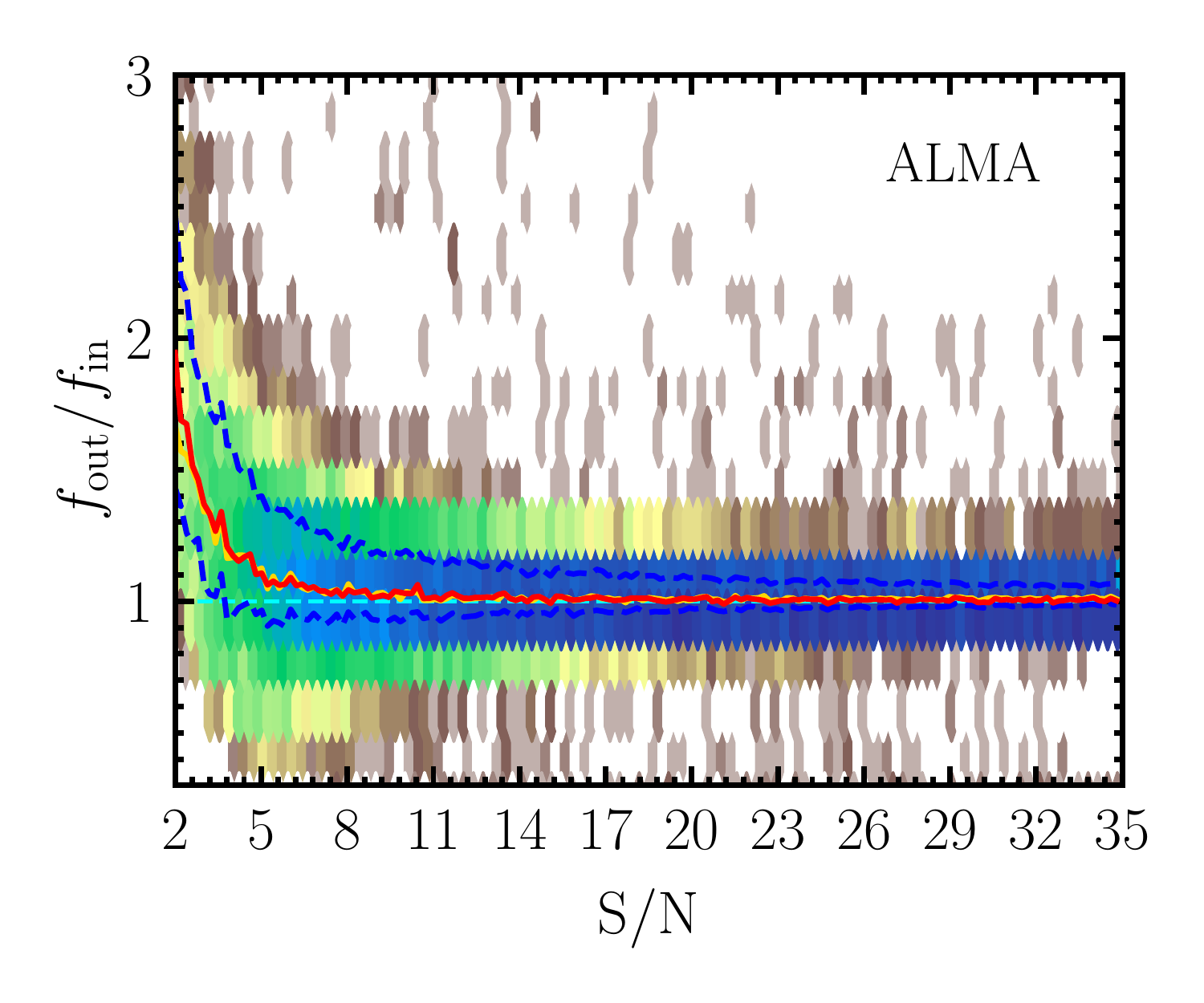}
\caption{The ratio between the output and the input flux densities of mock sources as a function of the input S/N for the ALMA continuum observations. 
The data-points obtained from 5000 realizations are shown in 2D hexagonal bins (high density of points in blue; low density in brown).
We show the mean (red) and the median (yellow) values of the flux ratio. The blue dashed curves enclose the $1\sigma$ range relative to the mean curve. To help guiding the eye, the cyan line indicates the ratio $f_{\rm out}/f_{\rm in}=1$. In this work we correct the flux densities of the detected sources using the mean curve (see Table~\ref{tab:Interf}).} 
\label{ALMA_fluxBoosting}
\end{figure}

We quantify the flux boosting for the ALMA data following exactly the same approach as for the SMA data, though using the algorithm for detection outlined in Section~\ref{sec:ALMAcont}, and the noise map obtained 
in Section~\ref{sec:ALMA}. 
Figure~\ref{ALMA_fluxBoosting} shows the ratio between the recovered and input flux densities.
We find that sources with input S/N$=4$ are boosted on average (red curve) by 17\%, while the flux boosting becomes negligible for sources with S/N$>10$.
We used the average curve shown in Figure~\ref{ALMA_fluxBoosting} to correct the flux densities in our ALMA catalogue. 
We note that similar corrections are found in the literature in number counts studies conducted with ALMA, though at different wavelengths (e.g., \citealt{Simpson2015, Oteo2016a}). 

\section{Data-points for the spectral energy distributions of CO(5-4) detected sources}
\label{app:dataSED}

In table~\ref{tab:app_phot} we list for completeness the photometric data obtained from the literature and used in the SED fitting for QSO, QSO2, AGN1, and LAE1. 

\begin{table*}
\begin{center}
\caption{Data obtained from the literature for the SED fitting of QSO, QSO2, AGN1, and LAE1 (all units in mJy)$^{\rm a}$.}
\scalebox{1}{
\footnotesize
\setlength\tabcolsep{4pt}
\begin{tabular}{lcccccccc}
\hline
\hline\\[-2mm]
ID	&	$J$	&  $H$ & $K_s$  & W1      &   W2     &    W3    &     W4     &  1.4~GHz \\
\hline\\[-2mm]				       
QSO	&     $0.35\pm0.05$          &  $0.66\pm0.07$ & $0.50\pm0.08$  & $0.60\pm0.02$      &   $0.68\pm0.02$    &    $2.47\pm0.14$    &     $4.70\pm0.86$    & $<0.4$ \\
QSO2	&   	 -       &  - & -  & $<0.04$      &   $<0.09$     &    $<0.7$    &     $<6.9$     &  $<0.4$ \\
AGN1	&        -       &  - & -  & $<0.04$     &   $<0.09$     &    $<0.7$    &     $<6.9$     &  $<0.4$ \\
LAE1	&        -       &  - & -  & $<0.04$      &   $<0.09$     &    $<0.7$    &     $<6.9$     &  $<0.4$ \\
\hline
\hline
\end{tabular}
}
\flushleft{\scriptsize $^{\rm a}$ The data are from the following works: $J$, $H$, $K_s$ from 2MASS all-sky point source catalog (\citealt{Skrutskie2006}); W1,W2,W3,W4 from AllWISE Source Catalog
(\url{https://wise2.ipac.caltech.edu/docs/release/allwise/}); 1.4~GHz from VLA FIRST (\citealt{Becker1994}).}\\  
\label{tab:app_phot}
\end{center}
\end{table*}

\end{appendix}

\end{document}